\documentclass[reprint,aps,showpacs,pra,nofootinbib,superscriptaddress]{revtex4-2}
\usepackage{color}
\usepackage[dvipsnames]{xcolor}
\usepackage{newlfont}
\usepackage{graphicx}
\usepackage{amssymb}
\usepackage{amsmath}
\usepackage[caption=false]{subfig}
\usepackage{bm}
\usepackage{verbatim}
\usepackage[latin1]{inputenc}
\usepackage{bbm}
\usepackage{multirow}
\usepackage[thinlines]{easytable}
\usepackage{braket}
\usepackage{amsthm}
\usepackage{bbold}
\usepackage{subfig}
\usepackage{mathrsfs}
\usepackage[varg]{txfonts} 
\usepackage{amsfonts}
\usepackage{newtxtext}

\newcommand{\ba}{\begin{array}}
	\newcommand{\ea}{\end{array}}

\newcommand{\mc}{\mathcal}

\newcommand{\mb}{\mathbb}

\newcommand{\bift}[1]{\big( #1\big)}

\newcommand{\Bitd}[1]{\Big[ #1\Big]}
\newcommand{\BBtd}[1]{\Bigg[ #1\Bigg]}
\newcommand{\Bift}[1]{\Big( #1\Big)}
\newcommand{\BBft}[1]{\Bigg( #1\Bigg)}

\newcommand{\mbr}{\mathbf{r}}

\newcommand{\mh}{\mathcal{H}}

\newcommand{\td}{\text{d}}

\newcommand{\mk}{\mathcal{K}}
\newcommand{\ml}{\mathcal{L}}
\newcommand{\tiga}{\tilde{\gamma}}
\newcommand{\bg}{\bar{g}}

\usepackage[pagebackref=false, colorlinks=true]{hyperref}
\definecolor{redish}{rgb}{0.7,0.2,0.0}  
\definecolor{bluish}{rgb}{0.2,0.5,0.8}
\hypersetup{linkcolor=red,          
	citecolor=teal,        
	filecolor=magenta,      
	urlcolor=teal}          

\begin{document}
	\title{Geometry of quantum states in random matrix ensembles and the chaos-integrability transition}
\author{Ankit Gill}\email{ankitgill20@iitk.ac.in}
 \affiliation{Department of Physics, Indian Institute of Technology, Kanpur 208016, India}
\author{Keun-Young Kim}\email{fortoe@gist.ac.kr}
\affiliation{Department of Physics and Photon Science, Gwangju Institute of Science and Technology, 123 Cheomdan-gwagiro, Gwangju 61005, Republic of Korea}\affiliation{Research Center for Photon Science Technology, Gwangju Institute of Science and Technology, 123 Cheomdan-gwagiro, Gwangju 61005, Republic of Korea}

\author{Kunal Pal}  \email{kunal.pal@apctp.org} 
\affiliation{Asia Pacific Center for Theoretical Physics, Pohang 37673, Republic of Korea}
\author{Kuntal Pal}\email{kuntalpal@gist.ac.kr}
\affiliation{Department of Physics and Photon Science, Gwangju Institute of Science and Technology, 123 Cheomdan-gwagiro, Gwangju 61005, Republic of Korea}

\begin{abstract}
We consider the geometry of quantum states associated with random matrix Hamiltonians belonging to ensembles that exhibit an integrable-to-chaotic transition in terms of the nearest-neighbour energy level spacing distribution, focusing, specifically, on the $\beta$-Gaussian ensembles with generic Dyson index. In the case that the total Hamiltonian contains a single parameter, the distance between two states is captured by the fidelity susceptibility, whereas, when the total Hamiltonian contains multiple parameters, this distance is measured in terms of the quantum metric tensor. For the tridiagonal Gaussian $\beta$-ensemble, which shows chaos-to-integrability transition with varying Dyson index, we first calculate an analytical expression for the ensemble-averaged fidelity susceptibility for a two-by-two matrix representation of the Hamiltonian for generic values of the Dyson index, and show that it diverges as the total Hamiltonian, which is the sum of a diagonal matrix with independent elements and a tridiagonal $\beta$-matrix, goes over to the integrable phase. Next, for large-dimensional matrices, we numerically compute the fidelity susceptibility and the quantum metric tensor components, respectively, for a one-parameter and a two-parameter class of Hamiltonians that we construct from the $\beta$-ensemble,  and find scaling relations of these quantities for chaotic and integrable phases. We also consider other variations of the $\beta$-ensemble, such as the one that preserves the rotational invariance of the ensemble, and compute the relevant ensemble-averaged fidelity susceptibility to show that it has similar features as the tridiagonal ensemble in the integrable as well as the chaotic phase, thereby establishing the universality of these properties. Finally, the presence of non-vanishing non-diagonal elements, which arises due to the rotational non-invariance of the $\beta$-ensembles,  is a specific feature of the corresponding metric tensor, and we use this to illuminate the difference between the quantum state space geometry of these ensembles and the rotationally invariant ones, such as the classical Gaussian ensembles.
\end{abstract}
\maketitle
{\hypersetup{linkcolor=black}
\tableofcontents
}
\section{Introduction}
Geometric approaches have played an increasingly valuable role in quantum many-body physics, offering deep insights into the structure of quantum states beyond conventional symmetry-based classifications. The corresponding geometry of pure quantum states is naturally described on the complex projective Hilbert space, where the Hermitian inner product of the ambient Hilbert space induces both a Riemannian and a symplectic structure \cite{bengtsson2017geometry, chruscinski2004geometric}. Specifically, for quantum systems governed by a hermitian Hamiltonian, the real part of the quantum geometric tensor defines a real, symmetric metric that governs the infinitesimal distance between nearby states (the quantum or Fubini-Study metric), while its imaginary part gives rise to the Berry curvature, a real-valued antisymmetric two-form that encodes the underlying symplectic structure. These structures together endow the projective Hilbert space with the geometry of a K\"ahler manifold, capturing both the metric and topology \cite{Kibble1979, Ashtekar, Brodygeometric}. 

The quantum geometric tensor (QGT), which is a pullback of the Fubini-Study tensor on the parameter manifold, was introduced by Provost and Vallee and provides a unified framework for characterising both the geometric and topological aspects of quantum states in parameter space \cite{Provost1980riemannian}. Defined on the manifold of quantum states, the QGT decomposes into two components: the quantum metric tensor (QMT), which quantifies the distance between infinitesimally close states, and the Berry curvature, which governs geometric phases under adiabatic evolution \cite{Berry1984, Simon}. By capturing how quantum states change under smooth variations of parameters, geometry provides a natural language for exploring quantum phase transitions and topological phenomena.

In the context of quantum phase transitions, the QMT has emerged as a sensitive probe of criticality, capturing singular behaviour near critical points even when conventional order parameters are absent \cite{Zanardi:2007byf, you2007, venuti2007quantum, Kolodrubetz, TS1, Dey, Henriet, Gutierrez, Streleck, Alexandrov}. Meanwhile, the Berry curvature part of the QGT encodes the system's response to adiabatic perturbations and underlies a variety of topological invariants, such as Chern numbers, which, in the linear response theory formalism, can be attributed to the QGT contributions to the generalised susceptibilities and dynamical response functions, linking geometric properties of quantum states to measurable quantities \cite{Xiaorev, Ozawa, Mera21}.  
In this paper, we use the geometry of quantum states of a random-matrix Hamiltonian belonging to an ensemble of matrices to characterise the transition from integrability to chaos in these ensembles.

One of the most useful ways of understanding dynamical and spectral properties 
of quantum systems is to use random matrix theory to model their spectral statistics. The wide range of applications of random matrix theory in physics is due to the fact that it has many powerful mathematical tools that one can employ to study statistical properties of physical quantities associated with extremely complex quantum systems. The random matrices can describe universal properties that do not depend on the details of a given system, but rather are determined by a set of global symmetries that are commonly shared by all systems in a given symmetry class. 

The classical Gaussian random matrix ensembles studied by Wigner and Dyson \cite{dyson1962threefold} have been used to understand the characteristic properties, namely the level repulsion and rigidity of the energy spectrum 
of quantum systems having classically chaotic counterparts \cite{bohigas1984characterization, berry1985semiclassical}. 
On the other hand, in integrable systems, the energy eigenvalues are uncorrelated random numbers; hence, the distribution of the nearest neighbour level energy spacing follows the Poisson statistics \cite{berry77}. 

However, in many physical situations, quantum systems can consist of parts of different symmetries, or can have classically chaotic and integrable counterparts. If a parameter exists in such systems, such that changing it causes the system to make a transition from  
different symmetry classes, then these transitions would be reflected in the symmetry class of the corresponding random matrix. Therefore, in those cases, the quantum systems can also 
follow other level spacing statistics, which are intermediate between the Wigner-Dyson statistics of classical Gaussian ensembles and Poisson statistics of integrable systems. We refer to  \cite{bogomolny1999models, robnik1996energy, cheon1991signature, csordas1994transition, lenz1991reliability, shukla1997effect, rabsonPhysRevB.69.054403, PhysRevLett.80.1003, PhysRevLett.70.1089} for a collection of references dealing with the general topic of chaos-integrability transition and breaking of the symmetries in the classical Gaussian ensembles. 

The classical random matrix ensembles are defined with two conditions: 1. the matrix elements are sampled from independent distributions, and 2. the probability distribution ($P(H)$) of obtaining a particular instance of the matrix $H$ is invariant under a symmetry transformation. For example, in the case of the Gaussian unitary ensemble (GUE), the probability distribution is invariant under unitary transformation of $H$ \cite{Potters, Mehta}. Therefore, in order to capture the spectral statistics 
intermediate between the Wigner-Dyson and Poisson, one presumably needs to consider more general random matrix ensembles, where one of the above two constraints is relaxed. 

The Rosenzweig-Porter (RP) model, which was originally introduced to explain the properties of the spectrum of atoms, is one such random matrix ensemble where the 
assumption of invariance under a symmetry transformation is relaxed compared to the 
classical Gaussian ensembles by introducing a parameter that spoils the usual relation between the variance of the diagonal and off-diagonal elements \cite{rosenzweig1960repulsion, kravtsov2015random, cadevz2024rosenzweig, PhysRevB.109.024205, PhysRevB.106.094204}. By tuning this parameter suitably, one can obtain the Poisson and Wigner-Dyson statistics of the energy level spacing distribution as two limits of the RP model. 
In its most commonly used form, the RP model consists 
of a diagonal matrix whose elements are sampled from independent and identically distributed random variables (usually from Gaussian distributions with a zero mean), and a random matrix (usually drawn from one of the classical Gaussian ensembles) is added to the diagonal matrix as a perturbation term.\footnote{There appeared other generalisations of this particular form of the RP model in the literature. See, e.g., \cite{kravtsov2015random, von2019non, facoetti2016non}.} This form for the Hamiltonian of the RP model makes it particularly suitable for studying the geometry of quantum states, as the level spacing distribution makes a transition from Poissonian to Wigner-Dyson distribution. 

Another class of random matrix ensembles that exhibit the transition from integrability and chaos in terms of the level-spacing distribution is the so-called Gaussian $\beta$-ensembles, where the Dyson index $\beta$ can take any real positive value. A convenient 
sparse triadiagonal matrix form of these ensembles was constructed in \cite{dumitriu2002matrix}. This class of ensembles also breaks the rotational invariance of the classical Gaussian ensembles, and shows 
integrability to chaos transition when the Dyson index is varied \cite{buijsman2019random}. An important difference between these ensembles and the classical Gaussian $\beta$-ensemble lies in the nature of eigenstates. Namely, due to the rotational invariance of the classical Gaussian ensembles, the eigenstates of a Hamiltonian sampled from these are uniformly distributed on the surface of a multidimensional sphere and have mutually independent components in terms of a complete basis, whereas, for a Hamiltonian sampled from the  Gaussian $\beta$-ensembles, roughly, if the first row of the eigenvector matrix is specified, one can obtain the full expression for the eigenstates from the eigenvalues, and the elements of the tridiagonal form of the Hamiltonian \cite{dumitriu2002matrix}. 
In this paper, we use both of the above-mentioned generalised random matrix ensembles to analyse the geometry of quantum states associated with Hamiltonians drawn from them, with the broad goal of identifying generic features of the geometry of quantum states for quantum systems undergoing a chaos-to-integrability transition.

Before outlining the summary of the main results of this work, we mention that there exist several works in the literature that have studied different aspects of the geometry of quantum systems associated with random matrix Hamiltonians. For a Hamiltonian sampled from a given random matrix ensemble, the probability distribution function of the QGT was introduced in \cite{berry2020quantum}, and was studied from random matrices of small sizes to predict the correct asymptotic formula for this distribution. Ref. \cite{Penner:2020cxk} obtained the exact joint probability distribution function of the QGT for large random matrix Hamiltonians belonging to the Gaussian ensembles, and compared with the numerical results obtained for a tight-binding Anderson model. In a recent work, \cite{Sharipov:2024lah}, the authors derived exact analytical expressions for the components of the ensemble-averaged QMT for a two-parameter family of Gaussian random matrix Hamiltonians, and also numerically computed these components for deformed Gaussian ensembles, which is a generalisation of the RP random matrix Hamiltonian.

We begin in Section \ref{FS_QGT} by briefly reviewing the definitions of the fidelity susceptibility (FS), QGT, QMT and ensemble-averaged QMT, as well as their connection with the time correlation functions of physical operators. In this section, we derive a relation between the ensemble-averaged FS and a correlation function, suitably averaged over the ensemble.
In the central part of the paper (Sections \ref{sec_beta_metric} - \ref{sec_invt_beta_ensemb}), we focus on the 
geometry of eigenstates belonging to the tridiagonal Gaussian $\beta$-ensembles, which also show integrability to chaos transition with varying Dyson index. To the best of our knowledge, the geometry of quantum states of the Gaussian $\beta$-ensemble, in terms of the FS and the QMT, has not been explored in the literature before. To this end, first, by assuming that the total Hamiltonian is a sum of two terms,  a $2 \times2$ matrix belonging to this ensemble, which is added to a diagonal Hamiltonian,  we obtain an analytical expression for the  FS for any generic value of the Dyson index, and find its behaviour with respect to the change of the integrability parameter ($r$) (\S  \ref{ssec_grr_trid_der}). We show that the corresponding ensemble-averaged FS (EFS) indeed shares generic features with those of the generalised RP model (considered in \cite{Sharipov:2024lah}), and specifically, the EFS diverges in the limit when $r \rightarrow 0$ for any generic values of the Dyson index $\beta>1$ of the integrability-breaking term in the Hamiltonian. We also derive the scaling law $1/r^4$ for the EFS for large $r$ from the analytical expression, and verify this from numerical simulations for large dimensions. A semi-analytical explanation of this behaviour is presented in Section \ref{sec_EFS_larger}, where we utilise the relation between the EFS and a time-correlation function of observables. 

Next, to understand the finite size effects, we numerically compute the components of the QMT for finite large values of $N$, $N$ being the rank of the Hamiltonian, for a two-parameter family of Hamiltonians (one parameter measures the radial deformations,  and the other angular deformation) we introduce in \S \ref{ssec_beta_N}, where each component of the total Hamiltonian is drawn from the Gaussian $\beta$-ensembles, and can have different Dyson indices. From the numerical results, we obtain the scaling laws for the QMT components (here, due to the breakdown of the rotational invariance in the $\beta$-ensemble, there are non-diagonal components in the QMT) and show that these laws are true irrespective of the value of the Dyson indices of the tridiagonal ensembles from which the component operators of the total Hamiltonian are drawn.  These scaling laws also indicate that the metric becomes approximately degenerate in the specific limit where the radial deformation parameter becomes large. We find out the approximate direction on the geometry of quantum states along which the displacement vanishes in this limit, and illustrate the difference from the geometry corresponding to the GUE Hamiltonians considered in \cite{Sharipov:2024lah}.

One natural question one can ask next in the context is the following: What is the effect of the rotational invariance of an ensemble on the associated geometry of quantum states? Since the tridiagonal representation of the $\beta$-ensemble breaks the rotational invariance of the classical Gaussian ensembles, to address this question and hence, to complement the analysis of Section \ref{sec_beta_metric}, an invariant $2\times2$ random matrix ensemble with an effective Dyson index is also studied, and we show that the EFS still captures the essential signature of the chaos-to-integrability transition (see Section \ref{sec_invt_beta_ensemb}). Finally, in Section \ref{sec_level_curvature} we analyse the sensitivity of the eigenvalues, rather than the eigenstates, of these systems, by studying the ensemble-averaged curvature of the energy eigenvalues (which is the second derivative of the energy levels with respect to the parameters present in the Hamiltonian). Indeed, the sensitivity of the energy levels of a complex many-body quantum system to the parameters is one of the most important approaches developed in the past few decades towards understanding why the energy levels of chaotic quantum systems exhibit random matrix universality \cite{Pechukas, Yukawa, haakebook, stockmann2007quantum}. In this section, we connect the second and third derivatives of the energy levels with different spectral functions (which, to the best of our knowledge, have not been identified in the literature before) and compute the analytical expression for the energy level curvature for the tridiagonal $\beta$-ensemble.
Our main findings and conclusions are summarised in Section \ref{sec_conclusions}, where we also list a few interesting directions along which the results of the present manuscript can be extended. 

This paper also contains several appendices where we provide some important details that are left out of the main text, but are integral parts of the paper. Specifically, in Appendix \ref{qmt_generic}, we provide the expression for the QMT for a generic superposition of states, and in Appendix \ref{gue_qmt_derivation}, we have outlined the steps of the derivation of the components of the QGT for the GUE Hamiltonian. 
An analytical approximation for the density of states of the Hamiltonian introduced in \S \ref{ssec_rp_trid} based on a series expansion of the Cauchy transform is discussed in Appendix \ref{sec:DOS_approx}.
Finally, in Appendix \ref{sec:beta_ensm_haar},  the expression for the FS of a specific random matrix ensemble, whose eigenvalues are drawn from a Gaussian $\beta$-ensemble with generic $\beta$ and has Haar random eigenstates,  is obtained, and its properties are elaborated. 

\section{Fidelity susceptibility and Quantum metric tensor for random matrix ensembles}\label{FS_QGT}

In this section, we first review the definition of the \textit{fidelity susceptibility} (FS) associated with a change in one of the parameters of the Hamiltonian of a quantum system, and then discuss its generalisation when more than one of the Hamiltonian parameters are changed -  a situation where a metric structure naturally appears in the space of eigenstates. We also discuss a connection between the FS and the time-correlation function of physical operators, which will be useful later in the paper.

\subsection{Ensemble and spectral averaged fidelity susceptibility}\label{ssec_EFS}

To begin with, we consider the following generic scenario, where the total Hamiltonian of a quantum system is given by\footnote{We assume that the Hamiltonian $H$ has no degeneracy in the spectrum. }
\begin{equation}\label{Ham_Roz_Por1}
    H(r)=H_0+r ~\mathcal{H}~,
\end{equation}
with $r$ being a positive parameter. Now consider the change in the fidelity of the $n$th eigenstate $\ket{n(r)}$ of the Hamiltonian $H(r)$ in \eqref{Ham_Roz_Por1}, i.e., change in $\mathcal{F}(r, dr)=|\braket{n(r)|n(r+\delta r)}|^2$.\footnote{Some authors define the fidelity as the absolute value of the overlap between the eigenstate at a parameter value $r$, and at $r+\delta r$.} Expanding the fidelity
for small values of $\delta r$, and using the normalisation of the 
states $\ket{n(r)}$, we have, in the limit $\delta r \rightarrow 0$, the following expansion
\begin{equation}
    \mathcal{F}(r, \delta r \rightarrow 0) \simeq 1- (\delta r)^2 g_{rr}^n(r)~.
\end{equation}
Here $g_{rr}^n(r)$ denotes the  FS of the $n$-th eigenstates, and a straightforward calculation shows that its expression can be written as\footnote{In this paper, we use the convention that $\sum_{j(\neq n)}$ denotes sum over all $j$, expect $n$, and no sum over the index $n$. On the other hand,  $\sum_{j\neq n}$ denotes the sum over both $j$ and $n$, excluding the cases when $j=n$.} \cite{Zanardi:2007byf}
\begin{align}\label{fid_sus_rr}
    g_{rr}^n(r) &=\braket{\partial_r n(r)|\partial_r n(r)}-\braket{\partial_r n(r)| n(r)}\braket{n(r)|\partial_r n(r)}~~\nonumber\\
    &= \sum_{j(\neq n)} \frac{\mathcal{H}_{nj}\mathcal{H}_{jn}}{(E_j(r)-E_n(r))^2}~.
\end{align}
Here $\mathcal{H}_{nj}=\braket{n|\mathcal{H}|j}$ denotes the matrix elements of $\mathcal{H}$ in the basis of $\ket{n}$, the eigenstate of $H$. Note that this quantity is invariant under an overall constant scaling transformation of the Hamiltonian $H(r)$.

Summing over the expressions for the FS $g_{rr}^n(r)$ for all the energy eigenstates, we define an \textit{spectral-averaged FS} as follows 
\begin{align}\label{av_fid_sus_rr}
    g_{rr}(r) := \frac{1}{N}\sum_{n=1}^{N} g_{rr}^n(r)~=\frac{1}{N}\sum_{j\neq n} \frac{\mathcal{H}_{nj}\mathcal{H}_{jn}}{(E_j(r)-E_n(r))^2}~,
\end{align}
where $N$ denotes the dimension of the Hilbert space of the Hamiltonian $H$. Summing over the FS of all the eigenstates in the spectrum removes the strongly varying nature of $g_{rr}^n(r)$, which fluctuates irregularly when $n$ is varied \cite{Skvortsov}. Note that this quantity may not have any interpretation as the FS of a single generic pure state, which is a linear combination of the energy eigenstates. 

Now we assume that $H_0$ and $\mathcal{H}$ are independent random matrices of rank $N$, so that their joint distribution is given by 
\begin{align}
    P(H_0,\mh) = \frac{1}{\mathcal{Z}_{\beta,N}} \exp\bigg[-\frac{\beta N}{4} \text{Tr}~\bigg(V_0(H_0)+\mathcal{V}(\mh)\bigg)\bigg]~,
\end{align}
where $\mathcal{Z}_{\beta,N}$ is a normalisation constant whose form also depends
on the expressions for the potentials $V_0(H_0)$ and $\mathcal{V}(\mh)$, and $\beta$ is the Dyson index of the ensemble.
For later purposes, here we note that the well-known expression for the joint probability distribution (JPD)  of the eigenvalues ($E_i$) of a rationally invariant $N \times N$ random matrix ensemble with a confining potential $V(H)=\sum_n \alpha^n \text{Tr} (H^n)$ is given by \cite{Mehta, Potters}
\begin{equation}
\label{joint_eig_dist}
    P(E_1, \cdots E_N) = \frac{1}{Z(\beta,N)} e^{-\frac{\beta N}{4}\sum_n \alpha_n\sum_{i=1}^N E_i^n} \prod_{j<k}\big|E_j-E_k\big|^{\beta}~,
\end{equation}
where $Z(\beta,N)$ is a normalization constant, and $\alpha_n$s are a set of real constants. 
In this paper, we shall compute the the ensemble-averaged FS (EFS) which we define as the average of $g_{rr}(r)$ in \eqref{av_fid_sus_rr} over the joint distribution of $H_0$ and $\mh_1$ as\footnote{Since, in this paper we always consider random matrix Hamiltonians, all the relevant quantities are obtained with an average over the ensemble (denoted by the symbol $\mb{E}$). Hence, we often omit explicitly referring to the quantifier `ensemble-averaged'.} 
\begin{widetext}
\begin{equation}\label{en_avg_FS}
\begin{split}
    \bar{g}_{rr}(r) := \mathbb{E}\big(g_{rr}(r)\big) 
    = \frac{1}{N}\sum_{j \neq n} \int \frac{\mathcal{H}_{nj}(r)\mathcal{H}_{jn}(r)}{(E_j(r)-E_n(r))^2} P(H_0) 
    P(\mh) ~ \td H_0 ~ \td \mh~.
    \end{split}
\end{equation} 
\end{widetext}
Here $P(H_0)$ and $P(\mathcal{H})$ are the distributions for the independent random matrices $H_0$ and $\mathcal{H}$, respectively. In the next section, we consider Gaussian random matrices, for which the potential function is $V(H)=H^2$.\footnote{For a discussion on the distribution of the FS for the Gaussian random matrix ensembles and its comparison with that of the disorder many-body systems, see \cite{Sierant1, Maksymov}.}

\subsection{Fidelity susceptibility from correlation functions}\label{sec_fs_cor}

An important property of the FS is that it can be connected to the time correlation function
of physical operators. To see this explicitly,  first consider the following trick to relate the  Kallen-Lehmann representation of an operator to its unequal time correlation function, 
\begin{equation}\label{E_to_Delta}
    \frac{1}{(E_j(r)-E_n(r))^2} = \int_{-\infty}^{\infty}~\frac{\td \omega}{\omega^2} \int_{-\infty}^{\infty}~\frac{\td t}{2 \pi}~ e^{-i (E_j(r)-E_n(r)-\omega)t}~.
\end{equation}
Using this relation, we can write the FS in eq. \eqref{fid_sus_rr} as \cite{ hauke2016measuring, kolodrubetz2017geometry}
\begin{equation}\label{FS_to_S}
    g_{rr}^n(r) = \int_{-\infty}^{\infty}~\frac{\td \omega}{\omega^2} ~\mathcal{S}_n(\omega) ~,
\end{equation}
where we have defined\footnote{The standard definition of this function in literature usually does not have the $1/ 2\pi$ factor. Here we have modified the standard definition so that the subsequent formulas for the FS and the moments are free from the $2\pi$ factor. }
\begin{equation}\label{S_n_omega}
\begin{split}
    \mathcal{S}_n(\omega) = \frac{1}{2\pi}\int_{-\infty}^{\infty}~\td t~ e^{i \omega t}~ \Big(\braket{n|\mh (t)\mh|n}-|\braket{n|\mh|n}|^2\Big)~,\\
    \end{split}
\end{equation}
with $\mh(t) = e^{i H t} \mh e^{-i H t}$. This function has the following spectral representation,
\begin{equation}
    \mathcal{S}_n(\omega) = \sum_{j (\neq n)} |\mathcal{H}_{nj}|^2 ~\delta\big(E_j-E_n-\omega\big)~,
\end{equation}
and has the moments
\begin{equation}\label{moments_S_n}
    M^{(n)}_k(r)=\int_{-\infty}^{\infty} ~\td \omega~\omega^k ~\mathcal{S}_n(\omega) =
    \sum_{j(\neq n)}  |\mathcal{H}_{nj}|^2~\big(E_j(r)-E_n(r)\big)^k~,~~k \in \mathbb{Z}~.
\end{equation}
These moments have concrete physical meaning: e.g., for $k=-2$, the moment $M_{-2}$ is just the FS of the $n$-th eigenstate \cite{you2015generalized}.  The meaning of the moment with $k=-1$ will be discussed in Section \ref{sec_level_curvature}. 

Now we work out the relation between the EFS and an ensemble-averaged correlation function.  Summing $\bar{g}^n_{rr}(r)$s over all the eigenstates $\ket{n}$ and averaging over the random matrix ensemble, in the case $H(r)$ belongs to one, we have the following relation\footnote{In this context, we note that for a specific perturbation term where $\partial_r H(r)$ is proportional to a local density operator, Ref. \cite{Skvortsov} obtained a similar expression for the FS averaged over all eigenstates. We also refer to this work for a thorough discussion of the scaling of FS in different phases of the RP model and its different variants.}
\begin{equation}\label{FS_to_correl}
    \bar{g}_{rr}(r) = \int_{-\infty}^{\infty}~\frac{\td \omega}{\omega^2} ~\mathcal{S}(\omega) ~,
\end{equation}
where we have defined the spectral function (in an appropriate ensemble-averaged sense)
\begin{equation}\label{S_omega}
    \mathcal{S}(\omega) = \frac{1}{2\pi}\int_{-\infty}^{\infty}~\td t ~e^{i \omega t}~ \Big(\braket{\mh (t)\mh}-\frac{1}{N} \mathbb{E}\sum_n|\braket{n|\mh|n}|^2\Big)~,
\end{equation}
with the symbol $\mathbb{E}$ denoting an ensemble average over the probability distribution of the components of the Hamiltonian; and we have used the following notation to denote the map which combines the trace and the ensemble average,
\begin{equation}
    \braket{\mathcal{O}}=\varphi_{N}(\mathcal{O})= \frac{1}{N} \mathbb{E}~\Big(\text{Tr}(\mathcal{O})\Big)~.
\end{equation}
We continue to use this notation in the rest of the paper.  This map is also known as the \textit{normalised expected trace} in the mathematics literature, and appears naturally in the context of free probability theory approaches to random matrices \cite{mingo2017free, nica2006lectures, voiculescu1992free}. For a fixed realisation of the Hamiltonian (i.e., without the ensemble-average), it can be thought of as the inner product with respect to the infinite-temperature thermal state.

The relation in \eqref{FS_to_correl} is the generalisation of \eqref{FS_to_S} for EFS, and indicates that an alternative way of obtaining the EFS would be to calculate the following ensemble-averaged two-point correlation function of the operator $\mh$, 
\begin{equation}\label{correlation_1}
   G(t)= \braket{\mh(t) \mh}=\varphi_{N}(\mh(t) \mh)= \frac{1}{N} \mathbb{E}~\Big(\text{Tr}(\mh(t) \mh)\Big)~,
\end{equation}
and use it in the eq. \eqref{FS_to_correl} along with \eqref{S_omega}. As an example, when $H(r)$ belongs to one of the classical Gaussian ensembles, it is possible to explicitly evaluate this correlation function in terms of the spectral form factor, and subsequently derive the expression for the EFS by employing the last relation \cite{Gill:2025upp, Gill:inprep}.

Note that  we can rewrite the expression for the spectral function $\mathcal{S}(\omega)$ as 
\begin{equation}
    \mathcal{S}(\omega) = \frac{1}{N} \mathbb{E}~\sum_{n \neq j} |\mathcal{H}_{nj}|^2 ~\delta\big(E_j-E_n-\omega\big)~.
\end{equation}
Considering the moments of this function,
\begin{equation}\label{moments_S}
    M_k(r)=\int_{-\infty}^{\infty} ~\td \omega~\omega^k ~\mathcal{S}(\omega) = \frac{1}{N} ~\mathbb{E}~
    \sum_{n \neq j}  |\mathcal{H}_{nj}|^2~\big(E_j(r)-E_n(r)\big)^k~,
\end{equation}
$k$ being integers, we see that these also have direct physical meanings. For example, $M_{-2}(r)$ is nothing but the EFS. 
We also note that the function $\mathcal{S}(\omega)$  is the frequency space expression of the connected correlation function, and is related to the dynamical susceptibility of the system through the Kubo linear response formula.

\subsection{Ensemble-averaged quantum metric tensor}
The expressions for the FS discussed so far can be generalised when the system Hamiltonian depends on more 
than one parameter $r^1, \cdots, r^k$, which we collectively denote as $\mbr$. 
In this case, we start from the so-called Fubini-Study distance 
for two very close-by states, $\ket{n}$ and $\ket{n+\delta n}$, given by \cite{Provost1980riemannian, Zanardi:2007byf, bengtsson2017geometry}
\begin{equation}\label{FS_line}
    d s^2:= d^2_{\text{FS}} (\ket{n}, \ket{n+\delta n})= \braket{\delta n|\delta n} - \braket{n|\delta n}\braket{ \delta n|n}~.
\end{equation}
The metric induced on the parameter manifold can be obtained by writing $\ket{\delta n} = \partial_a \ket{n(\mbr)} \td r^a $ (with the notation $\partial_a=\partial_{r^a}$), and using it in eq. \eqref{FS_line} we have 
\begin{equation}
    d s^2 =  g_{ab}^n(\mbr)~\text{d}r^a \text{d}r^b~,
\end{equation}
where $g_{ab}^n(\mbr)$, known as the QMT, is the real and symmetric part of the QGT of the $n$th eigenstate, defined as \cite{Provost1980riemannian, Zanardi:2007byf}
\begin{align}
        G^n_{ab} (\mathbf{r}) &= \braket{\partial_a n(\mbr)|\partial_b n(\mbr)} - \braket{\partial_a n(\mbr)|n(\mbr)}\braket{ n(\mbr)|\partial_b n(\mbr)} \nonumber\\
        &=\sum_{m(\neq n)}\frac{\braket{n(\mbr)|\partial_a H(\mathbf{r})|m(\mbr)}
    \braket{m(\mbr)|\partial_b H(\mathbf{r})|n(\mbr)}}{\big(E_m(\mbr)-E_n(\mbr))^2}~.
\end{align}
The second expression above can be obtained by using the eigenvalue equation for the $n$th eigenstate.

A related quantity studied in \cite{Sharipov:2024lah}  for Hamiltonians belonging to random matrix ensembles is called the averaged QGT, and is defined as\footnote{In Appendix. \ref{qmt_generic}, we have provided the generic expression for the QGT of a state that is a superposition of the eigenstates of the system Hamiltonian.} 
\begin{equation}\label{average_QGT}
    g_{ab}(\mbr):= \frac{1}{N}\sum_{n=1}^{N} G_{ab}^n(\mbr)~.
\end{equation}
For a Hermitian Hamiltonian, this quantity is real and symmetric in the indices $a$ and $b$ (its imaginary and antisymmetric part is zero). We call this the mean QMT; however, in the following, we shall often omit the quantifier `mean' and refer to it only as QMT for convenience. 

Finally, when the Hamiltonian $H$ belongs to some random matrix ensemble, we are interested in the average of the QMT components over the ensemble, so that we define the ensemble-averaged QMT components as 
\begin{equation}\label{en_avg_qmt}
    \bar{g}_{ab}(\mbr) := \mathbb{E}\big(g_{ab}(\mbr)\big) = \int \prod_j\td H_j ~P_j(H_j) ~g_{ab}(\mbr)~,
\end{equation}
where we have assumed that the total Hamiltonian is a sum of component Hamiltonians drawn from independent distributions, having probability weight $P_j(H_j)$.

Before moving on, a comment on the difference between the quantity defined in \eqref{fid_sus_rr} and a similar quantity studied in the context of 
random matrices in \cite{Penner:2020cxk} is in order. In defining the quantity \eqref{fid_sus_rr} we have assumed $H$ to be total Hamiltonian 
and consider applying a `perturbation' term of the form $\delta r ~ \mathcal{H}$ which give rise to the FS in \eqref{fid_sus_rr} and 
the expectation value of $\mathcal{H}$ are with respect to the eigenstates $ \ket{n}$ of the total Hamiltonian.  On the other hand, one can consider a scenario where $H_0$ is the initial Hamiltonian and the term $r \mathcal{H}$  acts as a `perturbation'. In that case, one can derive an identical-looking expression for the associated FS
where the matrix elements of $\mathcal{H}$ are with respect to $\ket{n_0}$ the eigenstates of $H_0$, and in the denominator, the energy difference term would be $(E^0_{n}-E^0_j)^2$.

\section{Geometry of quantum states in random matrix ensembles: chaos and integrability breaking}
\label{sec_RMT_geometry}

\subsection{Geometry of quantum states in a class of Gaussian random matrix ensemble}\label{ssec_GUE_QMT}

We first consider the scenario where the total Hamiltonian belongs to one of the classical Gaussian ensembles, which we take to be the \textit{Gaussian unitary ensemble} (GUE) here. 
When the total Hamiltonian belongs to the one-parameter family of the form \eqref{Ham_Roz_Por1}, with both the matrices $H_0$ and $\mh$ drawn from GUE with the same variance $\sigma^2=1/N$, the expression for the EFS is given by,
\begin{equation}\label{GUE_EFS}
    \bar{g}_{rr}(r)=\frac{N-1}{2(r^2+1)^2}~.
\end{equation}
This expression can be obtained by direct computation starting from the definition \eqref{en_avg_FS} \cite{Sharipov:2024lah}, or by calculating the correlation function $G(t)=\braket{\mh(t)\mh}$, and subsequently using the connection between the EFS and the correlation function derived in \S \ref{sec_fs_cor} (see eqs. \eqref{FS_to_correl} and \eqref{S_omega}) \cite{Gill:2025upp, Gill:inprep}.

Next, we move on to the case where the Hamiltonian has more than one parameter. 
Specifically, we parametrise the total Hamiltonian of the system such that it can be written as the following two-parameter family of Hamiltonians \cite{berry2020quantum, PhysRevLett.74.4055, austin1992statistical, haakebook, stockmann2007quantum}
\begin{equation}\label{Ham_Roz_Por_GUE}
    H(r, \phi)=H_0+r \cos \phi~\mathcal{H}_1+ r \sin \phi ~\mathcal{H}_2~,
\end{equation}
where $\mathcal{H}_i$, $i=0,1,2$ are three $N \times N$ matrices drawn independently from the GUE with zero mean and variance $\sigma^2$. 
These fix the variance of the total Hamiltonian ($H$) to be $\sigma^2(r^2+1)$.  In this paper, unless stated otherwise, we shall take $\sigma^2=1/N$. Furthermore, $0\leq r<\infty$ and $0\leq \phi \leq 2\pi$ are two parameters, which are coordinates in terms of which the components of the QMT will be written.

For this class of Hamiltonians, the expressions for the components of the ensemble-averaged QMT in eq. \eqref{en_avg_qmt} can be written as\footnote{Note that, since on average the total Hamiltonian does not have any preferred direction, the QMT components, averaged over the full ensemble, would be independent of the angular variable $\phi$ \cite{Sharipov:2024lah}. This is a concequence of the rotational invariance of the GUE.}  
\begin{equation}\label{gue_qmt_comp}
\begin{split}
    \bar{g}_{rr}(\mbr)= \frac{1}{N} \sum_{m \neq n} \mathbb{E}~\bigg[ \frac{(\tilde{\mh}_1)_{nm}(\tilde{\mh}_1)_{mn}}{(E_n(\mbr)-E_m(\mbr))^2}\bigg]~,\\~\bar{g}_{\phi\phi}(\mbr)= \frac{r^2}{N} \sum_{m \neq n} \mathbb{E}~\bigg[ \frac{(\tilde{\mh}_2)_{nm}(\tilde{\mh}_2)_{mn}}{(E_n(\mbr)-E_m(\mbr))^2}\bigg]~,\\
    \text{and}~~~~\bar{g}_{r\phi}(\mbr)= \frac{r }{N}\sum_{m \neq n} \mathbb{E}~\bigg[ \frac{(\tilde{\mh}_1)_{nm}(\tilde{\mh}_2)_{mn}}{(E_n(\mbr)-E_m(\mbr))^2}\bigg]~,
\end{split}
\end{equation}
where we have defined the following two operators,
\begin{equation}
    \tilde{\mh}_1=\cos \phi~\mathcal{H}_1+  \sin \phi ~\mathcal{H}_2~,~~\tilde{\mh}_2=-\sin \phi~\mathcal{H}_1+  \cos \phi ~\mathcal{H}_2~,
\end{equation}
and $\mbr=\{r,\phi\}$, while $E_n(\mbr)$ denotes the $n$-th energy of the total Hamiltonian $H$ corresponding to an eigenstate $\ket{n(\mbr)}$, with respect to which the matrix elements above are taken. Note that, the two new operators $\{\tilde{\mh}_1,\tilde{\mh}_2\}$ are related to the original operators $\{\mh_1,\mh_2\}$ by an orthogonal transformation.
Furthermore, after this change of variables, the expression for the total Hamiltonian becomes $H=H_0+r \tilde{\mh}_1$, and the new variables $\tilde{\mh}_i$ are GUE matrices having the same variance $\sigma^2=1/N$. 

Performing the average over the independent ensembles of $H_0$, $\mh_1$, and $\mh_2$,  the ensemble-averaged components of QMT  can be evaluated to be given by the following expressions\footnote{We have provided a brief outline of the procedure of obtaining the $rr$ component of the metric tensor in Appendix \ref{gue_qmt_derivation}. Other components can be derived by following an analogous procedure to that one.} \cite{Sharipov:2024lah} 
\begin{equation}\label{metric_rmt}
    \bar{g}_{rr}(r)=\frac{N-1}{2(r^2+1)^2}~, ~~\bar{g}_{\phi\phi}(r)=\frac{(N-1)r^2}{2(r^2+1)},~~ \bar{g}_{r\phi}=0~.
\end{equation}
The coordinate $r$ takes positive values. Note that the metric is diagonal in the $\{r,\phi\}$ coordinates.

For an ensemble of Hamiltonians with only a chaotic phase, we see that the QMT components do not diverge for any values of the parameters within the allowed ranges. This can be understood from the fact that due to the presence of the level repulsion, there is always a minimum value of the energy difference. This is, for example, different from what happens to the QMT components of the ground state of a quantum system that undergoes a quantum phase transition \cite{Zanardi:2007byf}. 
Furthermore, for large values of $r \gg 1$, the $rr$ component decays as $1/r^4$, whereas the $\phi\phi$ component approximately becomes constant. Physically, $g_{rr}$ measures the sensitivity of the states to the radial perturbation; therefore, for $r \gg 1$, where the GUE matrix $\tilde{\mh}_1$  dominates, the states become increasingly insensitive to the perturbation. On the other hand, since $\bar{g}_{\phi\phi}$ is a measure of the sensitivity 
of the states to the angular rotations in the parameter space, even when $\tilde{\mh}_1$, the
eigenstates change constantly with $r$.

\subsection{Geometry of quantum states in a class of integrability-breaking random matrix ensembles}\label{ssec_int_breaking_GUE}

\begin{figure}[h!]
		\centering
		\includegraphics[width=3.2in,height=2.3in]{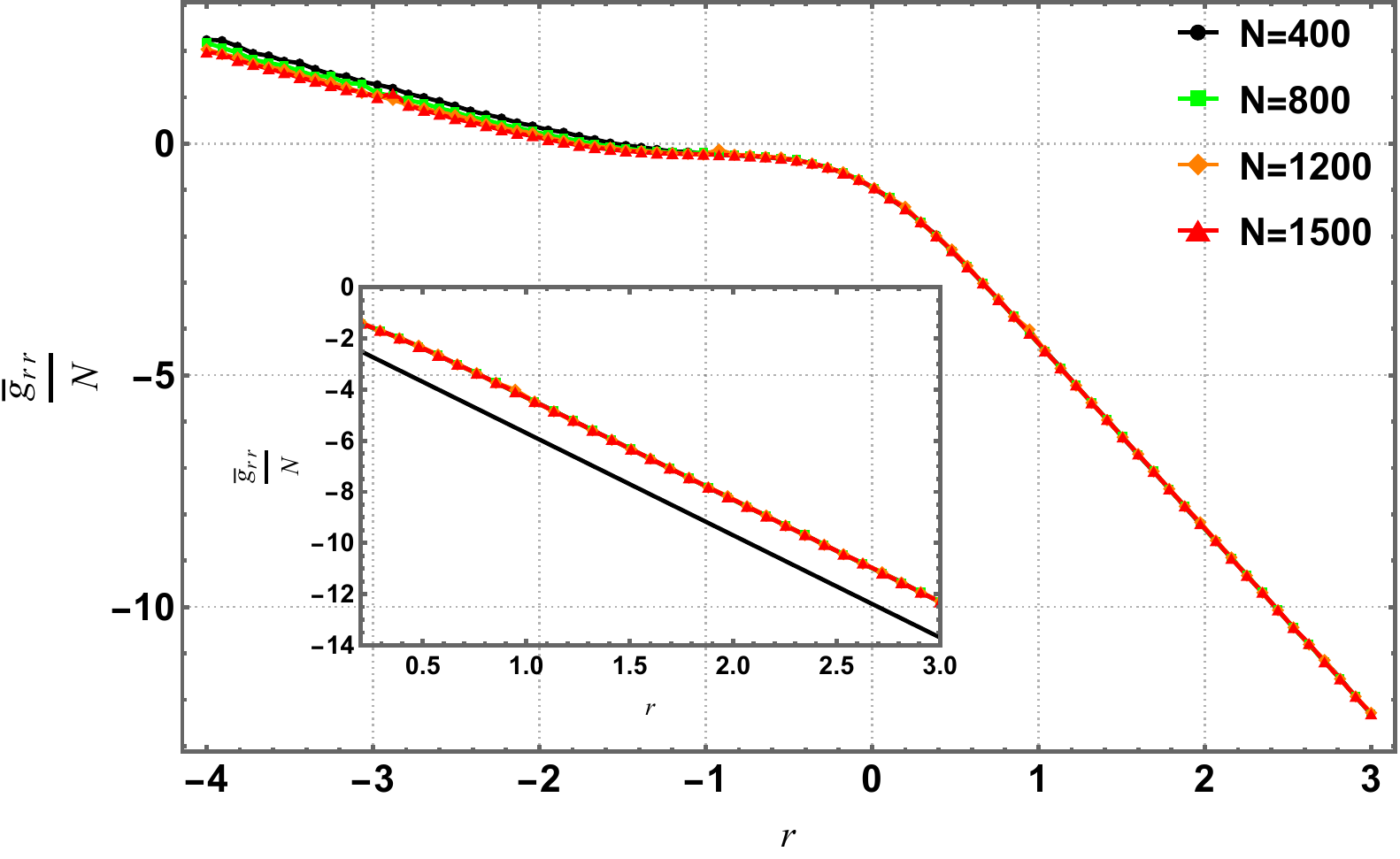}
		\caption{Log-Log (with base 10) plot of the numerical results for the EFS of the RP model for different values of $N$. The Hamiltonian is given in \eqref{Ham_Roz_Por1}, where $H_0$ is a diagonal matrix with i.i.d. Gaussian elements with variance unity, and $\mh$ is drawn from the GUE with variance $1/N$. The numerical results are shown after an average over 500 independent realisations of the Hamiltonian. The plot in the inset shows a $1/r^4$ scaling of $\bar{g}_{rr}/N$ for large $r$.}
		\label{fig:EFS_RP}
	\end{figure}

As we have discussed in the Introduction, one of the main focuses of this paper is to study the geometry of energy eigenstates in random matrix ensembles that show chaos-to-integrability transitions. For this purpose, in this section, we begin by considering the eigenstate space geometry of a class of the RP random matrix ensemble, whose Hamiltonian can be written as $H(r)=H_0+ r \mh$, where $H_0$ is a diagonal matrix with i.i.d. elements, which we draw from Gaussian distribution with zero mean and variance $\sigma_0^2$, and $\mh$ is a random matrix drawn from the GUE with variance $\sigma^2$. In Fig. \ref{fig:EFS_RP}, we have plotted the numerically evaluated expressions for the EFS for the RP model as a function of $r$ for different values of $r$. For large values of $r$ the EFS goes to zero with a scaling $1/r^4$, as shown in the inset of Fig. \ref{fig:EFS_RP}. On the other hand, for small values of $r$, the EFS diverges in the limit $r \rightarrow 0$ due to near-degeneracies in the Poissonian level spacing spectrum of the integrable Hamiltonian $H_0$. 

Next, to understand the geometry of quantum states for more than one parameter cases, consider a two-parameter family of Hamiltonians which can be written in the following way,
\begin{equation}\label{Ham_Roz_Por}
    H(r, \phi)=H_0+r \cos \phi~\mathcal{H}_1+ r \sin \phi ~\mathcal{H}_2~.
\end{equation}
This Hamiltonian is of the same form as the one considered previously for the case GUE Hamiltonian in \S \ref{ssec_GUE_QMT} (eq.\eqref{Ham_Roz_Por_GUE}), however, here the matrix $H_0$ is 
taken to be a diagonal matrix whose elements are drawn independently from Gaussian distributions with mean zero and unit variance. The level spacing statistics of this matrix is therefore given by the Poisson statistics, and hence it is an integrable Hamiltonian \cite{berry77, haakebook, stockmann2007quantum}.
This two-parameter family of Hamiltonians can be thought of as a generalisation of the standard version of the RP model \cite{rosenzweig1960repulsion}. 

\paragraph{QMT of $2 \times2$ integrability breaking Hamiltonians.}

Even though the limit $N \rightarrow \infty$ is usually more interesting compared to finite $N$ for random matrix theories, in many situations, finite $N$ cases can also provide interesting insights into different properties of the random matrix ensemble. In fact, conclusions obtained from the smallest matrix with size $N=2$ can have very useful information about their higher $N$ generalisations, one example being the famous Wigner's surmise of the 
level spacing distribution of random matrix eigenvalues, which is derived 
from $N=2$ matrices, and is  actually an excellent approximation 
of the level spacing distribution of larger matrices \cite{livan2018introduction} (see also \cite{PhysRevE.85.061130} and reference cited therein). 

In this spirit, a two-parameter family of $2 \times 2$ random matrices, where two GUE matrices are added to a random diagonal integrable matrix, was used in \cite{Sharipov:2024lah} to obtain an exact analytical expression for the components of the ensemble-averaged QMT. Specifically, the Hamiltonian is of a similar structure to that 
of the one in \eqref{Ham_Roz_Por_GUE}, where $H_0$ is a diagonal matrix whose elements are independent and identically distributed Gaussian random variables. 

The line element corresponding to the ensemble-averaged QMT (see eq. \eqref{en_avg_qmt}) obtained from a $2 \times2$ matrix representation of the integrability-breaking  Hamiltonian in  \eqref{Ham_Roz_Por} is given in the $\{r,\phi\}$ coordinates by the following expression\footnote{We use a tilde to distinguish the new quantities for the metric
in \eqref{metric_int_brk} from the analogous quantities for the random matrix case discussed in the previous section. However, to avoid cluttering the notation, we have removed the bar indicating an ensemble average. } \cite{Sharipov:2024lah}
\begin{align}\label{metric_int_brk}
ds^2 &=  \tilde{g}_{rr} \text{d}r^2+\tilde{g}_{\phi\phi}\text{d}\phi^2~,\\
\text{where},~~\tilde{g}_{rr}(r)&=\frac{1}{4} \Bigg(\frac{1}{\sqrt{2}r}\text{arccot}(r/\sqrt{2})-\frac{1}{2+r^2}\Bigg)~,\\~\text{and}~~\tilde{g}_{\phi\phi}(r)&=\frac{r}{2\sqrt{2}} \text{arctan}(\sqrt{2}/r)~.
\end{align}
Note that the QMT is again diagonal in the $\{r,\phi\}$ coordinates. As we shall show in \S \ref{ssec_grr_trid_der}, the $rr$ component of this line element (i.e., the FS, when the Hamiltonian contains a single parameter) is actually the same as the one obtained from a symmetric tridiagonal matrix representation of the GUE.

With the above line element in hand, consider two different limits of these metric components. In the limit $r 
\rightarrow 0$ (which corresponds to the limit where the Hamiltonian is completely integrable), it can be easily checked that the $rr$ component diverges, while the $\phi\phi$ component goes to zero. On the other hand, in the limit, $r \rightarrow \infty$ (this corresponds to the case where the Hamiltonian is far away from the integrable phase), $\tilde{g}_{rr}$ vanishes, but $\tilde{g}_{\phi\phi}$ takes a constant value 1/2. 

\paragraph{QMT of $N \times N$ integrability breaking Hamiltonians.} For a generic value of $N>2$, the analytical expressions for the QMT components for this integrability-breaking Hamiltonian ensemble are not known.
In \cite{Sharipov:2024lah}, these components were computed numerically, and it was shown that the metric components follow different scaling properties in different phases of this model. E.g., when $r \ll r^*=1/\sqrt{N}$ (corresponding to the localised phase of the model), the metric components follow a $1/r$ scaling, while in the ergodic phase, having $r \gg 1$, the metric components approach those of the GUE random matrix expressions in \eqref{metric_rmt}. Finally, in the non-ergodic extended phase ($r^* \ll r \ll 1$), the components $\tilde{g}_{rr}$ and $\tilde{g}_{\phi\phi}/r^2$ go to constant scaling as $N$.  

\section{Geometry of quantum states in Gaussian $\beta$-ensembles}
\label{sec_beta_metric}
In this section and the next, we study a generalisation of the integrability-breaking Hamiltonian considered in the previous section, such that the Dyson index ($\beta$) can take a range of real positive values, not just those of classical Gaussian ensembles. Specifically, in this section, we consider the tridiagonal matrix representation of the Gaussian $\beta$-ensembles, found in \cite{dumitriu2002matrix}.  A Hamiltonian belonging to this representation of the ensemble takes a real symmetric tridiagonal $N\times N$ matrix form
whose diagonal elements (denoted as $a_n$, $n=0, \cdots N-1$) are 
are independent Gaussian random variables with mean $0$, and variance as that of the ensemble, while the sub-diagonal elements
(which we call $b_n$, with $n=1, \cdots N-1$) are independently distributed random variables with the following distribution 
\begin{equation}\label{bn_distributions}
    p(b_n)= \frac{2}{\Gamma (k/2)} \Bigg(\frac{\beta N}{2}\Bigg)^{k/2} b_n^{k-1} e^{-\beta N b_n^2/2}~,
\end{equation}
where $k=(N-n)\beta$.  The JPD of the eigenvalues matches that of the one in eq. \eqref{joint_eig_dist}, with $V(H)=H^2)$ and $\beta$ can take any real positive value.  Note that, unlike the three classical Gaussian ensembles, the  Gaussian $\beta$-ensemble does not have rotational invariance. For a collection of works that studied different aspects of the Gaussian $\beta$-ensembles, we refer to \cite{baker1997calogero,le2007nearest, dumitriu2006global, desrosiers2006hermite, forrester2010log, Allez, witte2014moments, duy2016gaussian}. For discussion on the nonergodic phase in the ground state of the $\beta$-ensemble and its dynamical properties (such as the behaviour of the SFF, and its difference from other typical many-body quantum systems or random matrix models), we refer to the recent works \cite{das2, das3}. 

An important difference between the tridiagonal form of the Gaussian $\beta$-ensembles and the classical Gaussian ensembles is that the properties of the eigenvectors of the respective ensembles are quite different, even though both can have the same JPD of the eigenvalues. E.g., as a direct consequence of the rotational invariance of the probability distribution, the eigenvectors of, say, the Gaussian orthogonal ensemble (GOE) are uniformly distributed on the surface of $N$-dimensional unit sphere and have components which are mutually independent. On the other hand, let $T$ be a tridiagonal matrix, and $T=Q \Lambda Q^T$ be the eigendecomposition of $T$, with $\Lambda$ denoting the diagonal matrix constructed from the eigenvalues ($\lambda$) of $T$, then one can show that it is possible to uniquely reconstruct the matrices $T$ and $Q$ starting from the eigenvalues of $T$ and the first row of the matrix $Q$ \cite{parlett1998symmetric, dumitriu2002matrix}. 


A natural property of the ensemble of Hamiltonians with general $\beta$ is that, by controlling the value of the Dyson index, one can control the strength of the repulsion between the energy levels. In the limit $\beta \rightarrow 0$, the energy levels can be arbitrarily close to each other without any level repulsion, and therefore the level spacing statistics is Poissonian, associated with integrable systems \cite{berry77}. Increasing the value of $\beta$ from this value, one gradually gets ensembles with different degrees of level repulsion, including those of the classical Gaussian ensembles (with $\beta=1,2,4$), as well as ensembles having intermediate values of $\beta$.   Therefore, the Gaussian $\beta$-ensembles can be thought to be showing a chaos-to-integrability transition as $\beta$ is decreased below the GOE value ($\beta=1$) towards zero. Similar to the RP model, apart from the chaos to integrability transition, the $\beta$-ensembles also show Anderson localisation 
transition for a specific value of the Dyson index. Rewriting $\beta=N^{-\gamma}$ for a real parameter $\gamma$, it has been shown recently that $\beta$-ensembles have three different phases, respectively, an ergodic phase ($\gamma \leq0$), a nonergodic extended 
phase (for $0 < \gamma<1$) and a localised phase when $\gamma \geq1$ \cite{das2022nonergodic}. As pointed out in \cite{das2022nonergodic}, an interesting feature of the spectral properties of the Gaussian $\beta$-ensemble, which is quite different from that of the RP model is that, for the $\beta$-ensemble, the chaos-integrability transition (in terms of the level spacing statistics) occurs at the point of transition from the ergodic to non-ergodic extended phase at $\gamma=0$, whereas, in the RP random matrix ensemble, it occurs at the Anderson localisation transition.

Our goal is to explore the geometry of quantum states during the integrability-to-chaos transition in the context of these Gaussian $\beta$-ensembles, using the EFS as a probe. In some sense, here, compared to the previous section, we have more options to study the EFS as a probe for the integrability-to-chaos transition, since,
for this class of ensembles with generic Dyson index, one can, a priori, follow two different procedures to move to the integrable phase: 1. first, as was the case in \S  \ref{ssec_int_breaking_GUE}, we can add a term proportional to random matrix drawn from an ensemble with a fixed general value of the Dyson index $\beta>1$ to a diagonal matrix having independent random variables as the entry. As one takes the proportionality constant (denoted as $r$)  to zero, the total Hamiltonian makes a transition from chaotic to the integrable phase as $r \rightarrow0$. 2. Secondly, we can consider a Hamiltonian which is a sum of two terms, each of which is taken from these generalised $\beta$-ensembles and has, in general, different Dyson indices ($\beta_1$ and $\beta_2$). Then, as one takes either or both of these $\beta_i$ close to zero and takes $r \rightarrow 0$, the total Hamiltonian again makes a transition from chaotic to an integrable phase.  

In the next subsection, we follow the first option, i.e.,  we draw a matrix sampled from the tridiagonal Gaussian $\beta$-ensembles, add it to a diagonal integrable Hamiltonian, and derive the corresponding EFS.\footnote{One reason for following this procedure is that, as we shall see, the integral for the EFS does not converge for the tridiagonal Hamiltonian when $\beta<1$.} To have some analytical understanding of the EFS, we consider the simplest case of rank-2 matrices, and derive the exact expression for the EFS for any generic value of $\beta>1$.\footnote{In the next section \ref{sec_invt_beta_ensemb}, we have provided the computation of the EFS of a class of random matrices with general Dyson index, which, in contrast to the tridiagonal form used in this section, is rotationally invariant. Also in Appendix \ref{sec:beta_ensm_haar} we present a computation of the EFS for a random matrix ensemble with a generic Dyson index $\beta$.}  As we shall see, for $\beta=2$, the EFS matches with the $rr$ component of the line element in \eqref{metric_int_brk}.
Next, we compute the EFS for large $N$ matrices numerically, and discuss its dependence on $N$ and the parameter $r$.

\subsection{A one-parameter Hamiltonian from tridiagonal $\beta$-Gaussian ensemble}
\label{ssec_rp_trid}

We consider a system whose total Hamiltonian is in a form similar to the one in eq. \eqref{Ham_Roz_Por1} where $r$ is a positive parameter, and $H_0$ is a diagonal `integrable' Hamiltonian whose elements are independent random variables drawn from a given probability distribution, say, $p(x)$ (which we take to the Gaussian distribution in the following), and $\mathcal{H}$ is a Hamiltonian that we sample from the Gaussian $\beta$-ensemble.
For our purposes here, we now therefore introduce the following one-parameter class of Hamiltonians,
\begin{equation}\label{beta_int_break}
    H(r)=H_0+r ~\mathcal{H}_\beta~,
\end{equation}
where $H_0$ is a diagonal `integrable' Hamiltonian, i.e., we assume that its eigenvalues are drawn from independent Gaussian distributions with zero mean and variance $\sigma_0$ (which we choose to be unity later on)
and $\mathcal{H}_\beta~$ is an element of the tridiagonal $\beta$-Gaussian ensemble. This Hamiltonian $H=H_0+r~\mathcal{H}_\beta~$ can be thought of as the generalisation of the standard version of the RP model (which is usually defined with $\mathcal{H}$
taken from the classical Gaussian $\beta$-ensembles) for general values of $\beta$.
We employ this Hamiltonian in the following to compute the EFS defined in \eqref{av_fid_sus_rr}. 


One difference from the situation considered in the two previous sections is that here we are considering the simplest situation where the total Hamiltonian $H$ has only one parameter $r$, rather than both $r$ and $\phi$. Hence, in this case, the line element will consist of a single element, which is the EFS associated with the parameter $r$. The QMT components in such a situation are evaluated numerically in the next subsection. 

The total Hamiltonian $H(r)=H_0+r ~\mathcal{H}_\beta$ is still tridiagonal, with the diagonal elements being i.i.d. Gaussian distributed with zero mean and variance equal to $1+ \frac{2 r^2}{N\beta }$ (when $\sigma_0^2=1$), and the sub-diagonal elements are independent chi distributions, i.e., 
\begin{multline}
    H_{jj}=a_j(r)~,~~~H_{j,j+1}=H_{j+1,j}=b_j(r)~,~~\text{with}~,~\\
    a_j(r) \sim \mathcal{N}\Bift{0, 1+ \frac{2r^2}{N\beta }}~,~~~b_j(r)\sim \frac{r}{\sqrt{\beta N}} \chi_{(N-j)\beta}~.
\end{multline}
Therefore, when $k=(N-n)\beta \gg 1$, and $r<N\beta$, the sub-diagonal coefficients approximately behave as normal distributions, $b_n \sim \mc{N} \Bift{r\sqrt{\frac{N-n}{N}},\frac{r^2}{2 N \beta}}$. This structure indicates that each of the off-diagonal elements has a non-zero mean, which decreases monotonically from the top-left corner of the matrix towards the bottom-right corner.

We also note that the condition $(N-n)\beta \gg 1$ can be satisfied in different ways: 1. When $\beta \gg 1$, i.e., the initial tridiagonal part of the Hamiltonian has a very large degree of level repulsion to begin with (then the previous condition is satisfied for all $n<N$ and $N$); 2. When we have $N-n \gg 1/\beta$ for some fixed $\beta$. If $\beta>1$, then this is satisfied if we are not too close to the bottom right corner of the matrix. 3. Finally, if we consider the thermodynamic limit, $N \rightarrow \infty$, and are away from the edge, $n/N<1$, this condition is satisfied as well.  To summarise, if we consider the thermodynamic limit, and are away from the edge of the matrix, the above approximation for $b_n$ is valid.  Furthermore, note that, in this approximation, the variance of the $b_n$ coefficients, in the leading order, is independent of the position in the bulk.

    \begin{figure}
      \centering
       \subfloat[]{\includegraphics[width=0.5\columnwidth]{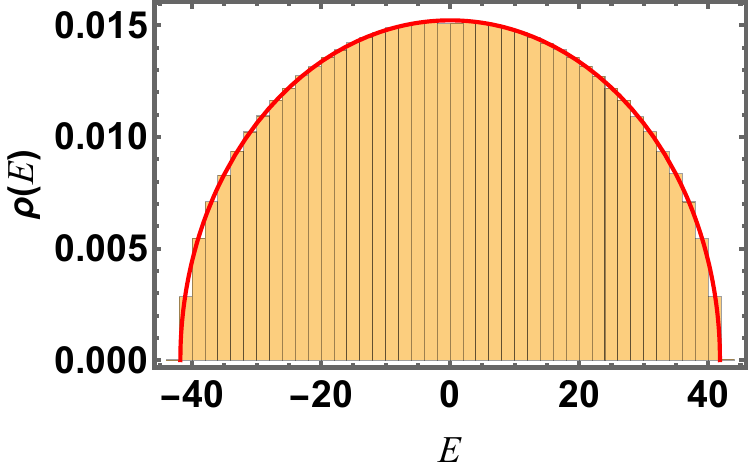}}
        \hfill
       \subfloat[]{\includegraphics[width=0.5\columnwidth]{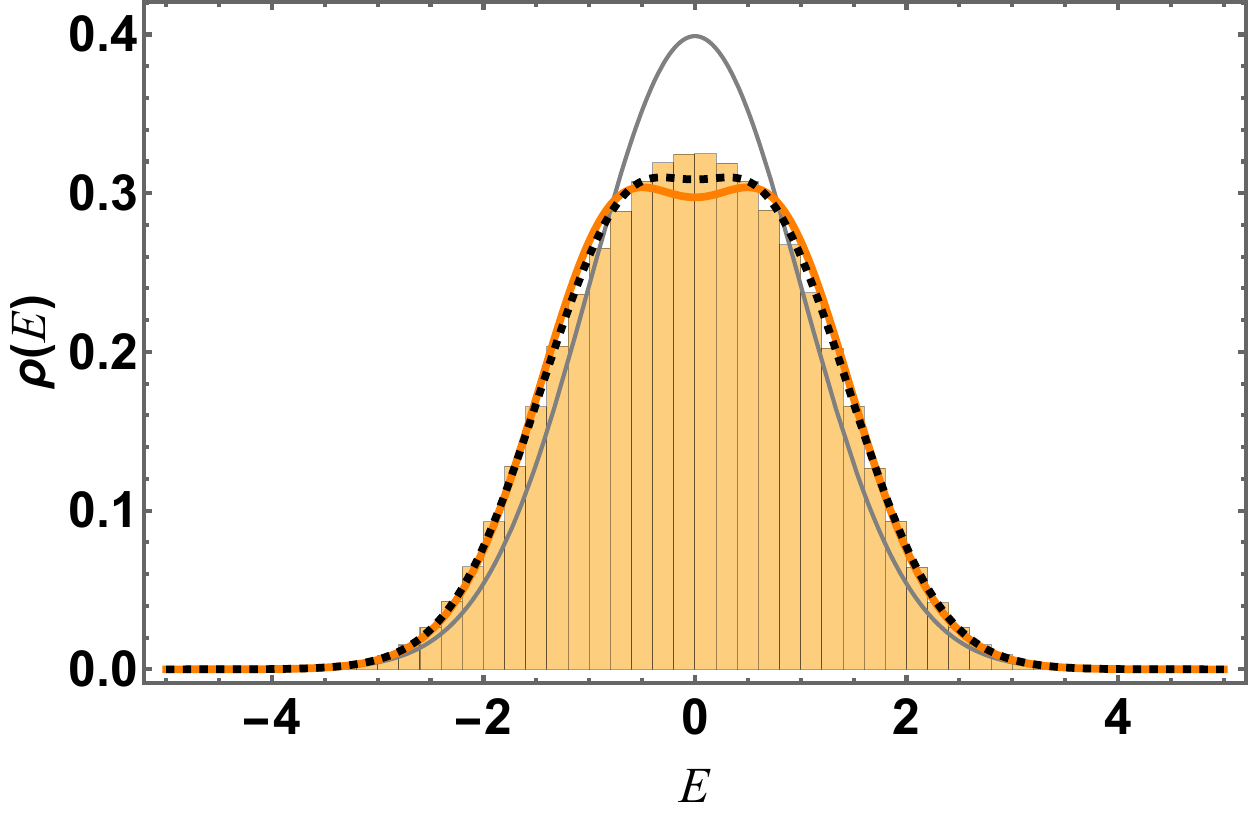}}
    \vspace{0.5em}
       \subfloat[]{\includegraphics[width=0.5\columnwidth]{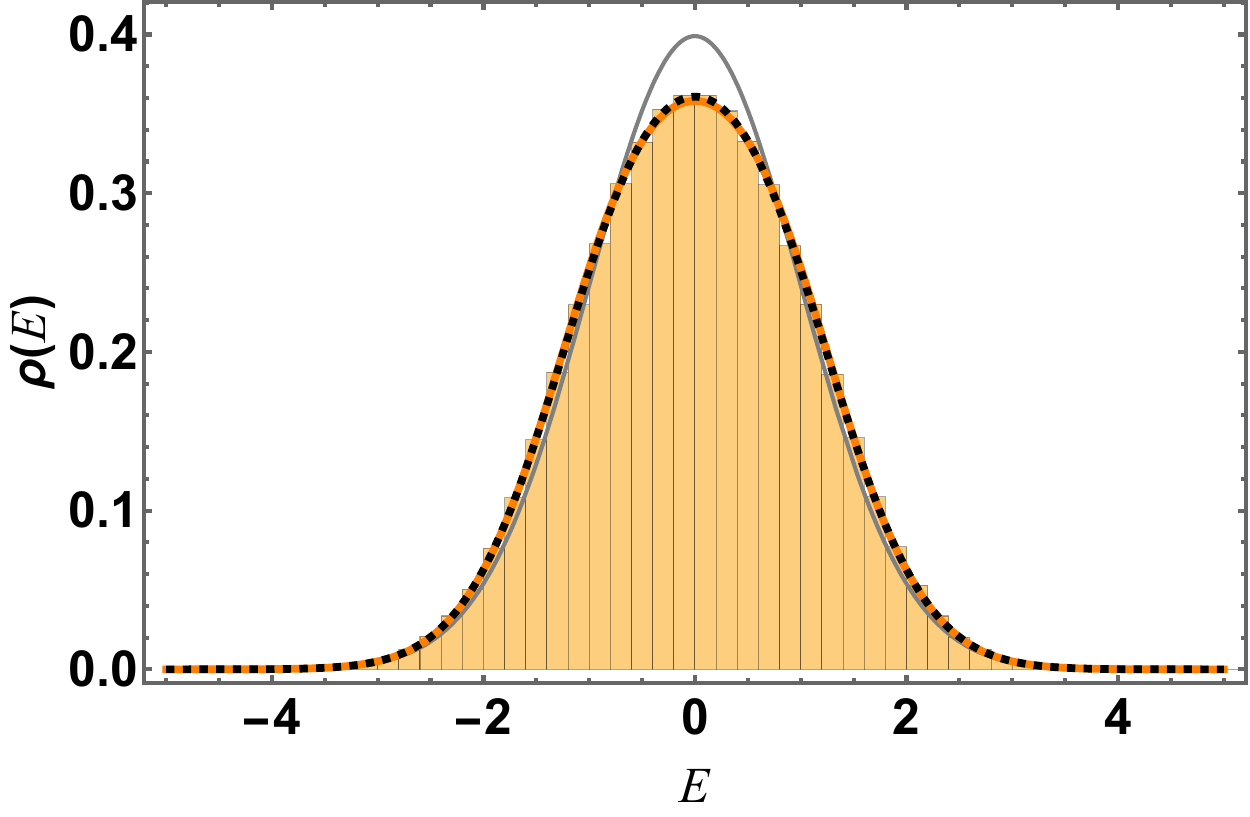}}
         \hfill
        \subfloat[]{\includegraphics[width=0.5\columnwidth]{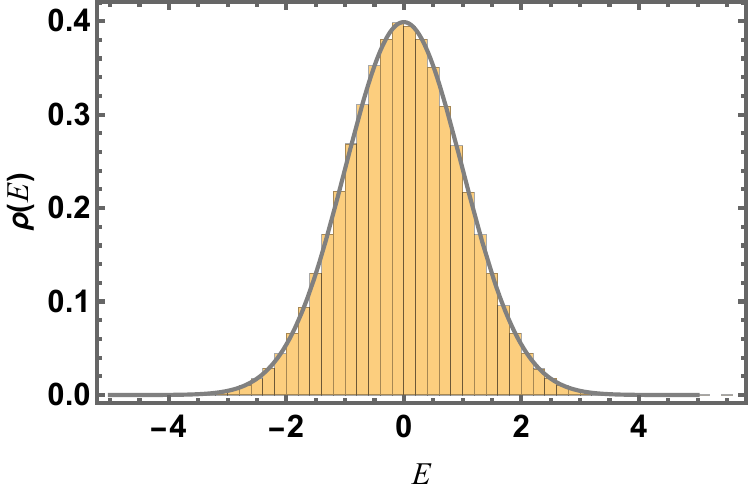}}
        \caption{DOS of the tridiagonal $\beta$-Gaussian version of the RP model (the Hamiltonian in \eqref{beta_int_break}) for different values of the parameter $\tiga$ (and hence $r=N^{-\tiga}$): $\tiga=-0.4$ (a), $\tiga=0.09$ (b), $\tiga=0.15$ (c), and $\tiga=1$ (d). The red curves represent the Wigner semicircle distribution with support from $-2r$ to $2r$. while the grey curves indicate a Gaussian distribution with mean zero and variance unity. Here we have set $N=2000$, $\beta=2$ and the results are shown after an average over $500$ independent realisations of the Hamiltonian. The orange and black dashed curves in panels (b) and (c) represent the analytical results derived from a perturbation expansion of the Cauchy transform of the DOS that we shall discuss in App. \ref{sec:DOS_approx}. }
      \label{fig:DOS_beta_RP}
   \end{figure}

Before moving on to study the properties of the eigenstates through the EFS, we first discuss the behaviour of the density of states (DOS) for the Hamiltonian in \eqref{beta_int_break}. To this end, it is useful to reparametrise the parameter $r$ as $r=N^{-\tiga}$, $\tiga \in \mb{R}$, and study the DOS for different $\tiga$. We have shown the plots for the DOS for different values of $\tiga$
and fixed $\beta$ in Fig. \ref{fig:DOS_beta_RP}. As can be seen from the plots, the DOS makes a clear transition from the semi-circle distribution to the Gaussian distribution as $\tiga$ is increased. Specifically, for $\tiga<0$, we expect the DOS to be well described by a Wigner semi-circle distribution since the $\mh_\beta$ term is dominant in the total Hamiltonian, while, for $\tiga>0$, for sufficiently large $N$, the first diagonal term starts to dominate, and the DOS is close to the Gaussian distribution. These expectations are confirmed by the numerical computation of the DOS shown in Fig. \ref{fig:DOS_beta_RP}, where the red curve shows the Wigner semicircle distribution 
\begin{equation}\label{semi-circle_dist}
    \rho_{sc}(E)=\frac{1}{2\pi r^2}\sqrt{(2r)^2-E^2}~,
\end{equation}
having support from $-2r$ to $2r$, the value of $r$ being determined from $r=N^{-\tiga}$, and the grey curve represents a Gaussian distribution\footnote{Note that changing the value of the index $\beta$ in the second term $\mh_\beta$, the DOS, for $\tiga<0$, remains unchanged, while for $\tiga>0$, it only quantitatively, since in the numeric we take the variance of the diagonal matrix $H_0$ to be $\sigma_0^2=2/(\beta N)$.} with zero mean and variance $\sqrt{\frac{2}{\beta N}}$. Between $\tiga_{c} < \tiga <1/2$, the DOS is given by a deformed version of the Gaussian distribution, where $\tiga_{c}$ is a lower cut-off value of this parameter, below which this approximation breaks down. In App. \ref{sec:DOS_approx}, we discuss an analytical approach to the computation of the DOS in this parameter range, by using a series expansion of the Cauchy transform of the DOS. From our numerics, where we have fixed $N=2000$, the cut-off value of $\tiga$ is determined to be $\tiga_{c} \approx 0.08$.

\subsection{Ensemble-averaged fidelity susceptibility}
\label{ssec_grr_trid_der}
Now we move on to the computation of the EFS for this model. We first consider the simplest case of $N=2$, where it is possible to obtain an analytical expression for the EFS for a generic value of the Dyson index. Numerical results for finite large $N$ are presented later in this section.

Therefore, we take $\mathcal{H}_\beta$ to be the following $2 \times 2$ matrix
\begin{equation} \label{H_beta}
	\mathcal{H}_\beta =\left(
		\begin{array}{ccc}
			a_0 & b_1 \\
			b_1  & a_1
		\end{array}
		\right)~,
\end{equation}
with the distributions of $a_i$ and $b_i$ the same as those specified at the beginning of this section, 
and assume that $H_0$ takes the following diagonal form in the same basis
\begin{equation}\label{H_0}
    H_0 =\left(
		\begin{array}{ccc}
			h_1 & 0 \\
			0  & h_2
		\end{array}
		\right)~,
\end{equation}
where $h_1$ and $h_2$ are sampled from independent Gaussian distributions with zero mean and variance $\sigma_0$.

The expression for $\bar{g}_{rr}$, averaged over the distributions of $H_0$ and $\mh_\beta$, is given by,\footnote{Here we have taken each of the distributions to be normalised, so that we do not need to divide by an overall constant, $Z$.}
\begin{widetext}
  \begin{equation}
\begin{split}\label{grr_beta_1}
    \bar{g}_{rr}(r, \beta) =\frac{1}{2 \pi \sigma_0^2}\int \frac{\mathcal{H}_{12}\mathcal{H}_{21}}{\big(E_n-E_m)^2 }
    e^{-\frac{1}{2\sigma_0^2} \text{Tr}\big(H_0^2\big)}~p(a_0) p(a_1)p(b_1)~\td H_0 ~ \td a_0 \td a_1\td b_1~.
\end{split}
\end{equation}  
\end{widetext}
Here $\mathcal{H}_{mn}=\braket{m|\mathcal{H}_{\beta}|n}$ are the matrix elements 
of $\mathcal{H}_\beta$ in the energy eigenbasis of the total Hamiltonian, $p(a_0)$, $p(a_1)$, and $p(b_1)$, respectively denote the probability distributions from which the elements $a_0,a_1$ and $b_1$ of the Hamiltonian \eqref{H_beta} are drawn. 

\textbf{Case.1: $\beta=2$.}
First, consider the case of $\beta=2$. Using the expression for the distributions of $a_n$, and $b_n$s with $\beta=2$ we get  (here we put $\sigma_0=1$)
\begin{widetext}
\begin{equation}
\begin{split}\label{grr_exp1}
    \bar{g}_{rr}(r, \beta=2) =\frac{1}{16  \pi^ 2 r}\int \frac{b_1^3 (z_1+z_2)^2}{\big(4 b_1^2 r^2 + z_1^2\big)^2}~
    \text{exp}~\Bigg[-\frac{1}{16} \Bigg(32 b_1^2+4x_1^2+8y_1^2+\frac{2}{r^2}(z_1-z_2)^2+(z_1+z_2)^2\Bigg)\Bigg]
    ~\td y_1 \td z_1 \td z_2 \td x_1 \td b_1 ~,
    \end{split}
\end{equation}
\end{widetext}
where we have performed the following coordinate transformations, 
\begin{equation}
\begin{split}
    h_{1,2} = (x_1 \pm x_2)/2~, ~~a_{0,1} = (y_1 \pm y_2)/2~,\\
    x_2=(z_1+z_2)/2~,~y_2=(z_1-z_2)/(2r)~.
\end{split}
\end{equation}

Performing the integrals in \eqref{grr_exp1}, we get back the expression for the $rr$ 
component in \eqref{metric_int_brk} implying that EFS in this case is the same as the GUE RP model. An important difference one should notice between this integral and the usual one we get from a complex Hermitian representation of the GUE (see \cite{Sharipov:2024lah})
is that, for the above integral, since all the matrix elements are real, the extra factor that comes from the radial part of the complex 
variable in a complex Hermitian representation (see e.g., the integral in \eqref{grr_integral} below) here comes from the distribution \eqref{bn_distributions} of the sub-diagonal elements itself. This is due to the fact that, like $b_1$, the distribution of the radial part can be thought of as the distribution of the square root of the sum of two squared independent Gaussian variables, hence a chi distribution.  Therefore, even though in the two cases the expressions for the EFS are the same, the nature of the contributions coming from the probability distributions of the individual metric elements is different. 

Note that the equality between the EFS for the GUE and the $\beta$-Gaussian ensemble with $\beta=2$ can be expected to be valid only for the case of $N=2$, since in that case, the joint eigenvalue distribution completely determines the spectral statistics, and as mentioned, the $\beta$-ensemble and the classical Gaussian ensembles share the same JPD of eigenvalues for the respective Dyson indices by construction. For $N >2$, the eigenvectors statics of the two ensembles are fundamentally different, hence the previous equality would not hold.

\textbf{Case 2: Generic values of the Dyson index.} We now move on to the case of general $\beta$. The tridiagonal matrix representation is particularly suitable for 
obtaining the EFS for a Gaussian ensemble with a generic value of the Dyson index for $\beta>1$.\footnote{For $0 <\beta\leq 1$, the integral for EFS does not converge. This can be understood from the fact that for $\beta<1$, the distribution of $b_1$ diverges in the limit $b_1 \rightarrow 0$, thereby making the integral for the EFS fail to converge. 
}
For any $\beta>1$, the analytical expression for $\bar{g}_{rr}(r) $ in terms of the hypergeometric function,
we get from \eqref{grr_beta_1} is given by
\begin{widetext}
\begin{equation}\label{EFS_beta_gen}
 \bar{g}_{rr}(r, \beta>1)  = \frac{\beta}{16 \Upsilon^2} \BBtd{\frac{(\beta \sigma_0)^{-\beta}\bift{\beta \sigma_0^2+r^2(\beta-1)}}{r\Gamma(\beta/2)}\BBft{\sqrt{\pi}\beta^{(1+\beta)/2}\sigma_0\Upsilon^{\beta/2}\Gamma\Bift{\frac{\beta-1}{2}}-2 r \bift{\beta \Upsilon}^{\beta/2}\Gamma(\beta/2)~_2F_1\BBft{\frac{1}{2},\frac{\beta}{2};\frac{3}{2}, \frac{-r^2}{\beta \sigma_0^2}}}-2\Upsilon}~,
\end{equation}
\end{widetext}
where, $\Upsilon=(\beta \sigma_0^2+r^2)$.
Once again, in the limit $r \rightarrow \infty$, this quantity goes to zero, while in the limit $r \rightarrow 0$ it can be expanded as
\begin{multline}
     \bar{g}_{rr}(r, \beta) \approx \frac{\sqrt{\pi \beta} \Gamma \big(\frac{\beta-1}{2}\big)}{16 \Gamma \big(\beta/2\big)\sigma_0~ }\frac{1}{r} -\frac{1}{4\sigma_0^2} + \frac{3\sqrt{\pi}  \Gamma \big(\frac{\beta-1}{2}\big)}{16 \sqrt{\beta}\sigma_0^3\Gamma \big(\frac{\beta}{2}-1\big)} r\\+\frac{\beta-3}{6 \beta \sigma_0^4}r^2 +\mathcal{O}(r^3)~.
\end{multline}
This shows that the EFS diverges in the limit $r\rightarrow 0$ as $1/r$ for any generic value of the Dyson index $\beta>1$.
For some specific integer values of $\beta$, the above long expression for $\bar{g}_{rr}$ reduces to particularly simple forms. E.g., for $\beta=3$ and $\beta=5$, we have,
\begin{equation}\label{EFS_beta3}
    \bar{g}_{rr}(r, \beta=3) =\frac{1}{8 \sigma_0^2} \BBft{\frac{3\sigma_0^2+2r^2}{r \sqrt{3\sigma_0^2+r^2}}-2}~.
\end{equation}
and 
\begin{multline}\label{EFS_beta5}
    \bar{g}_{rr}(r, \beta=5) = \frac{1}{60 r \sigma^2} \Bigg(5\sigma_0^2\sqrt{5\sigma_0^2+r^2}+4r^2\sqrt{5\sigma_0^2+r^2}\\-4r^3-15r\sigma^2\Bigg)~.
\end{multline}

	\begin{figure}[h!]
		\centering
		\includegraphics[width=3.1in,height=2.3in]{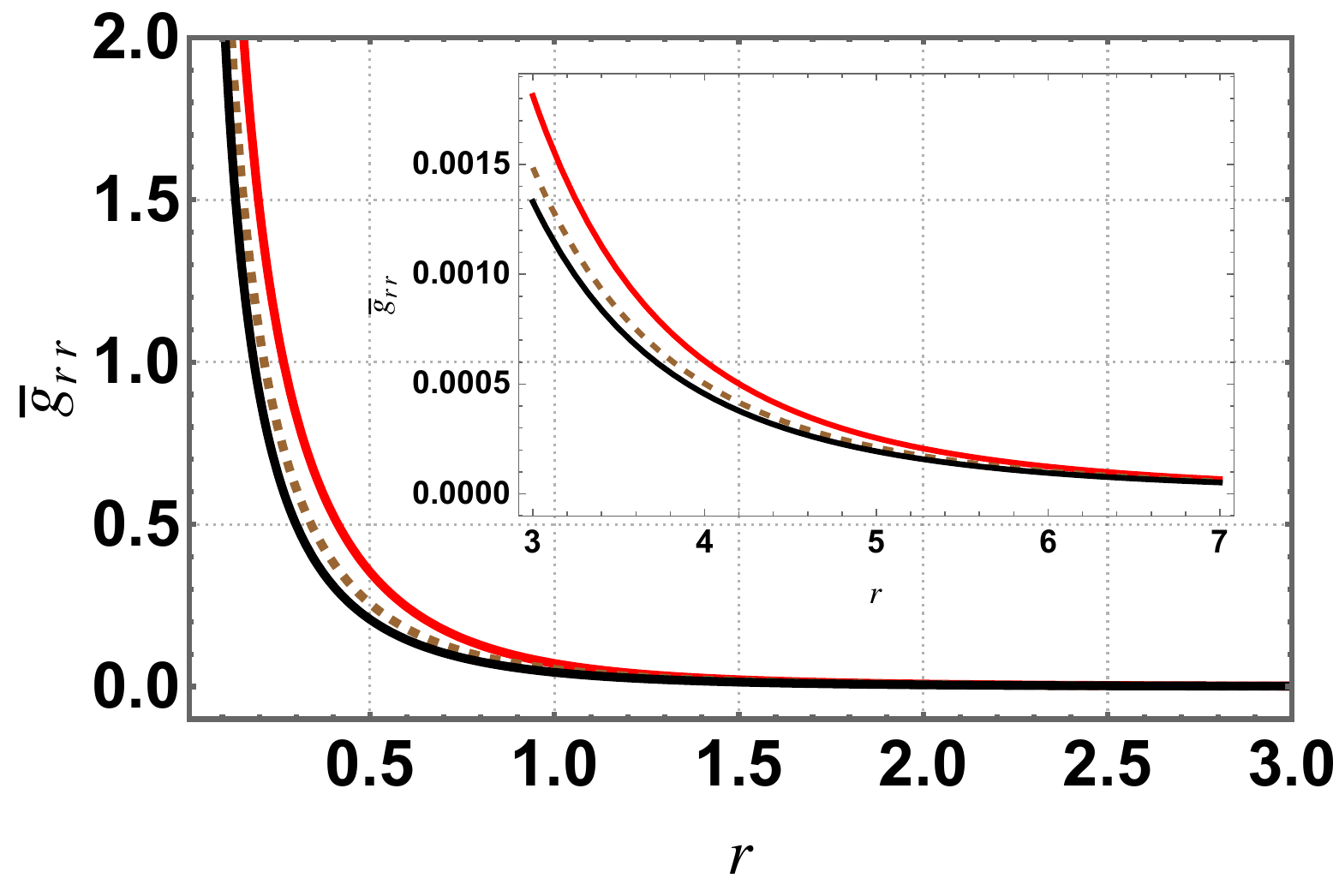}
		\caption{Plot of the EFS ($\bar{g}_{rr}(r, \beta)$) for a $2 \times 2$ representation of the Hamiltonian \eqref{beta_int_break}, where $\mh_{\beta}$ is the `tridiagonal' representation of the  $\beta$-Gaussian ensemble, for different values of the Dyson index: $\beta=2$ (red), $\beta=3$ (brown, dashed), and $\beta=5$ (black). We have fixed $\sigma_0^2=1/2$.  The plot in the inset shows the behaviour for large values of $r$.}
		\label{fig:grr_BETA}
	\end{figure}

In Fig. \ref{fig:grr_BETA} we have shown the variation of $\bar{g}_{rr}(r, \beta)$ with respect to $r$ for these values of the Dyson index, along with the expression for $\beta=2$. As can be seen, for higher values of $\beta$, $\bar{g}_{rr}(r, \beta)$ decays faster
for large $r$, and its general behaviour is quite similar to the case of $2 \times 2$ rotationally invariant effective $\beta$ ensemble studied next in Section \ref{sec_invt_beta_ensemb} (compare with the plots in Figs. \ref{fig:grr_ETA1} and \ref{fig:grr_ETA12}).
By expanding these expressions for $r \gg 1$, in the leading order, it can be seen that the EFS scales as $r^{-4}$, i.e., 
\begin{align}
    \bar{g}_{rr}(r, \beta=2)  &\sim \frac{ \sigma_0^2}{3} \frac{1}{r^4}~,~~\bar{g}_{rr}(r, \beta=3)  \sim \frac{9 \sigma_0^2}{32} \frac{1}{r^4}~,\nonumber\\
    &~~\text{and}~~\bar{g}_{rr}(r, \beta=5)  \sim \frac{25 \sigma_0^2}{96} \frac{1}{r^4}~.
\end{align}
In general, from the exact expression in \eqref{EFS_beta_gen} for the EFS for a generic value of the Dyson index ($\beta>1$), we get, by assuming $r> \sigma_0 \sqrt{\beta}$ (i.e., $r> \sqrt{2/N}$ for $\sigma_0=\sqrt{2/(\beta N)}$, here $N=2$), the following expression in the leading order 
\begin{equation}
    \bar{g}_{rr}(r, \beta>1) \sim \frac{\beta^2 \sigma_0^2}{4(\beta^2-1)}\frac{1}{r^4} +\mc{O}(r^{-6})~,
\end{equation}
thereby, establishing $1/r^4$ scaling of the EFS for this parameter range. 

\begin{figure}[h!]
		\centering
		\includegraphics[width=3.2in,height=2.3in]{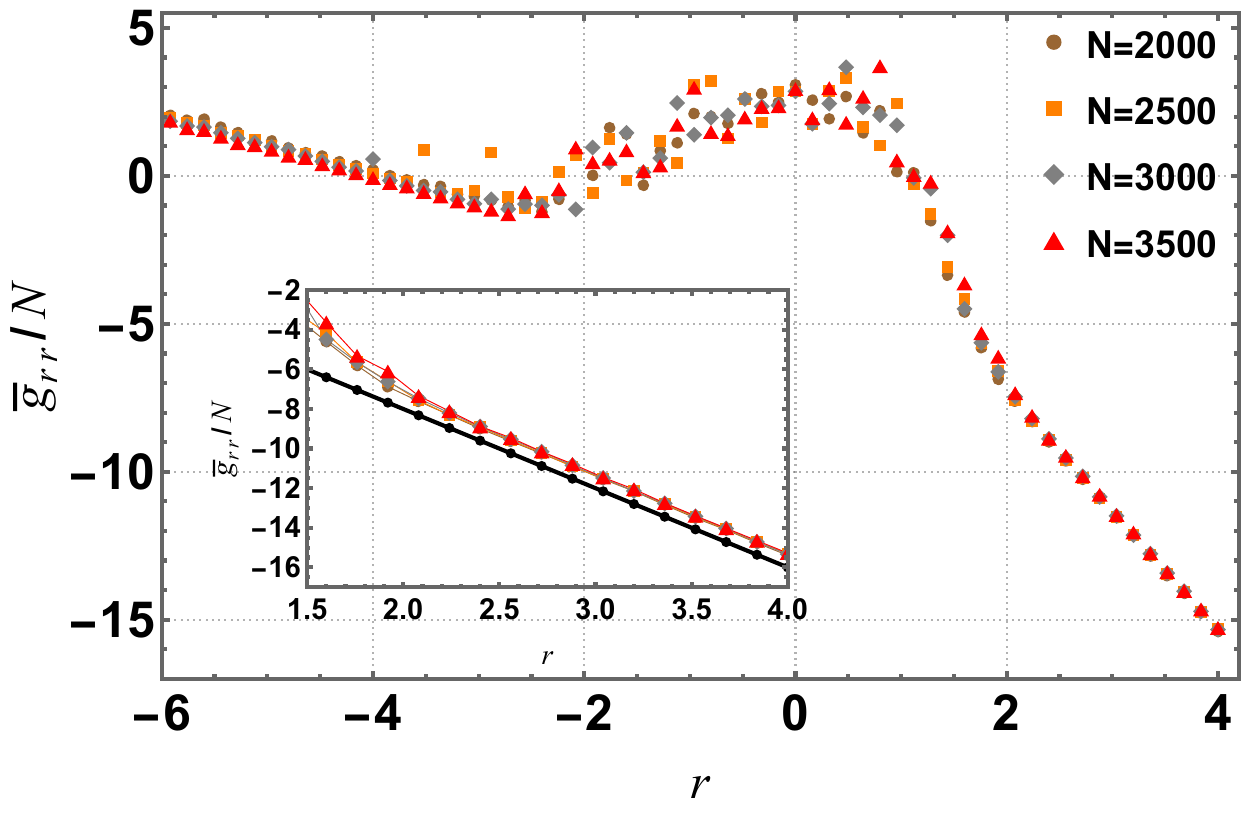}
		\caption{Log-Log plot of the numerical results for the EFS for the Gaussian $\beta$-ensemble with large $N$. The Hamiltonian is given in \eqref{beta_int_break}, where $H_0$ is a diagonal matrix with i.i.d. Gaussian elements, and $\mh$ is drawn from the tridiagonal $\beta$-ensemble with the Dyson index $\beta>1$. Here we have fixed $\beta=2$, and the numerical results are shown after an average over 500 independent realisations of the Hamiltonian. The plot in the inset shows a $1/r^4$ scaling of $\bar{g}_{rr}/N$ for large $r$.}
		\label{fig:EFS_TRID_BETA2}
	\end{figure}

Next, we move on to the computation of the EFS for large values $N$. In Fig. \ref{fig:EFS_TRID_BETA2}, we have shown the numerical results for the EFS computed for different values of $N$, and a fixed value of $\beta>1$. Clearly, there are three regimes for the EFS depending on the value of the parameter $r$.  For large values of $r>1$ (which corresponds to $\tiga<0$), where the $\beta$-Gaussian term dominates in the total Hamiltonian, the EFS regularly decreases with increasing $r$. Furthermore, the decay of $\bar{g}_{rr}/N$ is independent of $N$ in the thermodynamic limit, indicating that EFS scales with $N$ for large $r>1$. As we have shown in the inset of Fig. \ref{fig:EFS_TRID_BETA2}, it is possible to obtain a $1/r^4$ scaling law for the normalised EFS for large $r$.  This scaling for large $r$ is consistent with the analytical expressions in eqs. \eqref{EFS_beta3} and \eqref{EFS_beta5} for $N=2$, as we have discussed above. 

On the other hand, for small values of $r \ll 1$, a regime where the diagonal integrable term dominates, EFS shows fluctuation with $r$ due to the absence of strong level repulsion in the spectrum, and its value also depends on the value of $N$. However, it should be kept in mind that the scaling $1/r$ we found for the $N=2$ case analysed above is expected to be valid for $r<r^*=1/\sqrt{N}$, for finite values of $N$. In the intermediate region, $r^*<r<1$, the behaviour of EFS is qualitatively different, and not properly captured by the $N=2$ analytical expression we have derived above.

\subsection{Quantum metric tensor with $\beta$-Gaussian ensemble}\label{ssec_beta_N}
So far in this section, we have considered the simplest situation where the total Hamiltonian contains one parameter so that the metric for the quantum states space geometry is one-dimensional. We now discuss the more general situation 
where the total Hamiltonian has more than one parameter; therefore, the full geometry of states is captured by the ensemble-averaged QMT. To this end, we introduce the following two-parameter class of Hamiltonians constructed from matrices drawn solely from the tridiagonal $\beta$-ensemble having different Dyson indices, which also have deformations along an angular direction marked by the angle $\phi$, apart from the parameter $r$: 
\begin{equation}\label{Ham_Roz_Por_beta}
    H(r, \phi)=H_{0\xi }+r \cos \phi~\mathcal{H}_{1\beta}+ r \sin \phi ~\mathcal{H}_{2\beta}~.
\end{equation}
Here, all the $N\times N$ individual matrices are now being sampled from tridiagonal $\beta$-ensembles. Specifically, we assume that the matrix $H_0$ is drawn from a  Gaussian $\beta$-ensemble with a Dyson index $\xi<1$, and $\mh_1$ and $\mh_2$ are drawn from independent Gaussian $\beta$-ensembles with the same Dyson index $\beta>1$. The choice $\xi<1$ is made so that $H_0$ has Poissonian level spacing statistics.

An important point we notice here is that, since the tridiagonal $\beta$ ensemble is not rotationally invariant, in this case, unlike the cases with GUE matrices in Section \ref{sec_RMT_geometry}, the angular variable does change the spectrum, and as we shall see, the non-diagonal component of the metric is non-zero in this case. Note also that the total Hamiltonian $H(r,\phi)$ is still tridiagonal with the diagonal elements Gaussian distributed; however, the sub-diagonal elements no longer have chi distributions. 

\begin{figure}[h!]
		\centering
		\includegraphics[width=3.1in,height=2.1in]{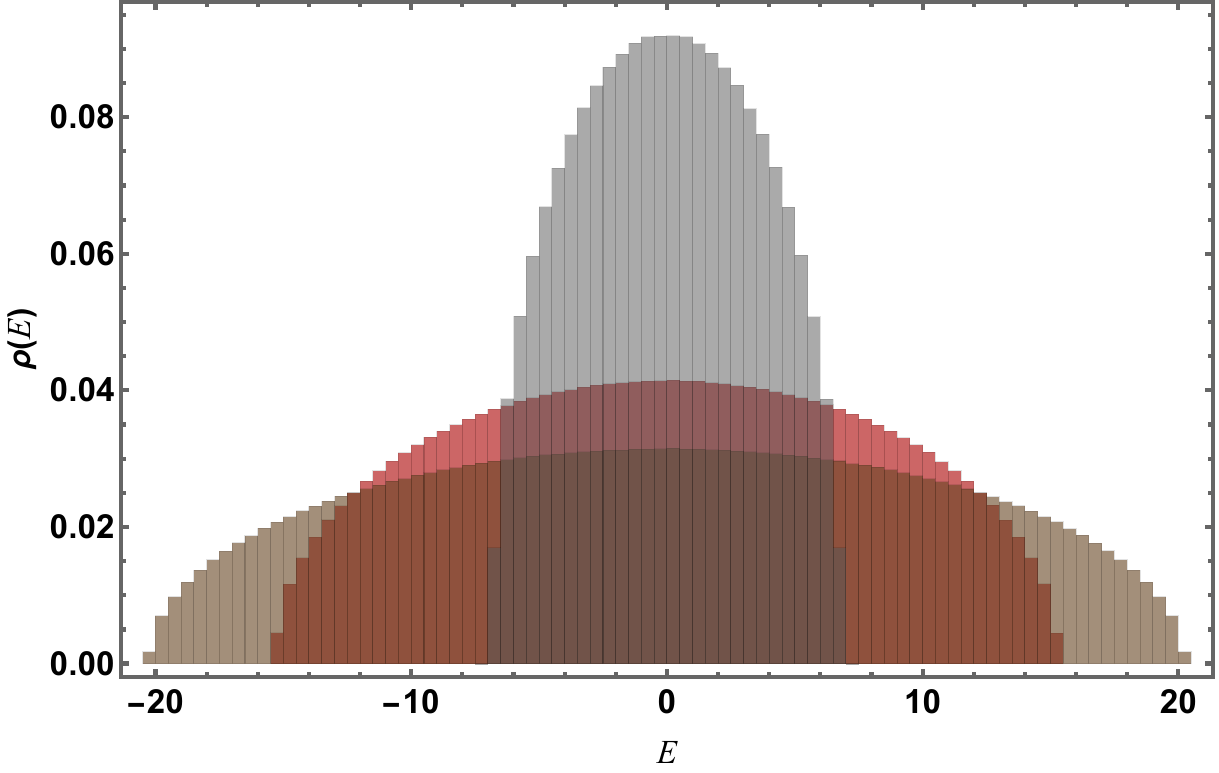}
		\caption{Plot of the DOS of the Hamiltonian in \eqref{Ham_Roz_Por_beta} for different values of the angle $\phi$, with all other parameters fixed: $\phi=\frac{\pi}{2}$ (red), $\phi=\frac{\pi}{2}-\frac{\pi}{6}$ (brown), and $\phi=\frac{\pi}{2}+\frac{\pi}{6}$ (grey). Here we have fixed $N=2000$, $\tiga=-0.25$ (where $r=N^{-\tiga}$), $\beta=2$, $\xi=0.5$, and the histograms are plotted with eigenvalues generated for 500 independent realisations of the Hamiltonian.}
		\label{fig:DOS_GRP_phi}
	\end{figure}

We have shown the variation of the spectrum of the Hamiltonian \eqref{Ham_Roz_Por_beta} with the angular variable by plotting the DOS for different values of $\phi$ in Fig. \ref{fig:DOS_GRP_phi}. The DOS takes the shape of a deformed semicircle distribution. Specifically, the increment of the angle $\phi$ (between $0 \leq\phi \leq \pi $)\footnote{As $\phi$ crosses $\pi$, the width of the DOS again starts to increase with increasing $\phi$. } decreases the width of the distribution, and hence, more and more eigenvalues concentrate near the centre of the spectrum. 

\begin{figure}[h!]
		\centering
		\includegraphics[width=3.4in,height=2.4in]{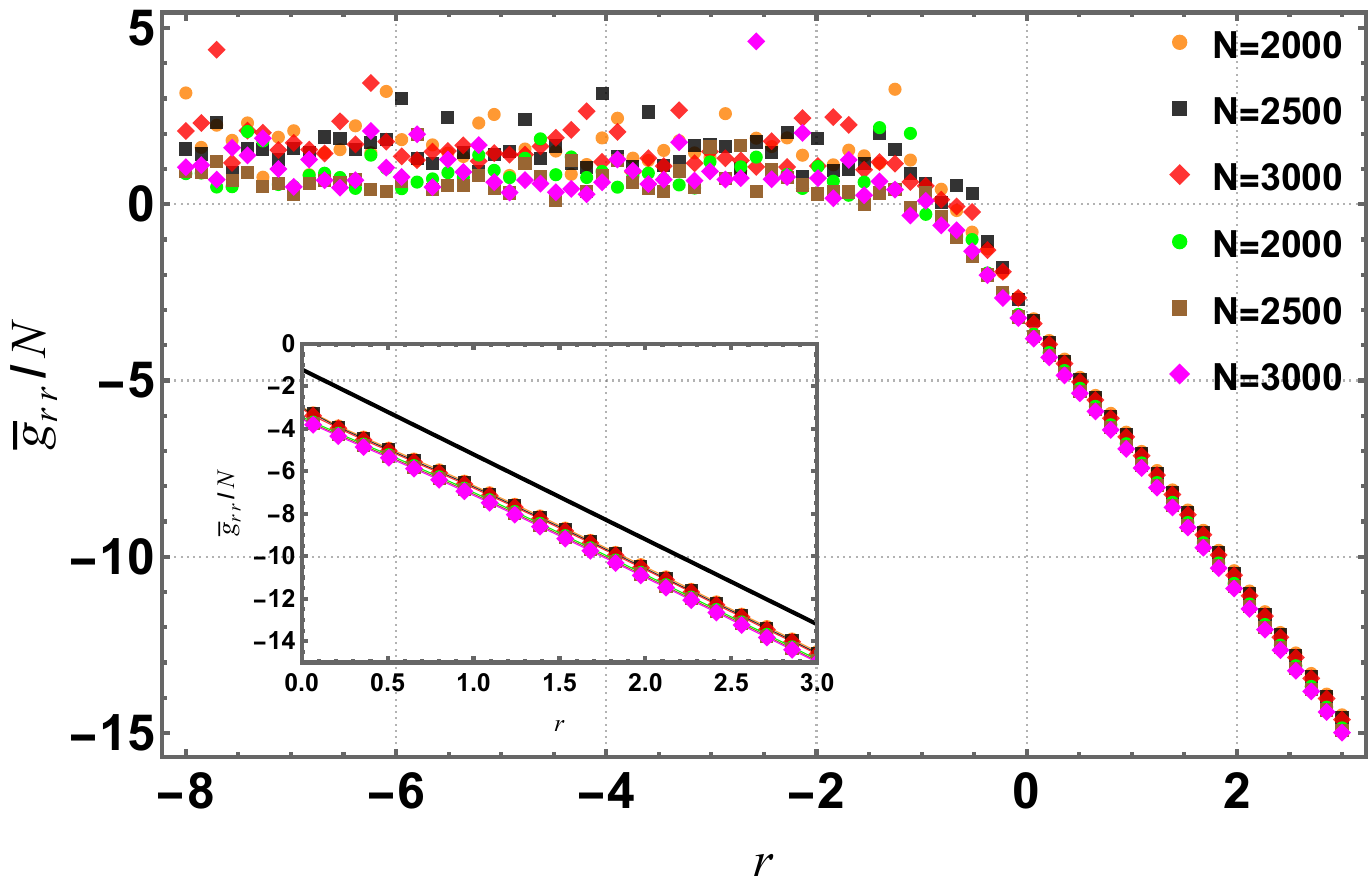}
		\caption{Log-Log  plot of the numerical results for the ensemble-averaged $rr$ component of the QMT associated with the Hamiltonian written in \eqref{Ham_Roz_Por_beta}, with the individual matrices being drawn from a $\beta$-ensemble with different Dyson indices ($H_0$ is from an ensemble with Dyson index $\xi<1$, while $\mh_1$ and $\mh_2$ are drawn from $\beta$-ensemble with the same Dyson index $\beta>1$). The first set of data (orange, black and red) shows the numerical values of the $\bar{g}_{rr}/N$ with $\beta=2$ and $\xi=0.5$, while the second set of data  (green, brown and magenta) represent this component when $\beta=3$ and $\xi=0.7$. The plot in the inset shows a $1/r^4$ fitting of the data for large $r$ (the black line). Here, we have fixed $\phi=\pi/4$, and the numerical results are shown after averaging over 1000 independent realisations of the ensemble.}
		\label{fig:qmt_BETA_grr}
	\end{figure}

\begin{figure}[h!]
		\centering
		\includegraphics[width=3.4in,height=2.4in]{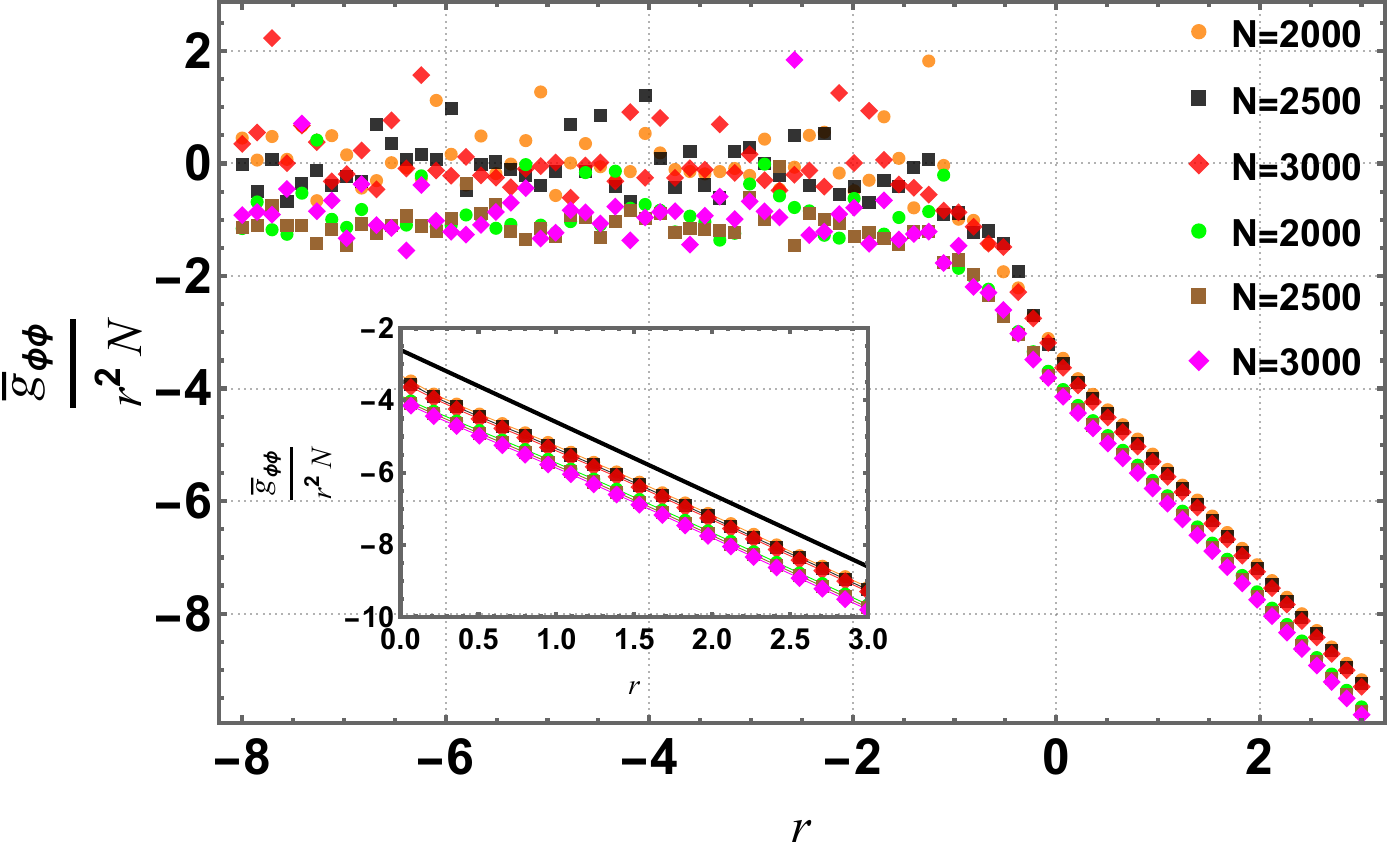}
		\caption{Log-Log  plot of the numerical results for the ensemble-averaged $\phi\phi$ component of the QMT associated with the Hamiltonian written in \eqref{Ham_Roz_Por_beta}. The numerical values of the parameters and the colour codings are the same as in Fig. \ref{fig:qmt_BETA_grr}. The plot in the inset shows a $1/r^2$ fitting of $g_{\phi\phi}/(r^2N)$ for large $r$. Here, we have fixed $\phi=\pi/4$, and the numerical results are shown after averaging over 1000 independent realisations of the ensemble.}
		\label{fig:qmt_BETA_gphiphi}
	\end{figure}

\begin{figure}[h!]
		\centering
		\includegraphics[width=3.4in,height=2.4in]{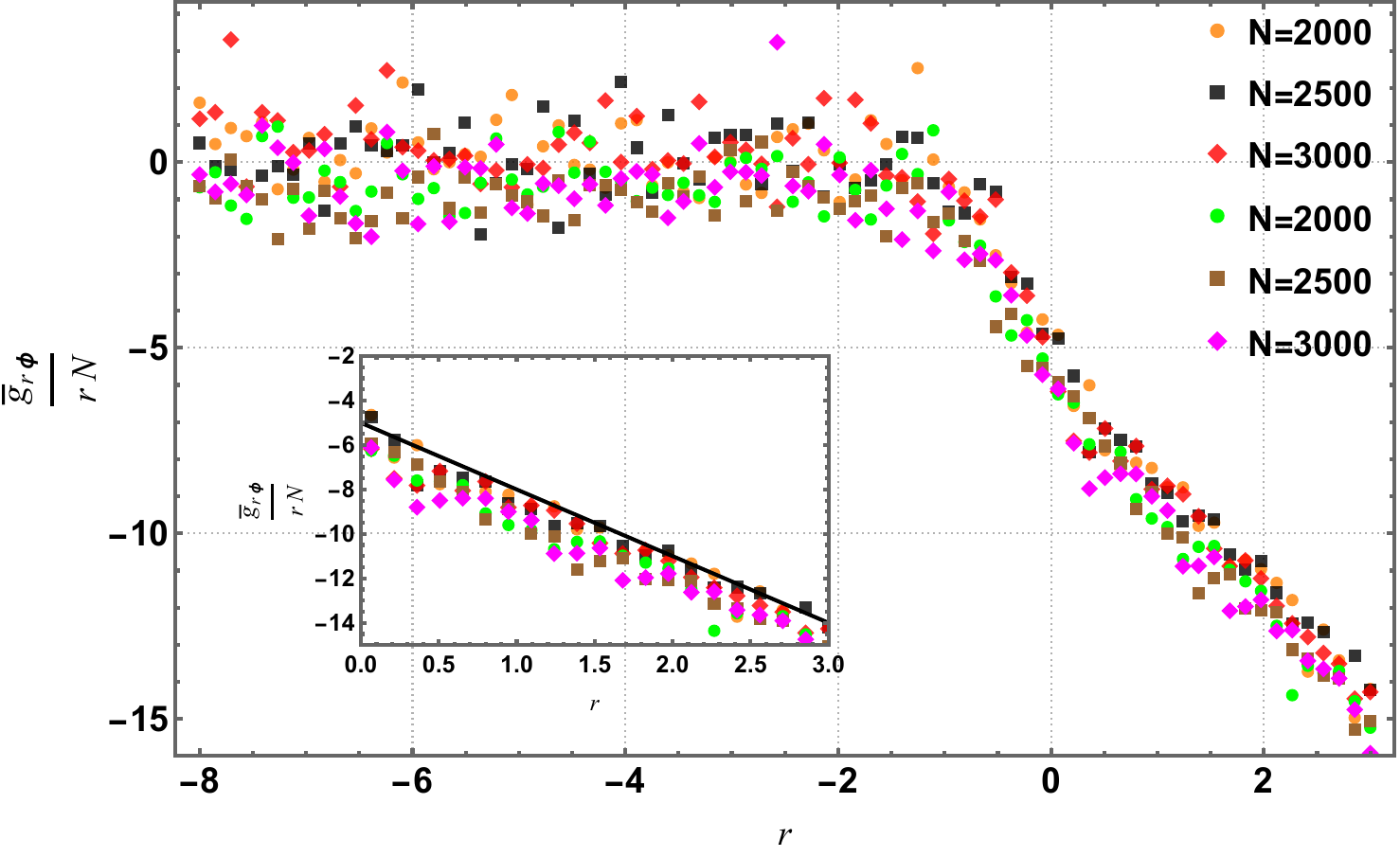}
		\caption{Log-Log plot of the numerical results for the ensemble-averaged $r\phi$ component of the QMT associated with the Hamiltonian written in \eqref{Ham_Roz_Por_beta}. The numerical values of the parameters and the colour codings are the same as in Fig. \ref{fig:qmt_BETA_grr}. The plot in the inset shows a $1/r^3$ fitting of $g_{r\phi}/(rN)$ for large $r$. Here, we have fixed $\phi=\pi/4$, and the numerical results are shown after averaging over 1000 independent realisations of the ensemble.}
		\label{fig:qmt_BETA_grphi}
	\end{figure}

In Figs. \ref{fig:qmt_BETA_grr}, \ref{fig:qmt_BETA_gphiphi}, and \ref{fig:qmt_BETA_grphi},  we have plotted the numerical values of the QMT components written in \eqref{gue_qmt_comp}, for different fixed values of $r$ (and with a fixed $\phi$).  Compared with the previous cases of GUE Hamiltonians, the mixed component $g_{r\phi}$ is non-zero, and the numerical values of this component are shown in Fig. \ref{fig:qmt_BETA_grphi}.\footnote{Since this non-diagonal component can be negative, to be consistent with the other plots and hence to plot in the Log scale we have shown the Log-Log plot of the absolute values of $\bar{g}_{r\phi}$. } We have shown the numerical results for two different combinations of the Dyson indices of the component tridiagonal $\beta$-Gaussian matrices: Case 1, when  $\beta=2$ and $\xi=0.5$, and Case 2, where we have set  $\beta=3$ and $\xi=0.7$. Therefore, in both cases, the matrix $H_0$ does not have energy level correlation. We can, of course, consider a more generic solution where one chooses the Dyson indices of $\mh_1$ and $\mh_2$ to be different. However, we do not 
consider this scenario here. It will be interesting to perform a detailed analysis of how the QMT components behave with different ranges of the Dyson index of the component matrices \cite{Gill:inprep}.

From these plots, we see that the overall behaviour of the three components is quite similar to each other (even though they approach zero for large $r$ in a different manner, as we discuss below); however, for small values of $r$, it is different from the EFS computed in the previous section, due to the fact that here $H_0$ is drawn from a $\beta$-Gaussian ensemble, rather than a diagonal matrix with i.i.d. random elements. 
Furthermore, there are two different regimes, i.e., the QMT components decrease as the parameter $r$ is increased,\footnote{The behaviour of the $r \phi$ component is slightly more irregular compared to the diagonal components of the QMT, even for large values of $r$. } and hence the system makes a transition from an integrable to a chaotic phase. This overall qualitative picture of the QMT component remains unchanged when the Dyson indices $\xi$ and $\beta$ are varied.

Let us now carefully analyse the large $r$ behaviour of the metric components, where the most important and interesting differences between the components lie. The scaling of the QMT components for large $r$ can be obtained from the fitting of the numerical data shown in the insets of the plots in \ref{fig:qmt_BETA_grr}, \ref{fig:qmt_BETA_grphi}, and \ref{fig:qmt_BETA_gphiphi}. Specifically, we have the following scaling of the QMT components for large values of $r$,
\begin{equation}\label{QMT_scaling}
    \bg_{rr} \sim \frac{N}{r^4}~,~~\bg_{r\phi} \sim \frac{N}{r^2}~, ~~\text{and}~~~\bg_{\phi\phi} \sim \frac{N}{r^0}~.
\end{equation}
Our numerical results indicate that these scaling laws remain unchanged when changing the Dyson indices of $H_{0\xi }$, $\mathcal{H}_{1\beta}$, and $\mathcal{H}_{2\beta}$, as well as varying the angular parameter $\phi$. We also note that the scalings of the diagonal components of the QMT for the Hamiltonian \eqref{Ham_Roz_Por_beta} for large $r$ are the same as the ones obtained from the analytical expressions for the QMT for the GUE given in \eqref{metric_rmt}. Since, the QMT must be positive semi-definite, its components should satisfy the condition $\bar{g}_{rr}\bar{g}_{\phi\phi} \geq (\bar{g}_{r\phi})^2$. This can be thought of as a bound on the diagonal components of the metric in a given set of coordinates in terms of the off-diagonal components. For the GUE case, the QMT in \eqref{metric_rmt}, since the off-diagonal component is exactly zero, and the diagonal components are positive, this condition is satisfied trivially. On the other hand, for the Hamiltonian with components drawn from $\beta$-ensemble, it can be seen that the scaling relations for $r \gg 1$, which we have written in \eqref{QMT_scaling}, imply $\text{det}(\bg) \sim 1/r^4$, thereby suggesting that the QMT is nearly degenerate in this limit. In other words, computing the eigenvalues (denoted as, say, $\lambda_{\pm}$) of the QMT from this scaling, we see that $\lambda_{+} \approx N\Bift{1+\frac{1}{r^4}}$, and $\lambda_{-} \approx \mc{O}(r^{-6})$. Therefore, for $ r \gg 1$, $\lambda_{-} \approx 0$, and hence, the displacement along this eigen-direction contributes negligibly in the total line element $\td s^2$ on the parameter manifold, which is approximately determined by a single direction - the projection along the eigenvector direction of $\lambda_+$.

As can be seen from the exact metric components in \eqref{metric_rmt}, the vanishing of the metric determinant in the limit $r  \rightarrow \infty$ is also true for the GUE case, i.e., the metric becomes degenerate.  However, the difference between the GUE and the $\beta$-ensemble is the eigen-direction along which the displacement vanishes. For GUE, which is already diagonal in the $r,\phi$ coordinates, this direction is the eigenvector $e_-^{(GUE)}=(1,0)^T/\sqrt{2}$, i.e., the $\td r$ direction. This is as expected, since for large $r$, the total Hamiltonian is approximately given by that of a GUE matrix (recall the Hamiltonian is as in \eqref{Ham_Roz_Por_GUE}, which is just equal to $H=H_0+r \tilde{\mh}_1$), hence changing $r$ only rescales the energy eigenvalues without changing the eigenstates appreciably in this limit. On the other hand, for the $\beta$-ensemble Hamiltonian we are considering  \eqref{Ham_Roz_Por_beta}, this statement is no longer true due to the absence of the rotational invariance of the ensemble. In fact, in this case, the direction along which the displacement vanishes (i.e., the eigenvector associated with $\lambda_{-} \approx 0$) is given by $e_{-} \propto (-r^2,1)  $ which corresponds to the direction $-r^2 \td r +\td \phi$ on the geometry corresponding to the metric tensor $\bg_{ab}$, and hence, is a combination of the basis one-forms $\td r$ and $\td \phi$.


\begin{figure}[h!]
		\centering
		\includegraphics[width=3.4in,height=2.4in]{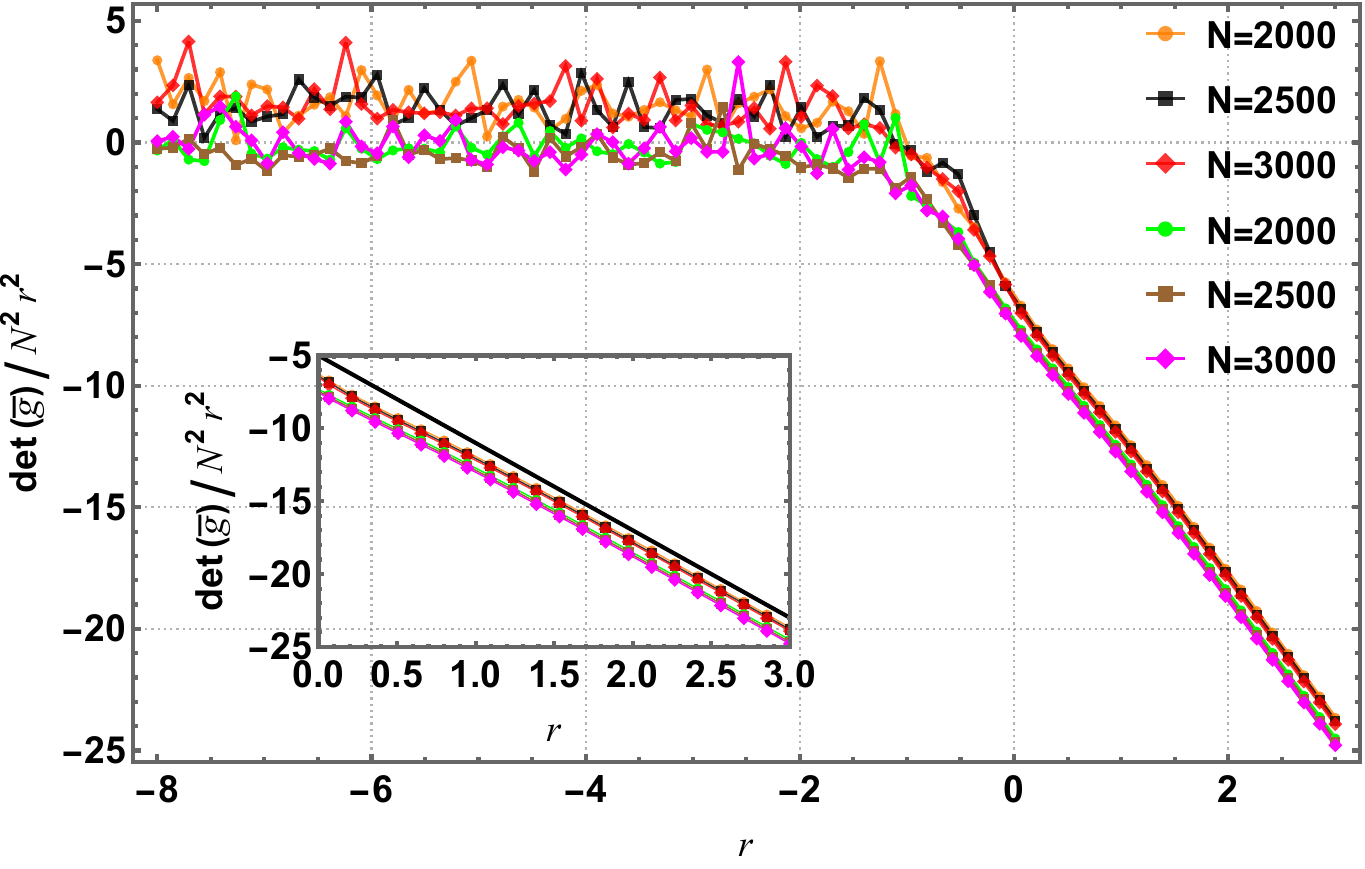}
		\caption{Log-Log plot of the numerical results for the determinant of the ensemble-averaged QMT computed for the Hamiltonian in \eqref{Ham_Roz_Por_beta}. The numerical values of the parameters and the colour codings are the same as in Fig. \ref{fig:qmt_BETA_grr}. The plot in the inset shows a $1/r^6$ fitting of $\text{det}(\bg)/(N^2r^2)$ for large $r$. Here, we have fixed $\phi=\pi/4$, and the numerical results are shown after averaging over 1000 independent realisations of the ensemble.}
		\label{fig:det_qmt_BETA}
	\end{figure}

In Fig. \ref{fig:det_qmt_BETA}, we have shown the plot of the numerically evaluated determinant of the QMT (after ensemble average) for the two sets of Dyson indices we have considered in the previous plots. The overall behaviour of the determinant is quite similar to the individual components. However, importantly, as we have seen above, the scalings of the QMT components for large $r$ (in eq. \eqref{QMT_scaling}) imply that the determinant vanishes as $1/r^4$ in the large $r$ limit. Therefore, $\text{det}(\bg)$ should scale as $1/r^4$ in the limit of large $r$, a fact which is confirmed by our numerical results, and is shown in the inset of Fig.  \ref{fig:det_qmt_BETA}.

We emphasise here once more that the non-vanishing of the non-diagonal component of the QMT for the $\beta$-ensemble is one of the main distinguishing features between this and the invariant ensembles, such as the GUE, in terms of the geometry of quantum states. Note that this conclusion is true in terms of the radial ($r$) and angular ($\phi$) parameters we have used. It will be interesting to find out a set of coordinates on the parameter space for which this QMT would be diagonal. Even though we do not have the exact expressions for the QMT components, it is possible to gain some important intuition on the nature of the underlying Riemannian geometry by using the scaling relations in \eqref{QMT_scaling}, and trying to diagonalise the metric. Since the parameter space is two-dimensional, we can always transform to a coordinate system where the metric is conformally related to the flat Euclidean space. However, rather than explicitly finding those coordinate transformations (since we do not know the exact QMT components in the $\{r,\phi\}$ coordinates), we perform the following specific transformations $u(r)=r$, $v(r,\phi)=\phi+f(r)$, where $u$ and $v$ are the new coordinates, and $f(r)$ is a function of $r$. Now using the well-known relations for the transformation of the metric components \cite{carroll2019spacetime, nakahara2018geometry}, we get the expression for the non-diagonal component of the metric in the new coordinates to be given by\footnote{Since we do not have exact expressions for the metric components, these manipulations are done with the scaling relations for them, and hence are only valid in an appropriate approximate sense and should not be taken as exact statements about the underlying Riemannian geometry of quantum state space. Nevertheless, we do hope that the lessons we gain from them are generic enough so that they illustrate some important characteristics of the said geometry. } 
\begin{equation}
    \bg_{uv}=\frac{1}{f^\prime(r)}\bg_{rr}+\bg_{r\phi}~,
\end{equation}
with $f^\prime(r)$ denoting the derivative of $f(r)$ with respect to $r$.
Demanding that the metric is diagonal in the $u,v$ coordinates, and using the scalings in \eqref{QMT_scaling}, we obtain the expression for the function to be: $f(r)=\frac{1}{r}+\text{constant}$. Hence, transformations (setting the integration constant in the last step to zero)
\begin{equation}
    u=r~,~~ v=\frac{1}{r} +\phi~,
\end{equation}
approximately diagonalises the QMT, at least, for large $r$. Since we have assumed $r \gg 1$, $u=r$, and $v \approx \phi$ diagonalises the QMT, thereby explaining the consistency with the GUE behaviour in this limit. 

We shall conclude this section by making the following comment on the non-diagonal component of the QMT that we have found to be non-zero above. As we have mentioned, \cite{Sharipov:2024lah}, which computed these components for a two-parameter generalisation of the RP model, found the off-diagonal component to be vanishing, and it was shown in this reference that, in the two-dimensional parameter space of the said model, the integrable point corresponds to a conical singularity when embedded in the three-dimensional Euclidean space, thereby distinguishing such embedding from a chaotic Hamiltonian, for which the corresponding embedding surface is a smooth one, as can be seen from the components of the QMT for the GUE Hamiltonian, given in \eqref{metric_rmt}. In a recent work, \cite{Gorsky:2025ylb}, which studied the geometry of the parameter space of the Russian doll model having a time-reversal symmetry-breaking parameter, the authors found a non-vanishing off-diagonal component of the corresponding metric tensor, and it was established that this indicates a deformed embedded cone. Since, for our case, only the scaling laws for the metric components for $r  \gg 1$ could be obtained, we can not directly compare with the results of \cite{Gorsky:2025ylb}, and see if the deformation of the embedding cone is a generic feature, and leave this question as an open problem for future work.

\section{The fidelity-susceptibility for an invariant $\beta$-ensemble with an effective Dyson index}\label{sec_invt_beta_ensemb}

In the previous section, 
we derived the EFS associated with a $2\times2$ random matrix ensemble with general Dyson index, by sampling the matrix from the tridiagonal $\beta$-ensemble. However, as discussed, in this representation of the Gaussian $\beta$-ensemble, the ensemble with tridiagonal form is no longer invariant under symmetry transformations (unlike the classical Gaussian ensembles). To see what effect this invariance has on the geometry of corresponding energy eigenstates, in this section, we compute the EFS from a one-parameter family of $2 \times 2$ random matrix ensembles introduced in \cite{vivo2008invariant}, which is rotationally invariant but has an effective real value of the Dyson index ($\beta_{e}$). As we show in the following, when a matrix drawn from this ensemble is added to a diagonal integrable Hamiltonian, the resulting EFS once again diverges as one moves to the integrable phase, suggesting that this feature can be a generic one in random matrix ensembles having an integrability-to-chaos transition, irrespective of whether the underlying ensemble is rotational invariant or not.

\subsection{A $2 \times 2$ matrix ensemble with an effective Dyson index}

Consider four random variables $u,v,x,y$ taken from a one-parameter ($\eta$) dependent JPD, $\mathbb{P}_{\eta} (u,v,x,y)$ and an 
ensemble defined by the following Hermitian and unitary invariant $2 \times 2$ matrices,
\begin{equation} \label{Ham_eta}
		\mathcal{H} = \left(
		\begin{array}{ccc}
			u & \frac{1}{\sqrt{2}} (x+ iy)  \\
			\frac{1}{\sqrt{2}} (x - iy)  & v
		\end{array}
		\right)~= \left(
		\begin{array}{ccc}
			u & s e^{i \varphi}  \\
			s e^{-i \varphi}   & v
		\end{array}
		\right)~.
\end{equation}
The expression for the JPD ($\mathbb{P}_{\eta} (u,v,x,y)$) is assumed to be  given by the following expression \cite{vivo2008invariant}
\begin{equation}\label{JPD_elements}
    \mathbb{P}_{\eta} (u,v,x,y) = Z_{\eta} e^{-\frac{1}{2} \text{Tr}\mathcal{H}^2} \Big(2\text{Tr}\mathcal{H}^2 - (\text{Tr}\mathcal{H})^2 \Big)^{-\eta}~,
\end{equation}
where $\eta \in [0,1]$, and $Z_\eta$ is a normalisation constant. The resulting  expression for the JPD of the eigenvalues ($\lambda_1, \lambda_2$) of the Hamiltonian $\mathcal{H}$ can be evaluated to be  
\begin{equation}\label{JPD_eigenval_beta}
    P_{\eta}(\lambda_1,\lambda_2) = \mathcal{N}_\eta e^{-\frac{1}{2} \big(\lambda_1^2+\lambda_2^2\big)} \big|\lambda_1-\lambda_2\big|^{2-2\eta}~,
\end{equation}
where the normalisation constant is $\mathcal{N}_\eta=\Big(\sqrt{\pi}2^{3-2\eta} \Gamma \big(3/2-\eta\big)\Big)^{-1}$.
As can be seen, by making a comparison with the general expression for the JPD of eigenvalues in \eqref{joint_eig_dist}, $P_{\eta}(\lambda_1,\lambda_2)$ represents the JPD of eigenvalues of a $2 \times2$ Gaussian $\beta$-ensemble with an effective Dyson index, $\beta_e=2-2\eta \in  [0,2]$.\footnote{It is important to note that the distribution in eq. \eqref{JPD_eigenval_beta} is not exactly that of Gaussian $\beta$-ensemble with $N=2$, since $\beta_e$ can take the value zero (for $\eta=1$) which is not included in the standard $\beta$ Gaussian ensembles \cite{vivo2008invariant}.} 

When $\eta=0$, the JPD of the matrix elements $\mathbb{P}_{\eta} (u,v,x,y) $
factorises (see eq. \eqref{JPD_elements}), and the ensemble is reduced to the GUE with a strong correlation between the eigenvalues. While on the other limit of the range, when $\eta=1$, as can be seen from eq. \eqref{JPD_eigenval_beta}, the JPD of the eigenvalues factorises, while matrix elements themselves are strongly correlated \cite{vivo2008invariant}. Next, we use these $2 \times 2$
matrix ensembles for different values of $\eta$ to compute the EFS.  

\subsection{Ensemble-averaged fidelity susceptibility for $2 \times 2$ invariant $\beta$-ensemble}
We now use this  $2 \times 2$ matrix ensemble with an effective Dyson index
to study the geometry of eigenstates when a matrix belonging to this ensemble is added to an integrable Hamiltonian. Specifically, the total Hamiltonian once again takes the form \eqref{Ham_Roz_Por1}, where the integrable term $H_0$ is taken to be of the form
\begin{equation} \label{Ham_int}
		H_0 =\left(
		\begin{array}{ccc}
			h_1 & 0 \\
			0  & h_2
		\end{array}
		\right)~,
\end{equation}
where $h_1$ and $h_2$ are two independently distributed Gaussian variables with the same mean and variance, and $\mathcal{H}$ is 
drawn from a $2 \times 2$ invariant matrix ensemble with a generic Dyson index (eq. \eqref{Ham_eta}). 

The expression for $g_{rr}$ averaged over the ensemble can be written as
\begin{widetext}
\begin{equation}
\begin{split}
    \tilde{g}_{rr}(r) = \frac{1}{Z(\eta)} \int \frac{(\mathcal{H})_{12}(\mathcal{H})_{21}}{\big(E_n-E_m)^2 }~\Big(2\text{Tr}\mathcal{H}^2 - (\text{Tr}\mathcal{H})^2 \Big)^{-\eta}~
    \exp\Bigg[-\frac{1}{2} \text{Tr}\Big(H_0^2+\mathcal{H}^2\Big) \Bigg]~
    ~\td H_0 \td \mathcal{H}~,
\end{split}
\end{equation}
\end{widetext}
where, the normalisation constant $Z$ is given by 
\begin{equation}
\begin{split}
    Z(\eta) =  \int
    \exp\Bigg[-\frac{1}{2} \text{Tr}~\Big(H_0^2+\mathcal{H}^2\Big) \Bigg]~
    \Big(2\text{Tr}\mathcal{H}^2 - (\text{Tr}\mathcal{H})^2 \Big)^{-\eta}~\td H_0 \td \mathcal{H}~.
\end{split}
\end{equation}
Now using the matrix representations of $\mathcal{H}$ and $H_0$ from eqs. 
\eqref{Ham_eta} and \eqref{Ham_int}, we get the following expression for the EFS $\tilde{g}_{rr}(r)$,
\begin{widetext}
\begin{align}
\label{grr_integral}
    \tilde{g}_{rr}(r)&=  \frac{1}{Z(\eta)} \int \frac{ s^2 x_2^2 }{\big(4 s^2 r^2 + (x_2+r y_2)^2\big)^2\big(4 s^2 + y_2^2\big)^\eta} ~
    \exp \bigg[-\frac{1}{4}\Big(4 s^2+x_1^2+x_2^2+y_1^2+y_2^2\Big)\bigg]~ s \td s ~\td x_1 ~\td x_2 ~\td y_1~\td y_2~\\
    &=\frac{1}{Z(\eta)} \int \frac{ s^2 x_2^2 }{\big(4 s^2 r^2 + (x_2+r y_2)^2\big)^2\big(4 s^2 + y_2^2\big)^\eta} 
    \exp \bigg[-\frac{1}{4}\Big(4 s^2+x_2^2+y_2^2\Big)\bigg]~ s \td s ~\td x_2 \td y_2~.
\end{align}
\end{widetext}
Here we have defined new coordinates, $h_{1,2}=(x_1\pm x_2)/2$,
$u=(y_1+y_2)/2$, and $v=(y_1-y_2)/2$. 

Let us first consider the case $\eta=0$, where the integral can be explicitly evaluated to see that the EFS is given by the expression, 
\begin{equation}\label{EFS_eta0}
    \tilde{g}_{rr}(r, \eta=0)=\frac{1}{4r}\Bift{\text{arccot} (r)-\frac{r}{r^2+1}}~.
\end{equation}
This is essentially the same expression for the $rr$ component of the QMT in \eqref{metric_int_brk}, as should be expected from the JPD in \eqref{JPD_elements}. The difference in the scaling of $r$ is due to the difference in variances in the Gaussian distributions.

\begin{figure}[h!]
		\centering
		\includegraphics[width=3in,height=2.2in]{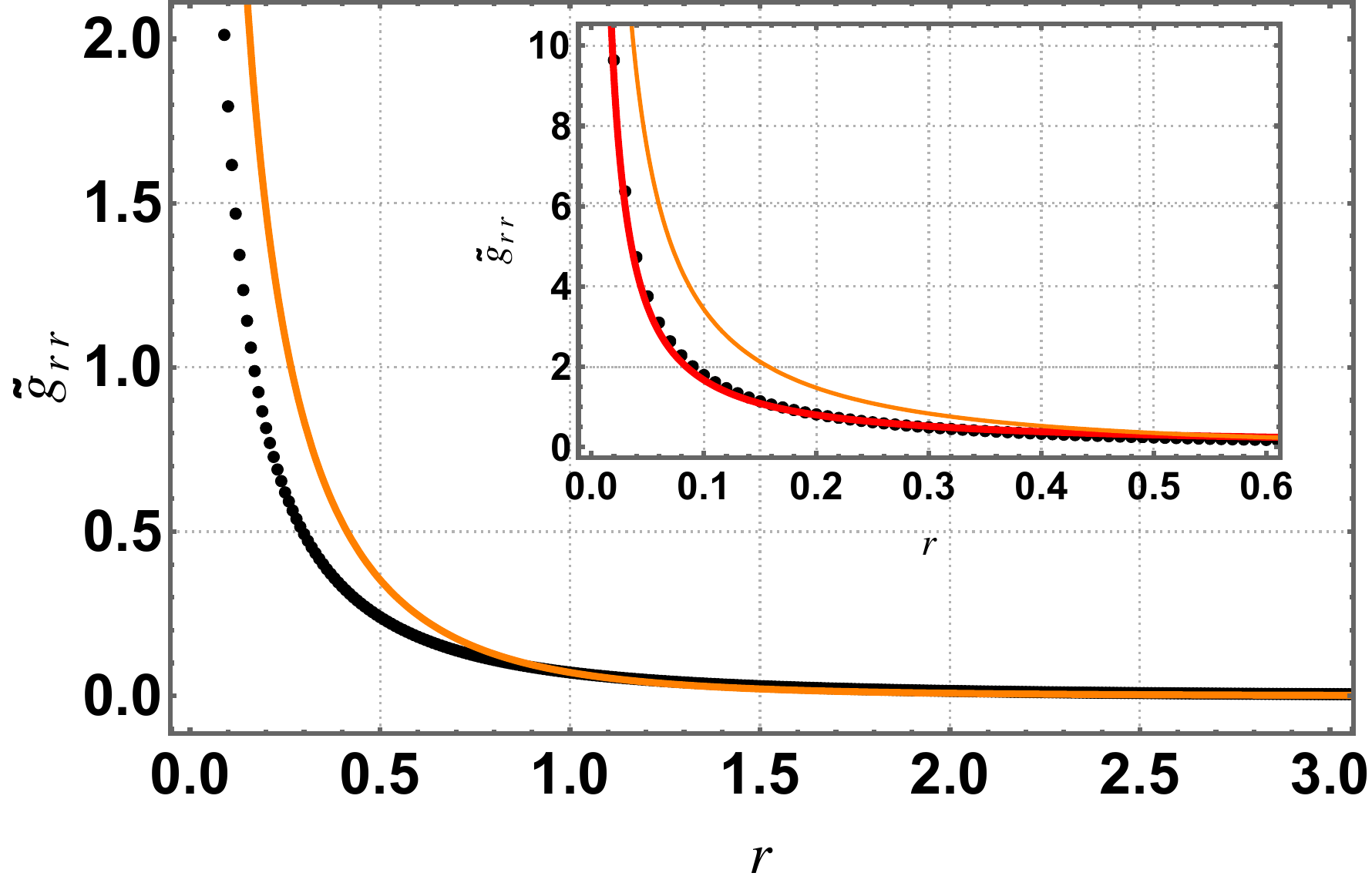}
		\caption{Plot of $\tilde{g}_{rr}(r)$ for the effective $\beta$-ensemble with $\eta=1$, i.e., $\beta_e=0$ (black dots).  The orange curve is the plot for the EFS with $\eta=0$ (see eq. \eqref{EFS_eta0}). The plot in the inset shows a fitting $\tilde{g}_{rr}(r) \approx 0.145/r^{1.06}$ (red curve) of the numerical data for small values of $r$.}
		\label{fig:grr_ETA1}
	\end{figure}

\begin{figure}[h!]
		\centering
		\includegraphics[width=3in,height=2.2in]{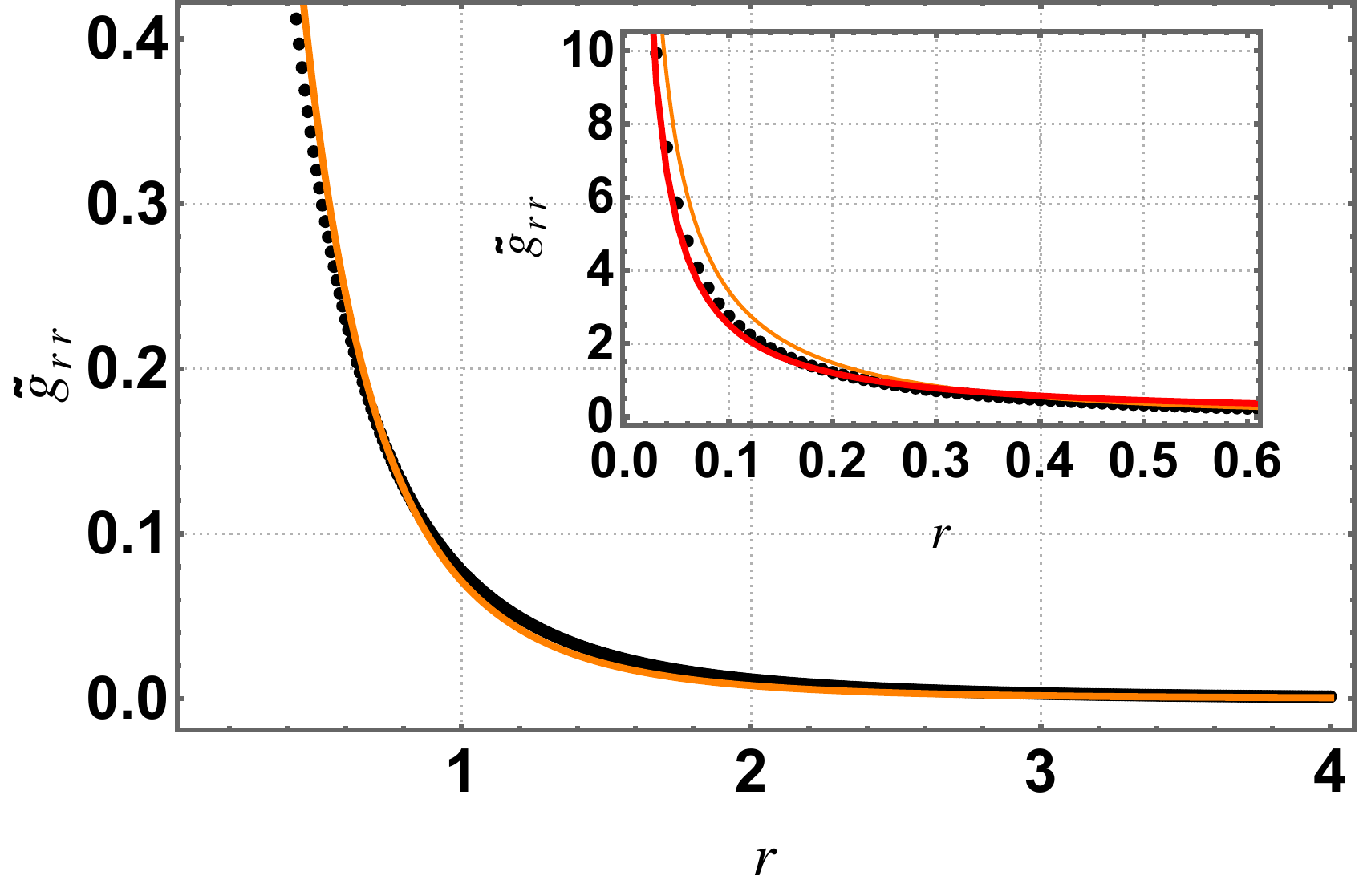}
		\caption{Plot of $\tilde{g}_{rr}(r)$ for the effective $\beta$-ensemble with $\eta=1/2$, i.e., $\beta_e=1$ (black dots).  The orange curve is the plot for the EFS with $\eta=0$ (see eq. \eqref{EFS_eta0}). The plot in the inset shows a fitting $\tilde{g}_{rr}(r) \approx 0.214/r^{1.07}$ (red curve) of the numerical data for small values of $r$.}
		\label{fig:grr_ETA12}
	\end{figure}

For generic values $\eta$, it is difficult to obtain an analytical expression for $\tilde{g}_{rr}(r)$ by performing all the above integrals.  Therefore, we evaluate it numerically for different fixed values of $\eta$. 
In Figs. \ref{fig:grr_ETA1} and \ref{fig:grr_ETA12} we have shown results of such numerical integration for $\eta=1$ (i.e., $\beta_e=0$)
and $\eta=1/2$ (corresponding to $\beta_e=1$), respectively. We observe that, for both cases, the metric component diverges close to $ r=0$, while in the limit $r \rightarrow \infty$, it approaches zero. This behaviour is quite analogous to the $rr$ component 
of the line element in \eqref{metric_int_brk}. This is expected, since,
if one directly considers the Hamiltonian $H=H_0+r \mh$, and computes the $rr$ component assuming $\mh$ is a matrix belonging 
to the GUE, then this is the same as the present case with $\eta=0$. 
For comparison, we have also plotted the $rr$ component of the metric in \eqref{metric_int_brk} in both Fig. \ref{fig:grr_ETA1} and Fig. \ref{fig:grr_ETA12} (magenta curve). As can be seen, the large $r$ behaviour for both $\eta=1$ and $\eta=1/2$ match quite well with the analytical expression in \eqref{metric_int_brk} (corresponding, in this section,  to $\eta=0$), however, they start to differ as $r$ is decreased (even though in all the cases the metric component diverges as $r \rightarrow 0$). 

Comparing plots with $\eta=1$ and $\eta=1/2$, another possible 
conclusion one can draw is that, even though there is a $(E_m-E_n)^2$ term in the denominator of the definition of $\tilde{g}_{rr}(r)$, its
decay with respect to $r$ may not be due to the correlation and repulsion between the eigenvalues but rather due to the fact that this energy difference term increases with increasing $r$. Thus, as $r \rightarrow \infty$, $\tilde{g}_{rr}(r)$ goes to zero even for $\eta=1$ - the case where there is no correlation between the eigenvalues of the Hamiltonian of this invariant $2 \times 2$ $\beta$-ensemble.

\section{Fidelity susceptibility in the chaotic phase: towards an analytical approximation}\label{sec_EFS_larger}
Consider again a quantum mechanical system with a total Hamiltonian of the form in eq. \eqref{Ham_Roz_Por1}: $H(r)=H_0+r \mh$, where both the component operators are random matrices. In the previous three sections, we discussed the behaviour of the EFS in three different scenarios: 1. Both the matrices $H_0$ and $\mh$ are drawn from independent classical Gaussian ensembles (e.g., GUE considered in \cite{Sharipov:2024lah}), see \S \ref{ssec_GUE_QMT}, where the expression for the EFS corresponding to the parameter $r$ is given by the $rr$ component of the metric in \eqref{metric_rmt}, and has $1/r^4$ scaling for large $r>1$. 2. $H_0$ is a random diagonal matrix with i.i.d. elements, and $\mh$ is a matrix belonging to the classical Gaussian ensembles, so that the total Hamiltonian is an element of the RP ensemble. The EFS, analytically computed for two-by-two matrix representations (eq. \eqref{metric_int_brk}), and numerically obtained for large-dimensional matrices (see Fig. \ref{fig:EFS_RP}), has a $1/r^4$ scaling for $r \gg 1$, i.e., in the ergodic phase of the system (see \S \ref{ssec_int_breaking_GUE}). 3. Similarly, when $H(r)$ is the sum of a random diagonal matrix, and a matrix drawn from the Gaussian $\beta$-ensemble with the Dyson index $\beta>1$, the EFS, computed analytically for $N=2$, as well as numerical results with higher-dimensional matrices also show the same scaling $1/r^4$ for $r \gg 1$ (see \S \ref{ssec_grr_trid_der} and Section \ref{sec_invt_beta_ensemb}). The purpose of this section is to provide a possible approximate semi-analytical explanation of this scaling behaviour by utilising the connection between the EFS and the two-point correlation function derived in \S \ref{sec_fs_cor}, and illuminate the role played by the DOS and the spectral form factor of the Hamiltonian in determining the EFS.\footnote{Unless specified otherwise, we shall assume for the purposes of this section that the limit of large dimension, $N \rightarrow \infty$ is considered.} 

We start by considering the generic relations in \eqref{FS_to_correl} and \eqref{S_omega}, from which we notice that in the expression for the ensemble-averaged spectral function $\mc{S}(\omega)$, the second term within the integral is nothing but the infinite-time average of the correlation function $G(t,r)=\braket{\mh(t)\mh}$, i.e., 
\begin{equation}\label{Cesaro_limit}
   G_\infty(r) := \lim_{T \rightarrow \infty} \frac{1}{T} \int _{0}^T \td t~ \braket{\mh(t)\mh} = \frac{1}{N} \mb{E} \sum_{n} |\braket{n(r)|\mh|n(r)}|^2~,
\end{equation}
where we have explicitly indicated that the infinite-time average depends on the parameter $r$ in the Hamiltonian \eqref{Ham_Roz_Por1}. 
This observation indicates that a useful way of writing this correlation function would be to separate out the infinite time average of the correlation function, and write the following ansatz for $G(t,r)$\footnote{We note that this may not be the most generic formula for the correlation function relevant for calculating the EFS. However, this is generic enough to capture, at least, some of the important features of the EFS we are interested in.  }
\begin{equation}\label{corre_ansatz}
    G(t,r)=\frac{1}{\mc{N}(r)} \bift{C_1(r)+C_2(r) F(t)}~,~~
\end{equation}
where $C_1, C_2$, and $\mc{N}$ are three time-independent constants (but they can depend on the parameter $r$) to be determined, and $F(t)$ is a time-dependent function, which we assume to be well-behaved and continuous. Now, we shall impose two conditions on the time-dependent function $F(t)$, one at the initial time as $t=0$, and the other at the limit of infinite time, $t \rightarrow \infty$.  Namely, we assume that 
$F(t=0)=1$ (i.e., the constant value of $F(t=0)$ being absorbed in the constant $C_2$), and in the limit $t \rightarrow \infty$ the function $F(t)$ decays to zero. One way of ensuring the second condition would be to assume that it is the Fourier transform of an absolutely integrable function defined on the real line, so that the condition $F(t \rightarrow \infty)=0$
is satisfied as a consequence of the Riemann-Lebesgue lemma  \cite{strichartz2003guide, rudin}.
We also note that, to be precise, the condition should be defined in terms of the infinite time average of the function $F(t)$. However, according to the Cesaro mean theorem for integrals, if the standard limit for the function $F(t)$ exists, then the Cesaro limit (eq. \eqref{Cesaro_limit}) must match it. Since, by our assumption, the standard limit for the function $F(t)$ at $t \rightarrow \infty$ vanishes, the infinite-time average is also zero.


The basic idea is to determine the two independent constants $C_1/\mc{N}$ and $C_2/\mc{N}$ in terms of $G(t=0)=G_0$ and $G_\infty(r)$, respectively, the initial value and the infinite time average of the correlation function.  The initial condition that $G(t=0)=\braket{\mh^2}$ determines the overall constant in terms of the initial value of the correlation function: $\mc{N}=(C_1+C_2)/\braket{\mh^2}$. 
On the other hand, denoting the infinite-time average of the correlation function by $G_{\infty}$, we have $G_{\infty}=C_1/\mc{N}$. This, in turn, implies, 
\begin{equation}
    G_{\infty}=\frac{C_1}{\mc{N}}=\frac{C_1}{C_1+C_2} \rightarrow  C_2=C_1\Bift{\frac{\braket{\mh^2}}{G_\infty}-1}~.
\end{equation}
Therefore, our final  ansatz for the correlation function is given by 
\begin{equation}\label{corre_ansatz}
    G(t)=G_{\infty}(r)+\Bift{\braket{\mh^2}-{G_\infty}(r)} F(t)~,
\end{equation}
and the expression for the EFS, obtained by substituting this ansatz in eqs. \eqref{FS_to_correl} and \eqref{S_omega} is
\begin{equation}\label{G_to_EFS_appr}
    \bar{g}_{rr}(r)= \Bift{\braket{\mh^2}-{G_\infty}}\int_{-\infty}^{\infty}~\frac{\td \omega}{\omega^2} \int_{-\infty}^{\infty}~\frac{\td t}{2 \pi} e^{i \omega t} F(t)~.
\end{equation}

To illustrate the use of this generic relation in practical applications, first, consider the case we discussed in \S \ref{ssec_GUE_QMT}, i.e., the Hamiltonian is a sum of two GUE matrices having the same variance (say, $\sigma^2$), so that the variance of $H(r)$ is $\sigma^2(r^2+1)$.\footnote{From now on we shall fix $\sigma^2=1/N$.} In this case, the correlation function $G(t)$ can be calculated exactly in terms of the spectral form factor (SFF), and its expression is given by \cite{Gill:2025upp},
\begin{equation}\label{GUE_coor_final}
 G^{GUE}(t,r)=\frac{1}{(r^2+1)}\Bigg(\frac{\mathcal{R}(t)}{N^2}+r^2\Bigg)~,
\end{equation}
where $\mathcal{R}(t)$ denotes the SFF associated with the total Hamiltonian, and is defined by 
\begin{equation}\label{SFF}
    \mathcal{R}(t) :=  \frac{1}{Z}\int \big|\text{Tr} \big(e^{- i H t}\big)\big|^2~ P(H)~ \td H~; 
\end{equation}
$Z$ being a constant that  normalises the distribution $P(H)$. In terms of the spectral two-point function, $\rho^{(2)}(E_1,E_2)$, the SFF is given by the following standard expression \cite{haakebook, Cotler:2017jue, stockmann2007quantum,PhysRevE.55.4067}
\begin{align}\label{SFF_twopt}
     \mathcal{R}(t)&= \int \mathcal{D}E~\sum_{i,j}e^{i(E_i-E_j)t}\\
     &= N+N(N-1) \int d E_1 dE_2 ~\rho^{(2)}(E_1,E_2) e^{i(E_1-E_2)t}~.
\end{align}

The above expression for the GUE correlation function is of the same form as our ansatz in \eqref{corre_ansatz}, where the time-dependent function $F(t)$ is just the SFF minus its constant part, with an overall $N$-dependent scaling factor which is $\mc{O}(1)$ in the limit of large $N$. Note that, here the infinite time average of the correlation function is given by, 
\begin{equation}
    G^{GUE}_\infty(r)= \frac{1+N r^2}{N(r^2+1)}~,
\end{equation}
a part of which comes from the SFF (the constant part of the SFF). Furthermore, in the limit $r \gg 1$, the infinite-time average value approaches unity, which is the initial value of the correlation function, for our choice of the variance of $\mh$. 
It was shown in \cite{Gill:2025upp, Gill:inprep} that the above expression for the GUE correlation function can be employed to derive the analytical expression for the EFS.

In summary, combining the relation \eqref{corre_ansatz} with \eqref{GUE_coor_final} and \eqref{SFF_twopt}, we propose the following ansatz for the correlation function $G(t,r)=\braket{\mh(t)\mh}$ for time evolution generated by the one-parameter class of Hamiltonians $H(r)=H_0+r \mh$:
\begin{align}\label{G_SFF_final}
    G(t,r) &= G_{\infty}(r)+\frac{\braket{\mh^2}-{G_\infty}(r)}{N(N-1)}\Bitd{\mc{R}(t)-N}
    \\&= G_{\infty}(r)+\Bift{\braket{\mh^2}-{G_\infty}(r)} \int \td E_1 \td E_2 ~\rho^{(2)}(E_1,E_2) e^{i(E_1-E_2)t}~.
\end{align}
We expect this approximation to be valid only for $r \gg 1$, even though for the RP model, as we shall show below, this is a good approximation even for $ r<1$.

The expression for the EFS obtained from this ansatz for the correlation function is therefore given by\footnote{We note that this integral is divergent in the limit $E_1 \rightarrow E_2$, so that to evaluate it we need to calculate the finite part of this integral. }
\begin{align}
    \bar{g}_{rr}(r)=  \Bift{\braket{\mh^2}-{G_\infty}(r)} \int \td E_1 \td E_2 ~\frac{\rho^{(2)}(E_1,E_2) }{(E_1-E_2)^2}~.
\end{align}
This establishes a direct connection between the EFS and the spectral two-point function - a measure for the short-range energy level correlation in the spectrum. 

Another observation made in \cite{Gill:2025upp, Gill:inprep}, and useful for our present purpose is that due to the $1/N^2$ factor with the SFF in the expression in \eqref{GUE_coor_final}, in the large $N$ limit, the time-dependent part of the correlation function can be approximated by the part of the SFF coming from the disconnected part of the energy correlation function, i.e, the DOS of the Hamiltonian $H(r)$. This is nothing but the absolute square of the ensemble-averaged trace of the time evolution operator, i.e., $|\braket{e^{-i t H(r)}}|^2$.  In fact, this approximation is enough to recover the correct scaling of the EFS with $r$ \cite{Gill:2025upp, Gill:inprep}. 
Using this fact in the last ansatz for the correlation function, we can write an approximate expression in terms of the ensemble average of the trace of the time evolution operator,
\begin{align}\label{G_SFF_disc}
    G(t,r) \approx  G_{\infty}(r)+\Bift{\braket{\mh^2}-{G_\infty}(r)} C(N) |\braket{e^{-i t H(r)}}|^2~,
\end{align}
where $C(N)$ is a positive constant, which is independent of the parameter $r$, and is of $\mc{O}(1)$ in the large $N$ limit. For the case of GUE matrices considered above, this is given by $C=N^2/(N^2-N)$. Since here we are interested in finding out the scaling with $r$, the exact expression of this constant will not be very important for our purposes in the following, and will be ignored in the numerical computations reported below. Note that, the first two expressions above are equivalent, as can be checked using the definition of the SFF.

Before proceeding to find out the scaling of the EFS using this correlation function, here we emphasise that, even though the expression for the correlation function in \eqref{G_SFF_final} is a reasonable formula, we do not expect it to be valid for any generic Hamiltonian of the form $H(r)=H_0+r \mh$. We shall only use it to calculate the scaling of the EFS for a Hamiltonian of the form $H(r)=H_0+r \mh$ for large values of $r$ when $\mh$ is a random matrix belonging to an ensemble. Furthermore, when $H_0$ is an integrable Hamiltonian, and $\mh$ is chaotic, we only expect the above formula to be valid for $r \gg 1$ (and in the limit $N \rightarrow \infty$, a condition, which we have assumed to be true throughout this section). 

With these lessons gained from the exact correlation function, and the approximate ansatz for the correlation function in \eqref{G_SFF_final} and \eqref{G_SFF_disc}, we turn to the cases of the RP model and $\beta$-Gaussian ensemble. The strategy we shall follow to derive the scaling of the EFS for large values $r$ for these Hamiltonians is the following: 1. Numerically evaluate the infinite-time average of the correlation function $G(t,r)$ using eq. \eqref{Cesaro_limit}, and find out the approximate scaling as a function of $r$ for $r>1$. 2. Plot the correlation function $G(t)$ numerically, and compare it with our proposed ansatz in eqs. \eqref{G_SFF_final} and \eqref{G_SFF_disc}. 3. Use the scaling of $G_\infty(r)$ with $r$ along with the following formula (obtained by combining eq. \eqref{G_SFF_disc} and \eqref{G_to_EFS_appr}) 
\begin{equation}\label{EFS_app_2}
    \bar{g}_{rr}(r) \sim \text{Abs} \BBtd{\Bift{\braket{\mh^2}-{G_\infty}}  C(N)\int_{-\infty}^{\infty}~\frac{\td \omega}{\omega^2} \int_{-\infty}^{\infty}~\frac{\td t}{2 \pi} e^{i \omega t} |\braket{e^{-i t H(r)}}|^2~},
\end{equation}
to derive the scaling of the EFS with $r$. Notice that, we have taken the absolute value of the expression on the right-hand side to emphasise that, since this is not the exact expression, even though it reproduces the correct scaling, it can give rise to the wrong sign \cite{Gill:2025upp}. 

The final expression for the EFS above can be written in a more meaningful form in terms of the DOS of the Hamiltonian by noticing that the expected trace of the time evolution operator is nothing but the Fourier transform of the DOS,
\begin{equation}
    \braket{e^{-i t H(r)}} = \int_\mc{S}~ \td E ~\rho_H(E) e^{-i t E(r)}~,
\end{equation}
where the integral is over the support of the DOS on the real line (denoted as $\mc{S}$).
Using this in \eqref{EFS_app_2}, we have the following final expression for the EFS \cite{Gill:2025upp, Gill:inprep}
\begin{widetext}
\begin{align}\label{EFS_from_DOS}
    \bar{g}_{rr}(r) &\sim \text{Abs} \BBtd{\Bift{\braket{\mh^2}-{G_\infty}(r)}  C(N) \int_{\mc{S}(r)}~~ \td E_1 \td E_2 \frac{\rho_H(E_1) \rho_H(E_2)}{(E_1-E_2)^2}}\\
    & \sim \text{Abs} \BBtd{\Bift{\braket{\mh^2}-{G_\infty}}  C(N)  \int_{\mc{S}(r)}~~ \td E_1 ~\rho_H(E_1)
    ~\text{Re} \Bitd{G_{H}(E_1+ i0^+)}}~,
\end{align}
\end{widetext}
where $G_{H}(z)$ denotes the ensemble-averaged trace of the resolvent operator associated with a Hamiltonian, which (in the $ N \rightarrow \infty$ limit) is also given by the Cauchy transform of the DOS,
\begin{equation}
    G_{H}(z) = \int_\mc{S}~ \td E~ \frac{\rho_H(E)}{z-E}~.
\end{equation}
In the above formula for the EFS, we have also explicitly indicated that, in general, the support of the DOS on the real line depends on the parameter $r$.

As an example, when the DOS of the Hamiltonian $H(r)$ is the Wigner semicircle with support from $-R(r)$ to $R(r)$, the expression for the Cauchy transform is given by \cite{Potters}, 
\begin{equation}
    G_{sc}(z)=\frac{2}{R(r)^2}\Bift{z-\sqrt{z^2- R(r)^2}}~.
\end{equation}
Therefore, in that case, the expression for the scaling of the EFS for large $r$, obtained from the formula \eqref{EFS_from_DOS} is 
\begin{equation}
    \bar{g}_{rr}(r) \sim  \Bift{\braket{\mh^2}-{G_\infty}(r)} \frac{2C(N)}{R(r)^2} 
\end{equation}
Hence, in this case, the scaling is determined by the radius of the semicircle and the infinite time average of the correlation function $G(t)$.

\begin{figure}[h!]
		\centering
		\includegraphics[width=3.4in,height=2.2in]{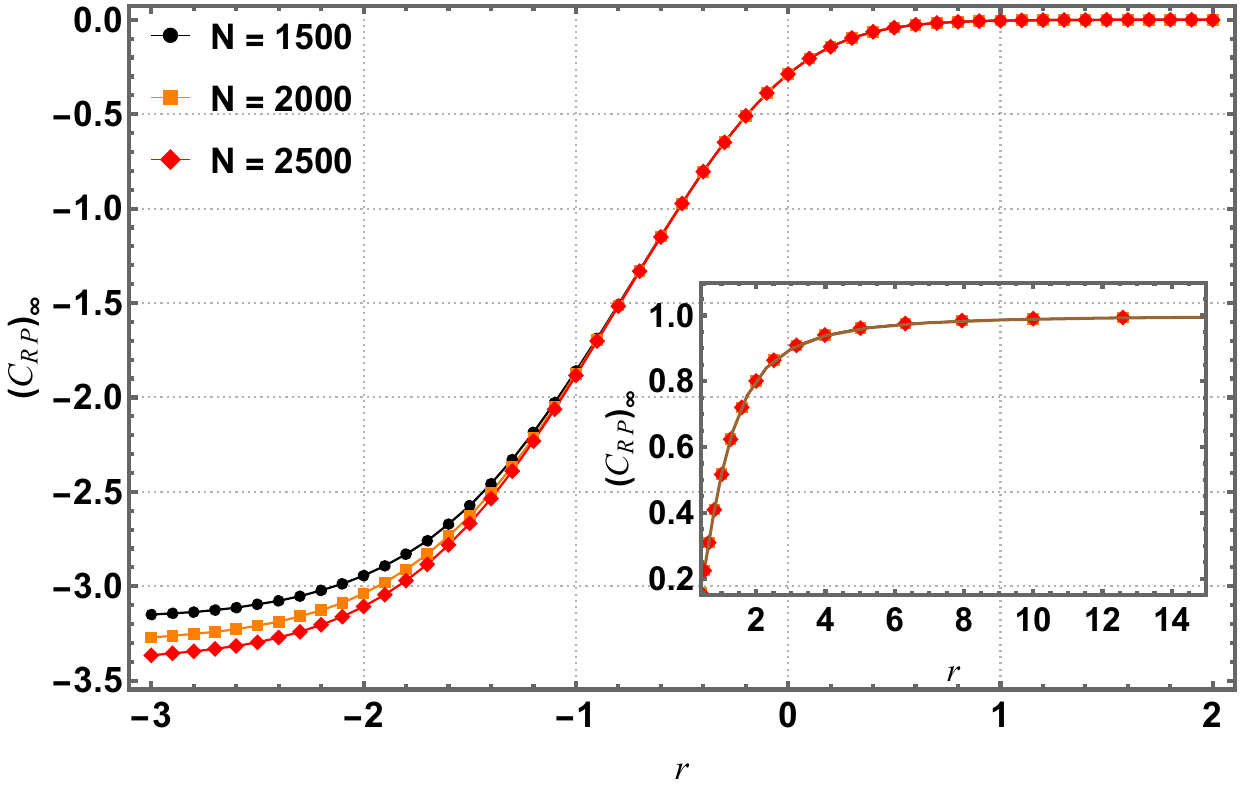}
		\caption{Infinite-time average of the correlation function computed from \eqref{Cesaro_limit} (shown in the main panel with Log-Log scaling), plotted after an average over $500$ independent realisations of the RP model Hamiltonian.  The plot in the inset shows the infinite-time average values for large $r>0.5$ in the linear scale. The brown curve indicates the approximate scaling for large values of $r$ given by the function in eq. \eqref{satur_fit_RP} with $\kappa \approx 1.91$, and $r_0 \approx 0.96$.}
		\label{fig:correl_RP_satur}
	\end{figure}

\paragraph{The RP model.} For the standard RP model  \cite{rosenzweig1960repulsion, kravtsov2015random}, the operator $\mh$ is a matrix drawn from the GUE with a variance which we fix here to be $1/N$. Since we need the infinite-time average of the correlation function $G(t)$ to use the analytical ansatz in either of the equations. \eqref{G_SFF_final} and \eqref{G_SFF_disc} (step 1 of the procedure described above), in Fig. \ref{fig:correl_RP_satur}, we have plotted the infinite-time average value of the correlation function numerically computed using \eqref{Cesaro_limit}. As can be seen, as $r$ is increased, the infinite-time average increases, and approximately saturates close to the initial value for $r \gg 1$.
The fitting curve (the brown curve in the inset of Fig. \ref{fig:correl_RP_satur}) is obtained from a fitting of the numerical data with the largest $N$, and is given by the following two-parameter family of functions (the superscript $n$ indicates that it is a numerically obtained fitting)
\begin{equation}\label{satur_fit_RP}
    G^n_\infty(r) = \frac{r^\kappa}{r^\kappa+r_0^\kappa}~.
\end{equation}
From the best-fitting of the numerical data for the infinite-time average with $N=2500$, we obtain $\kappa \approx 1.91$, and $r_0 \approx 0.96$.\footnote{For other values of $N$ we have shown Fig. \ref{fig:correl_RP_satur}, the fitting gives the same values of $\kappa$ and $r_0$ up to two decimal places. We also note that the fit given in \eqref{satur_fit_RP} is expected to be valid for $r \gg 1$, even though the numerical values of the two parameters $\kappa$ and $r_0$ reported are obtained by fitting the full set of data shown in Fig. \ref{fig:correl_RP_satur}. If one instead uses the saturation values only with $r>1$, then $\kappa$ gets closer to $2$. } 

\begin{figure}[h!]
		\centering
		\includegraphics[width=3.4in,height=2.2in]{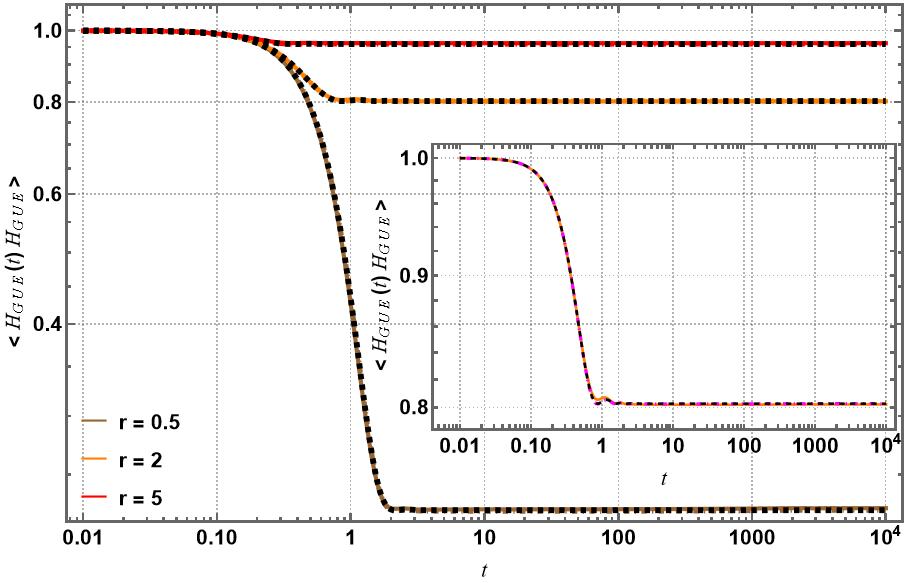}
		\caption{Time evolution of the correlation function $G(t)=\braket{\mh(t)\mh}$ (in the Log-Log scale), with the time evolution generated by the RP model Hamiltonian for different values of $r$, where $\mh$ is a matrix drawn from GUE with variance $1/N$. The numerical results are shown with an average over $500$ independent realisations of the ensemble, and we have fixed $N=1000$. The solid curves show the exact numerical correlation function, whereas the dot-dashed curves show the analytical correlation function written in eq. \eqref{G_SFF_final}. The black dashed curves show the expression in \eqref{G_SFF_disc} for the three values of $r$. As is seen in the main panel, the numerically computed correlation functions are practically identical to those obtained from the two analytical expressions. For a better visualisation of the differences between the numerical results and the two analytical expressions, in the inset we have shown the correlation function with $r=2$ separately, where the orange curve is the numerical result, the magenta dot-dashed curve is the expression in \eqref{G_SFF_final}, and the black dashed curve represents the expression in \eqref{G_SFF_disc}. }
		\label{fig:correl_RP}
	\end{figure}

In Fig. \ref{fig:correl_RP}, we have plotted the numerically evaluated correlation function $G(t,r)=\braket{\mh(t)\mh}$ with the time evolution generated by the RP model Hamiltonian (the solid curves), along with the analytical expressions in eqs. \eqref{G_SFF_final} and \eqref{G_SFF_disc}.  To evaluate the analytical expressions in eqs. \eqref{G_SFF_final} and \eqref{G_SFF_disc}, we have used the infinite-time average of the correlation function given in \eqref{satur_fit_RP} with the values of the constants given above. From this plot, we glean that both analytical approximations provide quite a good description of the numerical results. 

As was anticipated from the data for the infinite-time average in Fig. \ref{fig:correl_RP_satur}, the nature of decay of the correlation function strongly depends on $r$.  For small values of $r<1$, the correlation function decays from the initial values with time and approximately saturates to a value close to zero. However, as $r$ is increased beyond $r>1$, the decay of the correlation function becomes less pronounced, so that when $ r \gg 1$, it barely decays from the initial value. Of course, this can be understood from the fact that when $ r \gg 1$, the total Hamiltonian is dominated by $\mh$, hence, $G(t)$ is approximately given by $\braket{\mh^2}$. 

Now we move on to the derivation of the scaling of the EFS for large $r$. To this end, we use the well-known fact that in the ergodic phase of the RP ensemble, the expression for the DOS is well described by a Wigner semi-circle. Specifically, for $ r \gg 1$, the DOS is given by the semicircle distribution written in \eqref{semi-circle_dist}, with support from $-2r$ to $2r$ on the real line \cite{kravtsov2015random, guhr1996transitions}. Making use of the Cauchy transform of this DOS, given by \cite{Potters},
\begin{equation}
    G_{sc}(z)=\frac{1}{2r^2}\Bift{z-\sqrt{z^2-4 r^2}}~,
\end{equation}
in the expression for the EFS in \eqref{EFS_from_DOS}, along with the fact that, for our choice of variance $\braket{\mh^2}=1$, we have
\begin{equation}
    \bar{g}_{rr}(r \gg 1) \sim \Big|\BBft{\frac{r_0^\kappa}{r^\kappa+r_0^\kappa}} \BBft{\frac{-1}{2 r^2}} \Big|  \sim \frac{1}{r^{2+\kappa}}~.
\end{equation}
Since from the numerical fit we have $\kappa \approx 1.91$, this explains the approximate $1/r^4$ scaling of the EFS in the RP model. We expect that $\kappa \rightarrow 2$, in the thermodynamic limit. 

In this section, we have taken the first steps of understanding the behaviour of the EFS 
for random matrix Hamiltonians beyond the classical Gaussian ensembles from the time-correlation function of observables, and without directly using perturbation theory techniques. In the process, we have derived a generic ansatz for the said correlation function, and subsequently, using the connection between the time-dependent part of the correlation function and the DOS of the Hamiltonian, we were able to derive the approximate scaling for the EFS in the large $r$ limit. As should be clear, this is only a semi-analytic technique, which requires the information of the saturation value of the correlation function from numerical computations, and hence, this may fail to describe the scaling of the EFS for more generic Hamiltonians. However, we hope that this method reveals some of the very interesting connections between the EFS, SFF, and DOS of the system, thereby shedding light on the interrelations among common quantifiers of chaos and the chaos-to-integrability transitions in quantum many-body systems, which are more generic than the random matrix ensembles we have considered in this paper \cite{Gill:2025upp, Gill:inprep}.

\section{Energy level dynamics: ensemble-averaged level  curvature}\label{sec_level_curvature}

\subsection{Second derivative of the energy levels}

Another interesting quantity, which is very similar to the FS considered so far, however, is a direct quantifier of the nature of the dependence of the energy levels of a complex many-body quantum system on the parameters (rather than the eigenstates), is the so-called \textit{energy level curvature}. For a Hamiltonian of the form in eq. \eqref{Ham_Roz_Por1}, the curvature of the $n$th energy level $\mk_n(r)$ is defined as the second derivative of $E_n(r)$ with respect to the parameter $r$, i.e.,\footnote{Compared to the standard definition of the level curvature that exists in the literature \cite{haakebook, stockmann2007quantum, Pechukas, Yukawa, Zakrzewski:2023jpb}, we have absorbed a factor of 2 in the definition of $\mk_n$ for convenience. Also note that the expression in \eqref{E_curvature} is valid only for a Hamiltonian which can be written as $H(r)=H_0+r \mh$. The general expression for the second-order derivative of $E_n(r)$ of a generic Hamiltonian $H(r)$ contains an additional term which is proportional to $\braket{n(r)|d^2H(r)/dr^2|n(r)}$ \cite{Pechukas, Yukawa}. Furthermore, the sum of energy level curvatures over all the eigenvalues $E_n(r)$ is trivially zero, $\sum_n\mk_n=0$.} 
\begin{equation}\label{E_curvature}
    \mk_n(r)=\frac{1}{2}\frac{d^2E_n(r)}{dr^2}=\sum_{j(\neq n)} \frac{\mathcal{H}_{nj}(r)\mathcal{H}_{jn}(r)}{(E_n(r)-E_j(r))}~.
\end{equation}
This can be rewritten as 
\begin{equation}\label{EC_to_correl}
    \mk_n = -\int_{-\infty}^{\infty}~\frac{\td \omega}{\omega} ~\mathcal{S}_n(\omega) ~,
\end{equation}
where the function $\mathcal{S}_n(\omega)$ defined in eq. \eqref{S_n_omega}. Comparing with the expression for the moments of $\mathcal{S}_n(\omega)$ in eq. \eqref{moments_S_n}, we see that moment with $k=-1$ is just the $n$-th energy level curvature, i.e., $\mk_n=-M_{-1}^{(n)}$. 

Furthermore, defining the quantity $\chi_\mh(z)$ through the Cauchy transform of the function $\mathcal{S}_n(\omega)$,
\begin{equation}\label{Cauchy-S_n}
    \chi_\mh(z) = \int_{-\infty}^{\infty}~\frac{\td \omega}{z-\omega} ~\mathcal{S}_n(\omega)~,
\end{equation}
we see that the $n$-th level curvature can be written as $\mk_n= \chi_\mh(0)$. Note that the relation in \eqref{Cauchy-S_n}
is analogous to the relation between the dynamical susceptibility (which is an appropriate extension of the advanced or retarded response function in the complex plane) and the spectral function of a Hermitian operator in the linear response theory \cite{coleman2015introduction, altlandsimons}. 

\subsection{Energy level curvature for a $2 \times2$ tridiagonal $\beta$-ensemble}

We now compute the ensemble-averaged energy level curvature for an eigenstate of the Hamiltonian $H(r)=H_0+r \mh$, with $H_0$ and $\mh$ being the same matrices considered in \S \ref{ssec_grr_trid_der} (eqs. \eqref{H_0} and \eqref{H_beta}, respectively). Computation of the ensemble-average of the energy level curvature ($\bar{\mk}_n(r)$) for a specific value $n$ proceeds along a similar line as the EFS.  It is possible to obtain an exact expression for $\bar{\mk}_2 (r, \beta)$ (or $\bar{\mk}_1 (r, \beta)$) for a generic value of the Dyson index $\beta>0$ for the $2 \times2$ matrix ensemble by performing the relevant integrations. However, the final expression for it is quite lengthy, so we omit writing its expression here. For specific values of $\beta$, the expressions do get simplified; e.g., with $\beta=2$, we get the following expression for the ensemble-averaged second energy level curvature,
\begin{equation}
 \bar{\mk}_2 (r, \beta=2) = \frac{1}{4 \sqrt{\pi} \sigma_0} \Bigg(\text{arccosh}\bigg(\frac{4 \sigma_0^2}{r^2}+1\bigg)-\frac{2\sqrt{2}\sigma_0}{\sqrt{2\sigma_0^2+r^2}}\Bigg)~,
\end{equation}
while for $\beta=4$ the expression simplifies to  
\begin{multline}
 \bar{\mk}_2 (r, \beta=4) = \frac{1}{32 \sqrt{\pi} \sigma_0^3} \bigg((8 \sigma_0^2+3r^2)~\text{arccosh}\bigg(\frac{8 \sigma_0^2}{r^2}+1\bigg)\\-12\sigma_0\sqrt{4 \sigma_0^2+r^2}\bigg)~. 
\end{multline}

	\begin{figure}[h!]
		\centering
		\includegraphics[width=2.7in,height=2in]{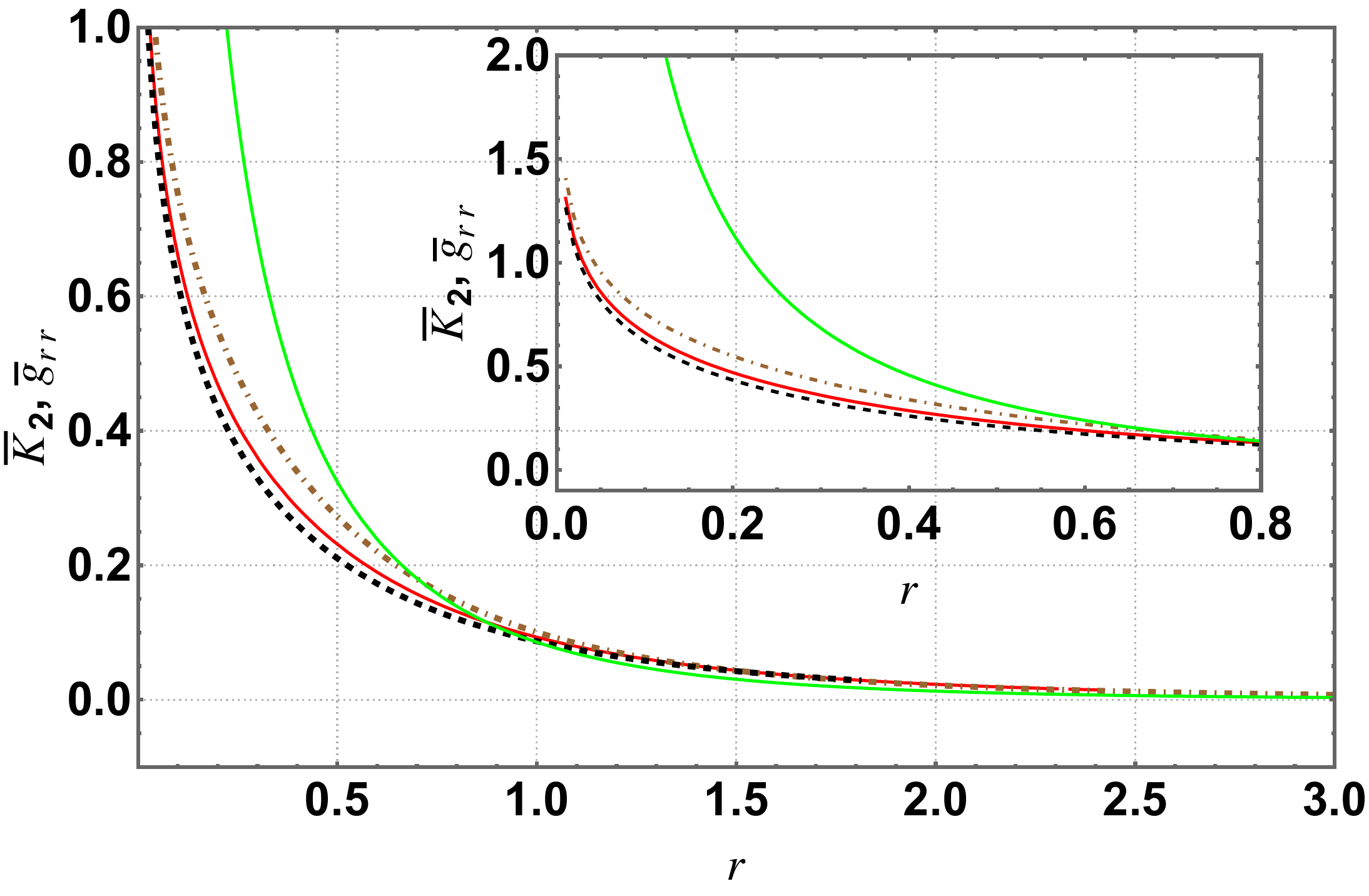}
		\caption{Plot of the second energy level curvature $\bar{\mk}_2 (r, \beta)$ for the total Hamiltonian $H(r)=H_0+r \mh$, where $H_0$ is a diagonal integrable matrix, and $\mh$ is sampled from tridiagonal $\beta$-ensemble with generic $\beta$: $\beta=1$ (dot-dashed brown) $\beta=2$ (red) and $\beta=4$ (dashed black), with $\sigma_0=1$. The eigen level curvature behaves as $\log r$ for small values $r$ and vanishes at large $r$ for generic values of $\beta>0$. For comparison, we have also plotted the EFS obtained in the same scenario with $\beta=2$ (the green curve), which scales as $1/r$ for small $r$. Plot in the inset shows the difference in the nature of divergence of the energy level curvature and the FS in the limit $r \rightarrow 0$.}
		\label{fig:E_CUV}
	\end{figure}

Series expansions of these quantities for small values of $r$  show that they behave as $\log r$ for small values of $r$, while for large values of $r$, they go to zero. In contrast to the EFS studied in \S \ref{ssec_grr_trid_der}, the energy level curvature scales as $1/r^3$ for large $r$. From the analytical expressions written above, we have for $r \gg 1$, the following approximations 
\begin{align}
    \bar{\mk}_2 (r\gg1, \beta=2)  &\sim \sqrt{\frac{2}{\pi} } \frac{\sigma_0^2}{3} \frac{1}{r^3}~,\nonumber\\
    ~~\text{and}~~\bar{\mk}_2 (r \gg1, \beta=4)  &\sim \sqrt{\frac{1}{\pi} } \frac{8\sigma_0^2}{15} \frac{1}{r^3}~,
\end{align}
which shows the $1/r^3$ scaling for large $r$.

In Fig. \ref{fig:E_CUV}, we have shown the plots of $\bar{\mk}_2 (r, \beta)$ for $\beta=1,2,4$ along with the plot of the EFS with $\beta=2$ derived in \S \ref{ssec_grr_trid_der} (see the expression of $\tilde{g}_{rr}(r)$ in eq. \eqref{metric_int_brk}), which diverges as $1/r$ in the limit $r \rightarrow 0$, and scales as $1/r^4$ for large values of $r$. This illustrates the difference between the behaviour of the ensemble-averaged level curvature and the EFS, both close to the integrability and deep in the chaotic phase.

\subsection{Third derivative of the energy levels}

Before concluding this section, let us now briefly consider the third derivative of the energy levels, which can be written, for a Hamiltonian of the form $H(r)=H_0+r \mh$, as 
\begin{align}
    \ml_n&=\frac{1}{6}\frac{d^3E_n(r)}{dr^3}\hspace{3 cm}\nonumber\\
    &=\frac{1}{6}\Big(\braket{n^{\prime\prime}(r)|\mh|n(r)}+\braket{n(r)|\mh|n^{\prime\prime}(r)}+2\braket{n^{\prime}(r)|\mh|n^\prime(r)}\Big)~.
\end{align}
Using the formula for the $\braket{n^{\prime\prime}(r)|k(r)}$ for $k \neq n$ obtained from the eigenvalue equation for $H(r)$, we get the following expression for the third derivative $\ml_n$, 
\begin{equation}\label{E_n_3rd}
     \ml_n(r)= \sum_{k (\neq n), m(\neq n)} \frac{\mh_{kn}\mh_{nm}\mh_{mk}}{E_{nk}E_{nm}} - \mh_{nn} \sum_{k (\neq n) }  \frac{|\mh_{nk}|^2}{E_{nk}^2}~,
\end{equation}
where $E_{nk}=E_n-E_k$.
Note that the second term in the above expression is proportional to the FS of the $n$-th eigenstate, eq. \eqref{fid_sus_rr}. 

We can obtain the connection with the correlation functions, as in the case of the FS and the second derivative of $E_n$, as follows. 
First note that since the second term above is proportional to the FS, it can be written in terms of the function $\mathcal{S}_n(\omega)$ (which is the Fourier transform of the connected two-point correlation function). Using a similar strategy to the manipulation of eq. \eqref{E_to_Delta}, we can write the first term in  \eqref{E_n_3rd} as 
\begin{equation}
    \int \frac{\td \omega}{\omega} \frac{\td \omega^\prime} {\omega^\prime} \mathcal{Q}_n (\omega, \omega^\prime)~
\end{equation}
where we have defined the function $\mathcal{Q}_n (\omega, \omega^\prime)$, which is a successive Fourier transformation of a three-point correlation function:
\begin{widetext}
    \begin{equation}
\mathcal{Q}_n (\omega, \omega^\prime)=  \int \frac{\td t}{2 \pi} \frac{\td t^\prime}{2 \pi} e^{-i(t \omega + t^\prime \omega^\prime)} \Big(\braket{n|\mh(t^\prime)\mh(0)\mh(-t)|n}-2\braket{n|\mh(t)\mh(0)|n}\mh_{nn}+\mh_{nn}^3\Big)~.
\end{equation}
\end{widetext}
To ensure the convergence of these integrals, sometimes one has to necessarily introduce factors of $\exp({-\epsilon |t|})$. 

Using this integral representation, the final expression for $\ml_n(r)$ in terms of the correlation functions is given by 
\begin{equation}
    \ml_n(r) = \int_{-\infty}^{\infty} \frac{\td \omega}{\omega} \frac{\td \omega^\prime} {\omega^\prime} \mathcal{Q}_n (\omega, \omega^\prime) - \int_{-\infty}^{\infty}~\frac{\td \omega}{\omega^2} ~\mathcal{P}_n(\omega) ~,
\end{equation}
where $\mathcal{P}_n(\omega)=\mh_{nn}\mathcal{S}_n(\omega)$. A detailed analysis of $\ml_n$ for classical Gaussian random matrix Hamiltonians, or the systems we have considered in the previous sections, is beyond the scope of this paper, and we hope to report on it in the future \cite{Gill:inprep}.

\section{Conclusions and Discussion}\label{sec_conclusions}

In this work we have analysed the geometry of the eigenstates for random matrix Hamiltonians that belong to some specific ensembles. The random matrix universality in the energy eigenvalues is commonly taken as a signature of chaos of many-body quantum systems in the sense that fine-grained properties of the energy spectrum of these complex quantum many-body systems are well approximated by the random matrix theory \cite{bohigas1984characterization, berry1985semiclassical}. Nevertheless, it is also important to understand the properties of the eigenstates of these many-body Hamiltonians, possibly, as a first approximation, by modeling them with random matrices. Instead of using conventional tools for this purpose, such as the inverse participation ratio of these eigenstates in a complete basis, we have employed the natural geometry of the space of quantum states induced by the Hermitian inner product, which measures the infinitesimal distance between two states. Here, we focus specifically on random matrix Hamiltonians, which are drawn from ensembles that are more general than those of the three classical Gaussian ensembles and show a chaos-to-integrability transition. In particular, we have focused our attention on two generalisations of the classical Gaussian ensembles, namely the RP random matrix ensemble and the tridiagonal Gaussian $\beta$-ensembles, both of which show chaos-to-integrability transition in the level spacing distribution of the energy eigenvalues and break the rotational invariance of the classical Gaussian ensembles.\footnote{We have also considered a specific form of the $\beta$-ensemble which keeps the invariance intact.} It is possible to obtain the (suitably averaged) QMT on the parameter space (in some cases analytically) and explore how it encodes different characteristic properties of the eigenstates of the random matrix Hamiltonian under consideration. 

We now briefly summarise the important conclusions and lessons obtained from the results presented in this paper.

To analyse the geometry of quantum states for chaotic and integrability-breaking Hamiltonians beyond the classical Gaussian ensembles and the RP model, we have introduced two classes of Hamiltonians based on the tridiagonal $\beta$-Gaussian ensembles. Specifically, the one-parameter class of Hamiltonians \eqref{beta_int_break}, where the matrix proportional to the parameter $r$ is drawn from a tridiagonal $\beta$-Gaussian ensemble with Dyson index $\beta$, has been employed to compute an exact analytical expression for the EFS for any generic value of the Dyson index $\beta>1$, and subsequently, show that the EFS diverges in the limit the system integrable (i.e., $r \rightarrow 0$ for our model). Furthermore, we derived the scaling relation for the EFS for large $r$ from this analytical expression, and showed that the $1/r^4$ is consistent with the numerical simulations for large-dimensional matrices.

On the other hand, the two-parameter class of Hamiltonians, \eqref{Ham_Roz_Por_beta}, where all the three component operators are drawn from independent tridiagonal $\beta$-Gaussian ensembles with generic Dyson indices, was used to compute the averaged QMT components numerically, and subsequently derive the scaling relations for the components in the chaotic phase with large $r$. 

The non-zero non-diagonal component of the QMT is one of the main distinguishing features between the cases with rotationally invariant ensembles (such as the GUE) and the tridiagonal $\beta$-Gaussian ensemble. This fact, along with the specific scalings of the QMT components for the two-parameter Hamiltonian constructed from the $\beta$-ensemble (listed in eq. \eqref{QMT_scaling}),  implies that the QMT becomes approximately degenerate in the large $r$ limit. Even though the last statement is true for the GUE as well, we have argued that the direction in the parameter manifold along which the QMT becomes degenerate (i.e., the direction specified by the eigenvector associated with the eigenvalue that approaches zero) is different from the GUE case, and hence is actually a 'mixture' of radial and angular directions.

Next, we extend these computations for a class of $\beta$-Gaussian ensembles with an effective Dyson index that is also rotationally invariant (unlike the tridiagonal $\beta$-Gaussian ensembles). We showed that the conclusion that the EFS diverges as one moves close to the integrable point is also true for this case, as well (see also the computation of the EFS 
in Appendix \ref{sec:beta_ensm_haar} for an ensemble having random eigenvectors and eigenvalues drawn from a Gaussian $\beta$-ensemble). 

Apart from the divergence of EFS as the pure integrable point is approached, another interesting point about the EFS for the integrability model we have studied, perhaps worth mentioning, is the generality of the expression for EFS for the $2 \times 2$ as a function of $r$ when the Dyson index of the coefficient of the parameter $r$ is taken to be $\beta=2$. In fact for all the three general class of such examples we have considered: for the tridiagonal $\beta$-Gaussian ensemble (in \S \ref{ssec_grr_trid_der}), an invariant $\beta$-ensemble with an effective Dyson index and another ensemble with Haar random eigenvectors (in Appendix \ref{sec:beta_ensm_haar}) all reproduce the expression for the $rr$ component of the metric in \eqref{metric_int_brk}.\footnote{This, for the second case, is not surprising due to the nature of the construction of this ensemble.}

Finally, in Section \ref{sec_level_curvature}, we have computed a closely related quantity to that of the FS, namely, the energy level curvature, which is the second derivative of the energy levels with respect to the parameters. We computed the explicit expression for the energy level curvature for the $ 2 \times2$ diagonal plus tridiagonal $\beta$-Gaussian matrix, and showed that, similar to the EFS, the energy level curvature also diverges (even though logarithmically, compared to the inverse power law for EFS) for this model in the limit of integrability.


We conclude by pointing out a few interesting future directions along which the results presented in this paper can be extended. 


1. In this paper, we mostly consider the Gaussian weight function, i.e., potentials of the form  $V(H)=H^2$. However, one can consider more general weight functions, for which the ensemble would still retain the rotational invariance of the classical Gaussian ensembles (even though the individual elements of the Hamiltonian matrix will not be independent) and study the ensemble-averaged QMT and its related quantities for such ensembles. 

2. Finally, in this work, we have assumed that the Hamiltonian governing the dynamics is Hermitian and consequently the `curvature' of the parameter manifold is induced by the Hermitian inner product, which was reflected in the behaviour of the geometry in the integrability to chaotic transitions. However, this feature will no longer be true in the cases where the dynamics of the many-body systems are governed by the non-Hermitian Hamiltonian, and in fact, the state-space geometry will be far richer than the one considered here.
Extending these results to the non-Hermitian cases, it would be interesting to investigate how the Ricci scalar of the QMT behaves under transitions between non-Hermitian integrable and chaotic regimes, in particular, since, unlike the standard Hermitian case, the bi-orthogonal nature of the inner product induces a non-Hermitian tensor structure on the base manifold, it would be interesting to study the curvature structure of such models.

\begin{center}
	\bf{Acknowledgments}
\end{center}
It is a pleasure to thank Hugo A. Camargo, Yichao Fu, and Viktor Jahnke for many insightful discussions and for comments on a draft version of the manuscript. The work of Kunal Pal is  supported by the YST Program at the APCTP through the Science and
Technology Promotion Fund and Lottery Fund of the Korean Government. This was also supported by the Korean Local Governments -
Gyeongsangbuk-do Province and Pohang City. Kunal Pal would like to thank the Gwangju Institute of Science and Technology for the hospitality where a part of this work was carried out.
This work was supported by the National Research Foundation of Korea(NRF) grant funded by the Korea government(MSIT)(RS-2026-25477372, RS-2025-02311201, RS-2025-02307394, RS-2024-00445164), and by the Creation of the Quantum Information Science R\&D Ecosystem (Grant No. 2022M3H3A106307411, RS-2023-NR068116) This work was also supported by GIST research fund (Future leading Specialized Resarch Project, 2026 and the ANCHOR program through the Gwangju ANCHOR Center, funded by the Ministry of Education(MOE) and the Jeonnam-Gwangju Special Metropolitan City, Republic of Korea.(2026-ANCHOR-05-001)
Ankit Gill, Kunal Pal, and Kuntal Pal contributed equally to this paper and should be considered as co-first authors. 

\appendix

\section{QGT for a generic superposition of eigenstates}\label{qmt_generic}
For completeness, here, we record the expression for the QGT 
\begin{equation}
    g_{ab} (\mbr) = \braket{\partial_a \Psi(\mbr)|\partial_b \Psi(\mbr)}-\braket{\partial_a \Psi(\mbr)|\Psi(\mbr)}\braket{\Psi(\mbr)|\partial_b \Psi(\mbr)}~,
\end{equation}
for a generic superposition of eigenstates, $\ket{\Psi(\mbr)}=\sum_nc_n \ket{n(\mbr)}$, where $c_n$s are complex numbers independent of the set of parameters $\{\mbr\}$ present in the total Hamiltonian.  A straightforward calculation shows that the expression for the QGT can be written as 
\begin{equation}
\begin{split}
    g_{ab} (\mbr) = \sum_{n,m} c_n^*c_m ~\Bigg(\sum_{k}(1-|c_k|^2) \braket{\partial_a n(\mbr)|k(\mbr)} \braket{k(\mbr)|\partial_b m(\mbr)} \\-\sum_{k \neq l} c_kc_l^* \braket{\partial_a n(\mbr)|k(\mbr)} \braket{l(\mbr)|\partial_b m(\mbr)}\Bigg)~.
\end{split}
\end{equation}
Specifically, for the state $\ket{\Psi(\mbr)}=N^{-1/2}\sum_n  \ket{n(\mbr)}$, the components of QGT are given by
\begin{equation}
\begin{split}
    g_{ab} (\mbr) = \frac{1}{N}\sum_{n,m} ~\Bigg(\sum_{k}\Bift{1-\frac{1}{N}}\braket{\partial_a n(\mbr)|k(\mbr)} \braket{k(\mbr)|\partial_b m(\mbr)} \\-\sum_{k \neq l} \frac{1}{N}\braket{\partial_a n(\mbr)|k(\mbr)} \braket{l(\mbr)|\partial_b m(\mbr)}\Bigg)~.
\end{split}
\end{equation}
As can be seen, the last two expressions, in general, represent different quantities from that of the average QGT over all the eigenstates, i.e., eq. \eqref{average_QGT}.

\section{Derivation of QMT component for GUE Hamiltonian}\label{gue_qmt_derivation}
In this Appendix, we briefly present the derivation of the $rr$ component 
of the ensemble-averaged  QMT ($\bar{g}_{rr}$ in \eqref{metric_rmt}) for the GUE Hamiltonian in \eqref{Ham_Roz_Por_GUE}.\footnote{ We note that the derivation 
presented in \cite{Sharipov:2024lah} contains a few typographical errors.} We start by writing the first expression in \eqref{gue_qmt_comp}
explicitly (since $H=H_0+r \tilde{\mh}_1$),
\begin{widetext}
 \begin{equation}
\begin{split}
   \bar{g}_{rr}(r)= \frac{1}{NZ}\sum_{m \neq n} \int ~ \frac{(\tilde{\mh}_1)_{nm}(\tilde{\mh}_1)_{mn}}{(E_n(r)-E_m(r))^2} ~
    \exp\Bigg[-\frac{1}{2 \sigma^2} \Big(\text{Tr} (H_0^2)+\text{Tr} (\tilde{\mh}_1^2)\Big)\Bigg]~\td H_0~\td \tilde{\mh}_1~,
    \end{split}
\end{equation}   
\end{widetext}
where we have performed the trivial integration over $\tilde{\mh}_2$. Now performing two successive variable changes: first using $H_0=H-r \tilde{\mh}_1$  
and subsequently $\tilde{\mh}_1=\frac{H_1}{\sqrt{r^2+1}}+\frac{r}{r^2+1} H$, we can rewrite the above expression as,\footnote{All the Jacobian factors associated with these transformations are constant, and hence cancel from the same factors form the integral in $Z$ in the denominator.} 
\begin{widetext}
  \begin{equation}
\begin{split}
    \bar{g}_{rr}(r)= \frac{1}{NZ(r^2+1)}\sum_{m \neq n} \int ~ \frac{(H_1)_{nm}(H_1)_{mn}}{(E_n(r)-E_m(r))^2} ~
    \exp\Bigg[-\frac{1}{2 \sigma^2} \Bigg(\frac{\text{Tr}(H^2)}{r^2+1}+\text{Tr} (H_1^2)\Bigg)\Bigg]~\td H~\td H_1~.
    \end{split}
\end{equation}  
\end{widetext}
Note that the motivation behind these particular coordinate changes is to write the total probability distribution in a form which factorises in two variables, one of which should be the total Hamiltonian $H$. 
Also note that, with the conventions fixed at the beginning of \S \ref{ssec_GUE_QMT}, the variance of the variable $H_1$ is $\sigma^2$. We can perform the integral over $H_1$ by using the standard formula for the correlator of the GUE matrices, written in terms of a complete orthonormal basis in the Hilbert space $\ket{\alpha}$: $\mathbb{E}\big((H_1)_{\alpha \beta}(H_1)_{\mu\nu}\big)=\sigma^2 \delta_{\alpha\nu}\delta_{\beta\mu}$. Since we have fixed $\sigma^2=1/N$, we have,
\begin{widetext}
  \begin{equation}
\begin{split}
    \bar{g}_{rr}(r)= \frac{1}{N^2Z(r^2+1)}\sum_{m \neq n} \int ~ \frac{1}{(E_n(r)-E_m(r))^2} ~
    \exp\Bigg[-\frac{1}{2 \sigma^2} \Bigg(\frac{\text{Tr}(H^2)}{r^2+1}\Bigg)\Bigg]~\td H~~
    =\frac{1}{N^2(r^2+1)}\sum_{m \neq n}  \mathbb{E}~ \Bigg[\frac{1}{(E_n(r)-E_m(r))^2}\Bigg]_{\text{GUE}(\sigma^2_\star)}~,
    \end{split}
\end{equation}  
\end{widetext}
where $~\sigma^2_\star=\sigma^2(r^2+1)$.
The expression in the final step above can be evaluated using a result obtained by Dyson \cite{dyson1962brownian}, known as the virial theorem associated with the Brownian motion approach to the eigenvalue distribution of a Gaussian random matrix. For a classical Gaussian random matrix ensemble with a Dyson index $\beta>1$, and variance $\sigma^2_\star$, the virial theorem states \cite{dyson1962brownian}, 
\begin{equation}\label{virial_th}
    \sum_{m \neq n}  \mathbb{E}~ \Bigg[\frac{1}{(E_n-E_m)^2}\Bigg]_{\text{GUE}(\sigma^2_\star)}~ = \frac{N(N-1)}{2\sigma^2_\star \beta(1-\beta^{-1}) }~.
\end{equation}
Now using this relation, along with the fact that here $\sigma^2_\star=\frac{r^2+1}{N}$, we get the required expression 
\begin{equation}
     \bar{g}_{rr}(r)= \frac{N-1}{2 (r^2+1)^2}~.
\end{equation}
By following a similar analysis, one can obtain the other components of the QMT in \eqref{metric_rmt}.

\section{Approximate density of states of the tridiagonal $\beta$-RP model }\label{sec:DOS_approx}
In this Appendix, we shall discuss an analytical approach for calculating the approximate DOS of the Hamiltonian in \eqref{beta_int_break}, using the Cauchy transform of the DOS, 
\begin{equation}
    G_{H}(z)=\int_{\mc{S}} \frac{\rho_{H}(E)}{z-E}~.
\end{equation}
In the large $N$ limit, assuming asymptotic freeness between $H_0$ and $H_\beta$, we can derive the following series expansion for the Cauchy transform,
\begin{equation}\label{G_H_series}
    G_{H}(z)= \sum_{n} \frac{(-1)^n}{n!}r^{2n}\Bitd{G_{H}(z)}^n \partial_z^nG_{H_0}(z)~,
\end{equation}
where $G_{H_0}(z)$ denotes the Cauchy transform of the DOS of $H_0$.  We refer to  \cite{Venturelli:2022hka, Jahnke:2025exd, DOSreview} for the derivation of this series expansion.

As can be seen from the last expansion, when $r<1$, it acts as a small parameter, and this series can be solved for the Cauchy transform $G_{H(z)}$ order-by-order in $r$. At the zeroth order, DOS of $H$ is the same as the DOS of $H_0$, i.e., a Gaussian distribution, whereas the first correction in $r$ to $G_{H(z)}$ due to the `perturbation' operator $H_\beta$ is given by
\begin{align}\label{GH_appr_1st}
    G_{H}(z) &\approx G_{H}^{(1)}(z) +\mc{O}(r^4)~, \nonumber\\
    & G_{H}^{(1)}(z)  = G_{H}^{(0)}(z) - r^2 G_{H}^{(0)}(z) \partial_z G_{H}^{(0)}(z)~,
\end{align}
where, $G_{H}^{(0)}(z)=G_{H_0}(z)$. Note that the first correction in the Cauchy transform of $H$ contains a term proportional to $r^2$, and there is no term linear in $r$. Hence, for $r \ll 1$, this would give quite a good approximation to the DOS of $H$. 
For our case, the analytical expressions for $G_{H_0}(z)$ and $G_{H}^{(1)}(z) $ can be written in terms of the imaginary error function as
\begin{equation}
    G_{H_0}(z)= \sqrt{\frac{\pi}{2 \sigma_0^2}} e^{-\frac{z^2}{2 \sigma_02}}\mc{A}(z)~,~~\mc{A}(z)=i+\text{erfi}\Bift{\frac{z}{\sqrt{2}\sigma_0}}~,
\end{equation}
and 
\begin{equation}
    G_{H}^{(1)}(z) = \frac{1}{2 \sigma_0^2}e^{-\frac{z^2}{ \sigma_02}} \mc{A}(z) \Bitd{\sqrt{2 \pi} (\sigma_0^2-r^2)e^{\frac{z^2}{\sigma_02}}+\pi zr^2 \mc{A}(z)}~.
\end{equation}

    \begin{figure}
      \centering
       \subfloat[]{\includegraphics[width=0.8\columnwidth]{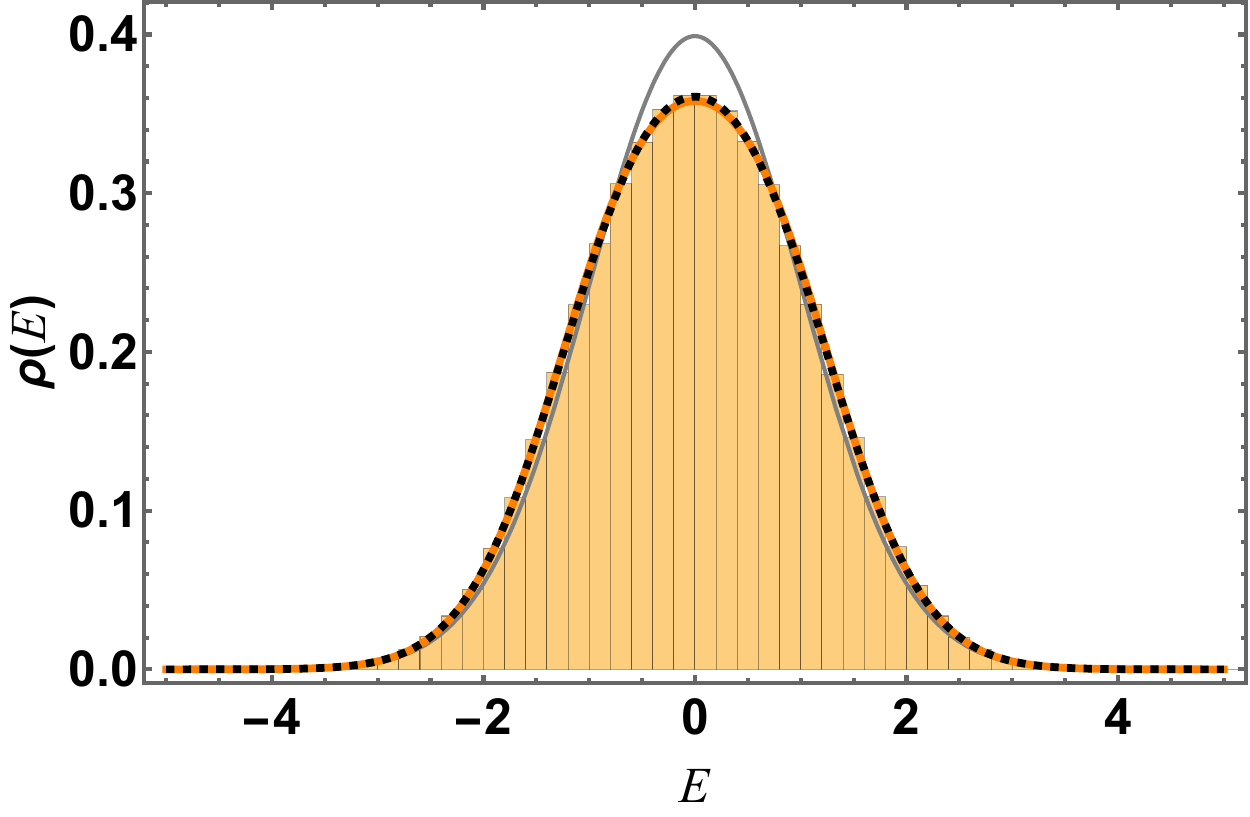}}
        \hfill
       \subfloat[]{\includegraphics[width=0.8\columnwidth]{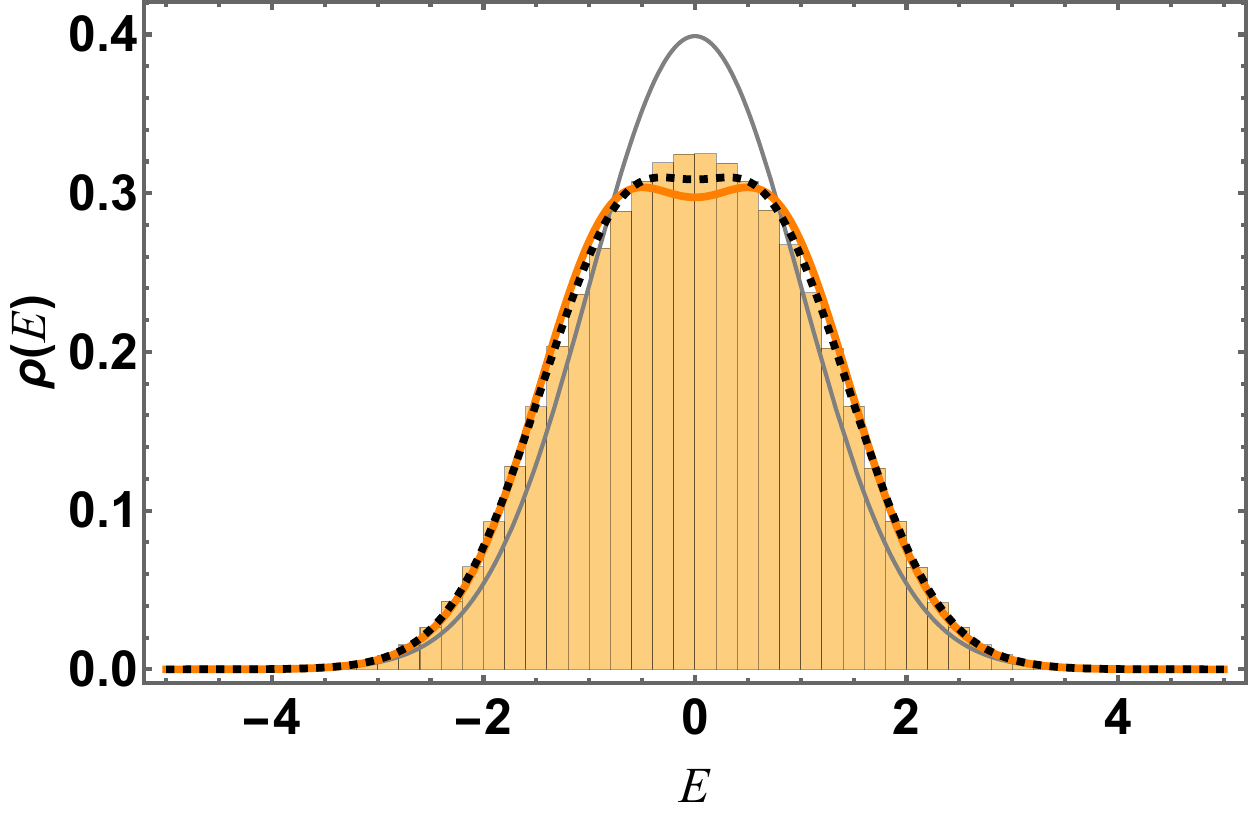}}
        \caption{DOS of the tridiagonal $\beta$-Gaussian ensemble version of the RP model (the Hamiltonian in \eqref{beta_int_break}) for different values of the parameter $\tiga$ in the range $0<\tiga <1/2$: $\tiga=0.15$ (a), and $\tiga=0.09$ (b). The grey curves indicate a Gaussian distribution with mean zero and variance unity. The orange and black dashed curves, respectively, represent the DOS obtained from the first and second corrections to the Cauchy transform. Here we have set $N=2000$, $\beta=2$ and the results are shown after an average over $500$ independent realisations of the Hamiltonian.  }
      \label{fig:DOS_approx}
   \end{figure}

We can use the first-order corrected Cauchy transform to derive the second correction to $G_{H}^{(1)}(z) $. From the series in \eqref{G_H_series}, this can be derived to be 
\begin{align}\label{GH_appr_2nd}
      G_{H}(z) &\approx G_{H}^{(2)}(z) +\mc{O}(r^6)~, \nonumber\\
     G_{H}^{(2)}(z)  &= \Bitd{G_{H}^{(0)}(z) - r^2 G_{H}^{(0)}(z) \partial_z G_{H}^{(0)}(z)} \nonumber\\
    &+\frac{r^4}{2} \Bitd{G_{H}^{(0)}(z) - r^2 G_{H}^{(0)}(z) \partial_z G_{H}^{(0)}(z)}^2 \partial_{zz}G_{H}^{(0)}(z) ~.
\end{align}
Here we have written all the terms in terms of the Cauchy transform of the DOS of $H_0$ to emphasise that knowing the DOS of this term can be used to determine the total Hamiltonian DOS. However, the explicit expression for $G_{H}^{(2)}(z) $ in terms of the imaginary error function is quite lengthy, and we do not provide it here for clarity. 

The comparison of the DOS obtained from the first and second corrections to the Cauchy transform (respectively, eqs.  using the \textit{Stieltjes inversion formula} 
\begin{equation}\label{dos_from_G}
	\rho_{H}(\lambda) = -\frac{1}{\pi}  \lim_{\epsilon \rightarrow 0^+} (\text{Im} ~ G_{H}(\lambda+ i \epsilon))\,.
\end{equation}
and the exact numerical simulation is shown in Fig. \ref{fig:DOS_approx} for two different values of $\tiga$. As can be seen, the approximations provide a very good description of the numerical DOS. Furthermore, as expected,  the approximations work better as $\tiga$ moves closer to $1/2$, specifically, as shown in panel (a) of Fig. \ref{fig:DOS_approx}, for $\tiga=0.15$, the DOS from the first and second corrections are almost identical, and they match exactly with the numerical result. On the other hand, for a slightly smaller value of $\tiga$, which is closer to the chaotic phase, e.g., $\tiga=0.09$ as in panel (b), the DOS obtained from the first and the second corrections differ in a small region in the middle of the spectrum. They also differ from the numerical histogram, even though the second correction does provide a better description of the numerical results, compared to the first-order correction.

As should be clear, the above-described approximation scheme successively computes the correction of the initial Gaussian distribution of the operator $H_0$ due to the perturbation operator $\mh_\beta$. For a finite value of $N$, if one sufficiently decreases  $\tiga$ below a certain threshold, such that $\mh_\beta$ can not be considered as a perturbation, this approximation scheme would fail, and the DOS can not be expressed as a deformation of the initial Gaussian. 

Before concluding this section, we emphasise that, although this procedure of approximating the DOS of Hamiltonian $H(r)=H_0+r \mh_\beta$ for $0 < \tiga < 1/2$ using free probabilistic techniques gives an analytical understanding of the same, we have not justified the assumption of freeness between $H_0$ and $H_\beta$. Due to the absence of rotational invariance in the $\beta$ ensemble, this is not obvious, since, roughly, the freeness requires that the eigenvectors of $H_\beta$ be in a generic position with respect to the eigenvectors of $H_0$ \cite{nica2006lectures, mingo2017free}. However, a detailed analysis of this question is beyond the scope of the present paper, and we hope to address it in future.

\section{Ensemble-averaged fidelity susceptibility for a $2 \times 2$ random matrix ensemble}\label{sec:beta_ensm_haar}

In this Appendix, we construct a specific random matrix ensemble, an element of which has the eigenvalue distribution of that of a Gaussian ensemble with generic Dyson index $\beta$ and has Haar random eigenstates, and we use it to compute the EFS, both for the case of ensembles with only chaotic phase, and also one showing a chaos-to-integrability transition. 

We consider the following two $2 \times 2$ diagonal matrices with real entries, 
\begin{equation}
    H_0 =\left(
		\begin{array}{ccc}
			h_1 & 0 \\
			0  & h_2
		\end{array}
		\right)~,~~ V =\left(
		\begin{array}{ccc}
			v_1 & 0 \\
			0  & v_2
		\end{array}
		\right)~,
\end{equation}
and consider rotating one of them by a random unitary matrix sampled from the Haar unitary group, 
\begin{align}
    U =\left(
		\begin{array}{ccc}
			e^{i u_1} \cos{\theta} & -e^{i u_2} \cos{\theta} \\
			e^{-i u_2} \cos{\theta}  & e^{-i u_1} \cos{\theta} 
		\end{array}
		\right)~.
\end{align} 
In the following, we assume that the JPDs of eigenvalues of the matrices $H_0$ and $V$ are independent, and are drawn from the JPD \eqref{joint_eig_dist} of the Gaussian $\beta$-ensemble. 
The total Hamiltonian of the ensemble we consider is given by
\begin{equation}
    H(r)= H_0 + r ~U V U^\dagger~.
\end{equation}
In this representation, therefore, the total Hamiltonian $H(r)$ and the term $\mh=U V U^\dagger$ are written, without loss of generality, in terms of the eigenbasis of the matrix $H_0$. We performed a random rotation of the matrix $V$ to make its eigenvectors generic with respect to those of $H_0$. Note that the total Hamiltonian can be decomposed into two parts, one of which is proportional to the identity matrix, and a generic non-diagonal part which depends on the parameters of the unitary matrix $U$. 

\begin{figure}[h!]
		\centering
		\includegraphics[width=3in,height=2.3in]{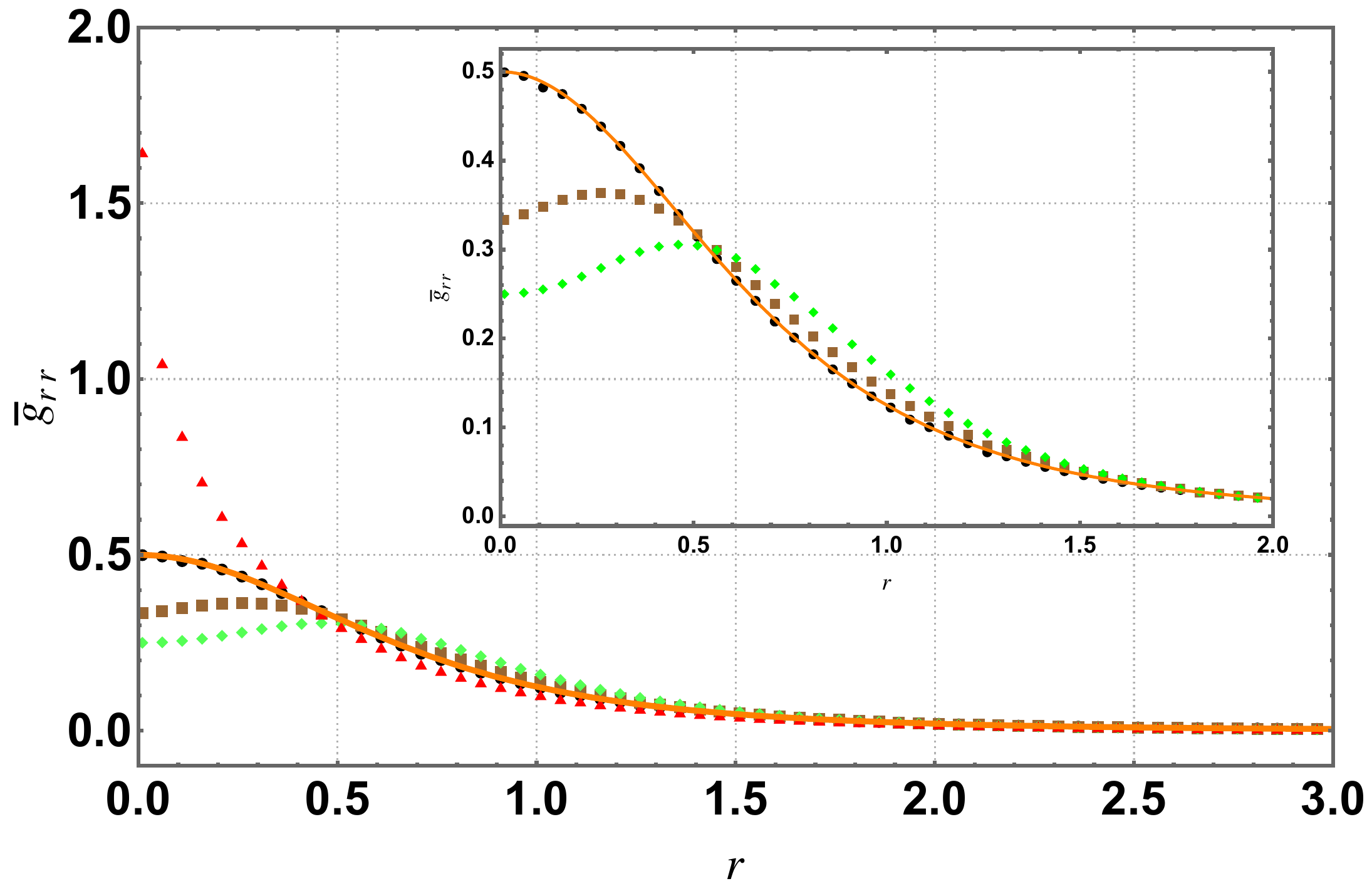}
		\caption{Plot of $\bar{g}_{rr}(r)$ for a $2 \times 2$  Hamiltonian of the form $H(r)= H_0 + r ~U V U^\dagger$, where eigenvalues of both the matrices $H_0$ and $V$ have correlations, and are drawn from JPD of the Gaussian $\beta$-ensemble.  We have shown the results when eigenvalues of both these matrices follow JPD in \eqref{joint_eig_dist} with the same Dyson index, and for several different cases, $\beta=1$ (red), $\beta=2$ (black), $\beta=3$ (brown), and $\beta=5$ (green). The solid orange curve shows the exact GUE expression in eq. \eqref{metric_rmt} with $N=2$. As the value of the Dyson index is decreased, and gradually the limit $\beta \rightarrow 0$ is approached, the $\bar{g}_{rr}(r)$ diverges when $r \rightarrow0$. The  plot in the inset shows more clearly the behaviour for large values of $\beta \geq 2$.}
		\label{fig:grr_BETA_haar}
	\end{figure}

\textbf{Case 1.} First, consider the case where the JPDs of both $h_i$ and $v_i$ are the same, i.e., the ensembles from which these matrices are drawn have the same Dyson index. 
The EFS computed from this Hamiltonian is given by (with the notation $h_{12}=h_1-h_2$ and $v_{12}=v_1-v_2$)
\begin{widetext}
    \begin{equation}
\begin{split}
    \bar{g}_{rr}(r)= \frac{1}{Z} \int~ \frac{h_{12}^2 v_{12}^2 \cos{\theta}^2 \sin{\theta}^2}{\Big(h_{12}^2+v_{12}^2 r^2 + 2 r h_{12} v_{12} \cos{2\theta} \Big)^2}~ P(h_1,h_2) ~ P(v_1,v_2) ~\td h_1 \td h_2 ~\td v_1 \td v_2 ~ \sin{2\theta} ~\td \theta~.
\end{split}
\end{equation}
\end{widetext}
Here, $P(h_1,h_2)$ and $P(v_1,v_2)$ are the JPDs of the eigenvalues in eq. \eqref{joint_eig_dist} 
having a general value of the Dyson index $\beta$. We assume that both matrices are drawn from ensembles with the same Dyson index. 
Performing coordinate transformations $h_1-h_2=\varrho \sin{\varphi}$ and $v_1-v_2=\varrho \cos{\varphi}$, we see that all the integrals, except the one over $\varphi$, can be performed analytically. Finally, evaluating the resulting integral over $\varphi$ numerically, we plot the results for different values of $\beta$ in  Fig. \ref{fig:grr_BETA_haar}. For $\beta=2$, the result matches exactly 
with the GUE expression in eq. \eqref{metric_rmt} with $N=2$. 
Changing values of the Dyson index, we see that decreasing the value of  $\beta$, the FS keeps increasing for small values of $r$, and in the limit $\beta \rightarrow0$ the FS diverges as $ r \rightarrow0$. This conclusion is consistent with the ones in Section \ref{sec_beta_metric}. 

	\begin{figure}[h!]
		\centering
		\includegraphics[width=3in,height=2.3in]{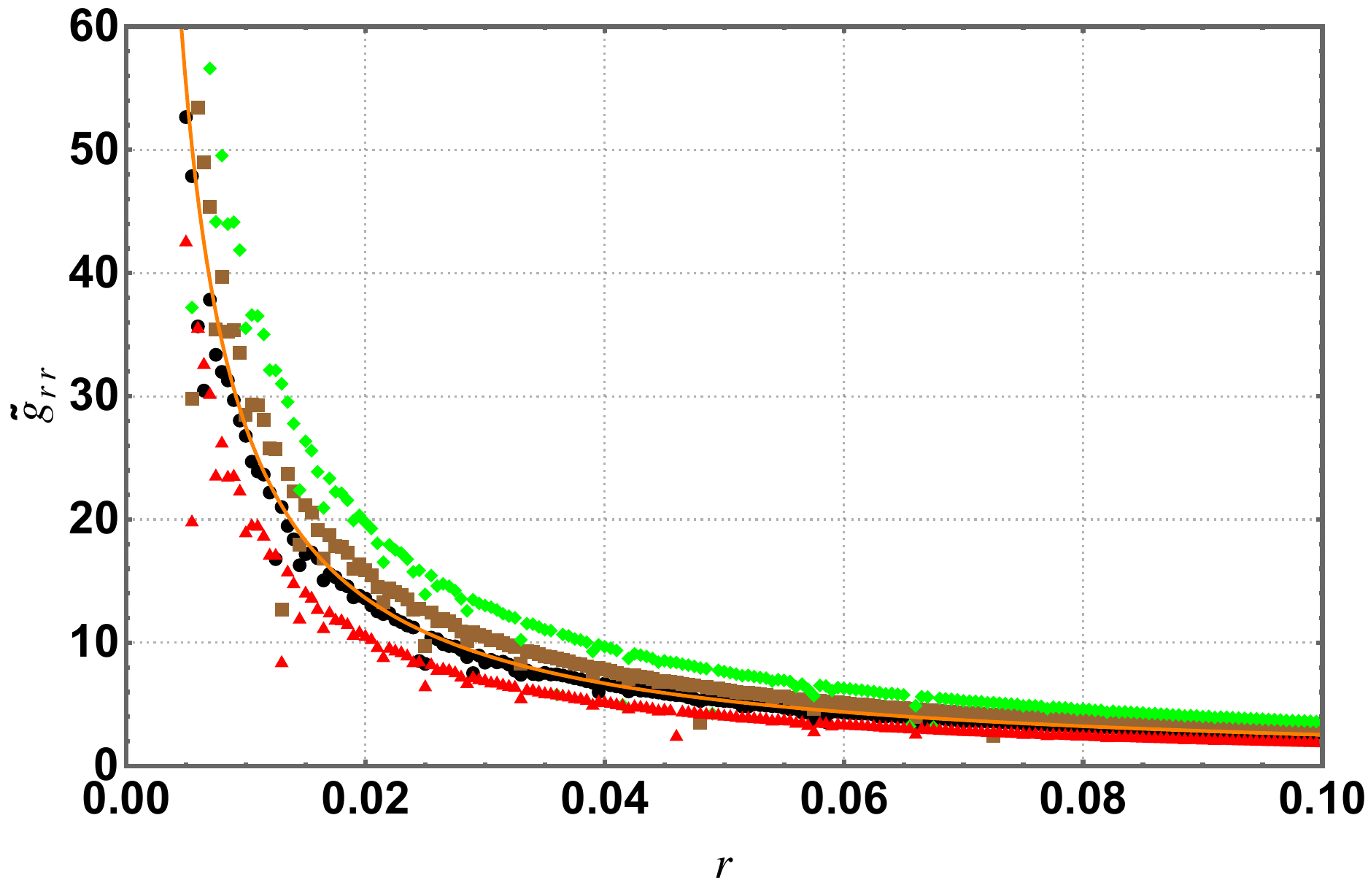}
		\caption{Plot of $\tilde{g}_{rr}(r)$ for a $2 \times 2$  Hamiltonian of the form $H(r)= H_0 + r ~U V U^\dagger$, where eigenvalues of the matrices $H_0$ are drawn from independent Gaussian distributions, while those of $V$ have correlations, and are drawn from JPD of the Gaussian $\beta$-ensemble.  We have shown the results for different values of the Dyson index,  $\beta=1$ (red), $\beta=2$ (black), $\beta=3$ (brown), and $\beta=5$ (green). The solid orange curve represents the exact expression in eq. \eqref{metric_int_brk}, which matches exactly with the $\beta=2$ case numerically obtained from this ensemble. 
        }
	\label{fig:grr_BETA_haar_int}
	\end{figure}

\textbf{Case 2.} Next, we consider the situation where the JPDs of $h_i$ and $v_i$ are different, in particular, we assume that the eigenvalues of the matrix $H_0$ are drawn from independent Gaussian distributions with mean zero and unit variance, while the JPD of eigenvalues of $V$ (i.e., $P(v_1,v_2)$) follows that of Gaussian $\beta$-ensemble with generic $\beta$. Once again,  by performing coordinate transformations $h_1-h_2=\varrho \sin{\varphi}$ and $v_1-v_2=\varrho \cos{\varphi}$, one can perform all the integrals appearing in $\bar{g}_{rr}(r)$, apart from the one over $\varphi$, which can be easily done numerically. The resulting FS is shown in Fig. \ref{fig:grr_BETA_haar_int}. When $\beta=2$, the result matches with the expression in \eqref{metric_int_brk}. Furthermore, as in the case of ensembles discussed in Sections \ref{sec_beta_metric} and \ref{sec_invt_beta_ensemb}, the EFS diverges in the limit $r\rightarrow0$, i.e., as the Hamiltonian approaches the integrable phase, while it goes to zero in the large $r$ limit. 

\pagebreak

\bibliography{reference}

\end{document}